\documentclass{aa}  

\usepackage{wasysym}  
\usepackage{graphicx}

\usepackage[maxfloats=256]{morefloats}
\usepackage[varg]{txfonts}
\usepackage[colorlinks=true, citecolor=blue]{hyperref}
\usepackage{natbib}
\bibpunct{(}{)}{;}{a}{}{,}

\begin{document}

   \title{Detection of open cluster rotation fields\\from Gaia EDR3 proper motions}
   \author{P. Guilherme-Garcia\inst{1}, A. Krone-Martins\inst{2,1}, A. Moitinho\inst{1}
        }

    \institute{
             CENTRA, Faculdade de Ci\^encias, Universidade de Lisboa, Ed. C8, Campo Grande, 1749-016 Lisboa, Portugal\\
          \email{andre@sim.ul.pt}
        \and
         Donald Bren School of Information and Computer Sciences, University of California, Irvine, 
         Irvine CA 92697, USA    
    }
   \date{Received 3 December 2021; accepted 31 March 2023}

\abstract{
Most stars from in groups which with time disperse, building the field population of their host galaxy. In the Milky Way, open clusters have been continuously forming in the disk up to the present time, providing it with stars spanning a broad range of ages and masses. Observations of the details of cluster dissolution are, however, scarce.  One of the main difficulties is obtaining a detailed characterisation of the internal cluster kinematics, which requires very high-quality proper motions. For open clusters, which are typically loose groups with tens to hundreds of members, there is the additional difficulty of inferring kinematic structures from sparse and irregular distributions of stars.
     }{
Here, we aim to analyse internal stellar kinematics of open clusters, and identify rotation, expansion, or contraction patterns.
     }{
We use Gaia Early Data Release 3 (EDR3) astrometry and integrated nested Laplace approximations to perform vector-field inference and create spatio-kinematic maps of 1237 open clusters. The sample is composed of clusters for which individual stellar memberships were already known, thus minimising contamination from field stars in the velocity maps. Projection effects were corrected using EDR3 data complemented with radial velocities from Gaia Data Release 2 and other surveys.
     }{
We report the detection of rotation patterns in eight open clusters. Nine additional clusters display possible rotation signs. We also observe 14 expanding clusters, with 15 other objects showing possible expansion patterns. Contraction is evident in two clusters, with one additional cluster presenting a more uncertain detection. In total, 53 clusters are found to display kinematic structures. Within these, elongated spatial distributions suggesting tidal tails are found in five clusters.
These results indicate that the approach developed here can recover kinematic patterns from noisy vector fields, as those from astrometric measurements of open clusters or other stellar or galactic populations, thus offering a powerful probe for exploring the internal kinematics and dynamics of these types of objects.
     }{}
     

\titlerunning{Detection of Open Cluster rotation fields from Gaia EDR3 proper motions}
\authorrunning{P. Guilherme-Garcia et al.}
\maketitle

\section{Introduction}  \label{Introduction}    


Explaining how galaxies build up is one of the central quests in astrophysics. In this context, most stars are believed to be formed in clusters \citep[e.g.][]{2003ARA&A..41...57L, 2012MNRAS.426.3008K}, which with time ultimately disintegrate, building up the galactic field population. 
In the Milky Way, globular clusters (GCs), which were formed in the early stages of our galaxy, are one of the contributors to populating the halo. In contrast, open clusters (OCs) and associations have been continuously forming in the disk for over the last $\sim$ 10 Gyrs, enriching it with stars spanning a broad range of ages and masses.

As groups of stars resulting from the gravitational collapse and fragmentation of a parent molecular cloud, cluster stars can remain bound for some time under the balance of their collective gravitational field and the pressure arising from their dynamics. Analytical and numerical N-body simulations \citep[e.g.][]{2, 2008MNRAS.389L..28G} have shed light on how clusters would then evolve, showing how factors such as encounters with giant molecular clouds and spiral arms, galactic tidal forces, and secular evolution (also referred to as evaporation) lead to quick disruption or gradual dissolution of star clusters. 

On the observational side, studying the disintegration process poses significant challenges.
On the one hand, stars that have been stripped or those that have escaped to the field and no longer belong to the cluster become distributed in low-brightness halos and tails, which are hard to detect observationally. 
On the other hand, detailed kinematic characterisation of the remaining cluster has traditionally been difficult due to the small stellar relative proper motions (except for the closest clusters), which limited studies to radial velocity line-of-sight studies and/or being seen in crowded fields.

Despite the challenges, some observational studies have yielded detections of these elusive patterns. Examples include the detection of tidal tails 
in OCs \citep[e.g.][]{2001A&A...377..462B, 
2010ApJ...711..559D, 2015MNRAS.449.1811D} 
as well as tens of GCs \citep[e.g.][]{1995AJ....109.2553G, 2000A&A...359..907L, 2010AJ....139..606C, 2010ApJ...721.1790C, 2010A&A...522A..71J, 2018MNRAS.474..683C},  
and rotation of GCs \citep[e.g][]{2000A&A...360..472V, 2003AJ....126..772A, 2006A&A...445..513V,
2013ApJ...779...81M, 2017ApJ...844..167B} in proper motions.


These difficulties are now being gradually overcome thanks to the European Space Agency (ESA) Gaia mission \citep{gaia}. Gaia is one of the most ambitious astronomical all-sky surveys from space today. The main objective of the mission is to bring a better understanding of the formation and evolution of the Milky Way. To this end, Gaia has already released a succession of the deepest, most accurate, and complete all-sky astrometric and photometric catalogues ever \citep{dr1, dr2, edr3}.

With Gaia, recent studies have now detected rotation patterns in over 20 GCs \citep[e.g.][]{2018MNRAS.481.2125B, 2019MNRAS.485.1460S, 2021MNRAS.505.5978V, 2021MNRAS.506..813D, 2021MNRAS.504.1144S}.  Open clusters, however, typically have many fewer members, ranging from tens of members to a few OCs with over a thousand identified members \citep{dias, upmask1, upmask2, rv_source}. This leads to sparser distributions, making the detection of spatial and kinematic patterns much harder. Even more so considering that OCs are often seen against the crowded background of the Galactic disk. It is thus impressive how the high quality of the Gaia data is now easily revealing tails and coronae in OCs \citep[e.g.][]{2019A&A...621L...3M, 2021A&A...645A..84M}. Still, even with Gaia, very few measurements of OC rotation have been accomplished: In their kinematic study of 28 OCs using Gaia Data Release~2 (DR2), \citet{Kuhn_2019} conclude that only one OC displayed signs of rotation; \citet{2020AN....341..638L} also using Gaia DR2 measured the rotation of Praesepe. Thorough searches in the literature have not revealed other examples, indicating that if there are more published determinations the number must be low.


Concurrently with the new availability of high-quality data in huge volumes, from Gaia and other surveys, we are also witnessing an explosion of advanced statistical and computation methods together with the necessary computing power. These new methods, or novel applications of older methods are both bringing new insights and enabling the analysis of very large data sets.


The focus of this work is to assess the dynamical state of large numbers of OCs, namely identifying signatures of rotation as well as expansion and contraction detectable with Gaia Early Data Release~3 \citep[][hereafter EDR3]{edr3}. For this, we developed a procedure for reconstructing the velocity fields of clusters based on the application of the integrated nested Laplace approximation (INLA) method \citep{inlabayes} to positional and kinematic measurements of cluster members. 
The analysis was performed on 1237 clusters for which suitable data are available.


We now follow with Sect.~\ref{Data}, in which we present the data sources and selection processes. Sect. \ref{vector field reconstruction method} details the methods developed for reconstruction of the proper motion vector fields. Analysis and results are shown in Sect. \ref{Analysis}. We conclude with a summary of results and conclusions in Sect. \ref{Conclusions}. Plots including the reconstructed fields of clusters with a detected kinematic structure are given in the \appendixautorefname{}.

\section{Data}  \label{Data}
In this article we use data from the Gaia EDR3, which contains proper motions precise at the hundreds of $\mu$as/yr level for more than a billion stellar sources, enabling kinematic and dynamic studies of OCs in large scales. These studies require membership lists, and here we use the detailed OC membership lists derived by \citet{upmask1, upmask2} for several clusters applying the UPMASK method \citep{upmaskmain} to Gaia DR2 proper motions \citep{dr2}. Based on these memberships, we extracted the following data from EDR3: positions $\alpha$, $\delta$, proper-motions $\mu_\alpha*$, $\mu_\delta$, parallaxes $\varpi$, the associated astrometric errors $\sigma_\alpha,\sigma_\delta,\sigma_{\mu\alpha*},\sigma_{\mu\delta},\sigma_\varpi$, correlations from EDR3, and cluster membership probabilities $p_{memb}$ for \citet{upmask1, upmask2} members available for 1275 OCs. In addition to astrometric information, cluster radial velocities were required to correct the effect of perspective acceleration on observed kinematics. Although this effect is small for most objects \citep[e.g.][]{1997MNRAS.285..479B, 2009A&A...497..209V, 2018A&A...616A..12G, Kuhn_2019}, it should be taken into account if the aim is to avoid systematics and probe into the noise limits. To account for the perspective effects in proper motions, we adopted radial velocity estimates from \citet{rv_source} for 965 clusters, and we further estimated bulk cluster radial velocities from the median of radial velocities of cluster member stars using Gaia DR2 \citep{2019A&A...622A.205K} for 265 clusters, LAMOST Data Release 4 \citep{2017ASInC..14...93W} for 33 clusters, RAVE Data Release 5 \citep{Kunder_2017} for seven clusters, and APOGEE Data Release 14 \citep{2018AJ....156..125H} for five clusters. As detailed in Appendix \ref{appendix:corrections}, we corrected the perspective and the projection effects in the measured proper motions following \citet{2009A&A...497..209V}.

\section{Vector field reconstruction method}  \label{vector field reconstruction method}

To study the internal kinematics of OCs, we searched the data for a statistically significant pattern defining the proper motion field shared by the cluster stars. To do so, we needed to reconstruct the underlying vector field and estimate its uncertainty, using the observed stellar proper motions and their uncertainties.

As OC members share a common motion and as external dynamical influences suffered by the cluster during its lifetime introduce spatially continuous perturbations, the underlying proper motion vector field can be considered mostly spatially correlated and continuous. However, observing this field is challenging due to the sparsity of stars, measurement errors, and the peculiar component of the stellar motions leading to internal motion dispersion, both at the order of hundreds of $\mu$as/yr, and thus at the same order or greater than the smooth signal that we have sought to study. Thus, it is natural to adopt  methods that can profit from the expected spatial correlation and continuity conditions to try to infer the underlying field from such noisy data.

The INLA method \citep{inlabayes} is one such method. It was created to model spatial data and it has been successfully used in different applications: from the mapping of the spread of disease \citep{inladoenca} to the prediction of heavy rainfalls \citep{inlarain}. INLA has also shown great potential in astronomy being used to reconstruct scalar fields of galaxy property maps from integral field unit observations \citep{inlaastro}.

INLA is a faster alternative to Markov chain Monte Carlo (MCMC) methods for Bayesian inference, as it approximates the solution in a fraction of the time MCMC requires. However, the posterior distribution must be assumed to be Gaussian (in which case the solution is exact), or nearly Gaussian (and thus INLA approximates the solution). This constitutes one of the assumptions we have made for our work. We have also assumed that internal cluster underlying proper motion fields are continuous and spatially correlated. Thus, we can take advantage of the fact that a continuous and correlated spatial field can be approximated by a Gaussian Markov random field when it is a solution of a stochastic partial differential equation (SPDE) with a Mat\'ern correlation function \citep{inlaspde}. This correlation function encodes how much one point in space is influenced by all other points depending only on their relative distances.

INLA allows the aforementioned assumptions to be considered, but it was created to analyse scalar fields while we are interested in vector fields. So we created a simple pipeline to reconstruct separate scalar fields in the projections of the vector field in the right ascension and declination coordinates, and then joining the inferred fields into a vector field. This reconstruction strategy provides a fast first approximation to the reconstruction, although it does not account for covariances between the proper motion components or some conditions as curl or divergence properties of the vector field.

Our starting point is an uncertain and non-homogeneous sampling of the underlying vector field, comprised by vector data measured at specific points in space and the uncertainty estimates for each position and vector measurement. Here we have used the positions of the stars in an equatorial coordinate system as covariates for the spatial model. The errors of the celestial sphere projected positions $(\alpha,\delta)$ have been ignored, as they are much smaller than the size of the clusters. One important aspect to retain is that a model can take the spatial correlation into account and local information is essential to study these fields, as this correlation represents a proxy function for signatures of rigid-body-like rotations of the cluster, gravitational bonds between the cluster stars, external gravitational influences from which the cluster may be suffering, etc. Here this has been achieved using a Mat\'ern function \citep{Matern1960}  that is a flexible correlation structure including Gaussian and exponential correlations as special cases \citep[e.g.][]{HandcockStein93,GuttorpTilmann06}.

Our pipeline to perform the vector field reconstruction was implemented in the R language \citep{Rlang} and adopted R-INLA\footnote{{\tt www.r-inla.org}} to reconstruct the individual scalar projections of each component of the proper motion vectors. It works as follows:

\begin{enumerate}
  \item For each cluster, we retrieved the relevant information for each star in its field: the right ascension, declination, proper motions in right ascension and declination as well as their errors, and the membership probability. To ensure the reconstruction focussed on stars that are more likely members of the cluster, stars with a membership probability $\le50\%$ were rejected.
  \item Then we removed the bulk cluster proper motion from the individual star proper motions by simply subtracting the cluster proper motion determined by \citet{upmask1, upmask2} using all the stars. This allowed us to focus on the analysis of the cluster internal kinematics.
  \item Next we created a two-dimensional triangular mesh in spatial coordinates, and in this mesh a representation of the scalar components of the vector field was subsequently estimated. The mesh covers the entire data region, with cutoff values preventing low mesh densities near observations (what could result in lower accuracy in the inference step), and maximum edges with respect to the distances between data points in spatial coordinates. Also, we refrained from extrapolating in the regions lacking data coverage.
  \item Then, we created a weight matrix representing the error of the data at the positions of each star on the mesh, and an SPDE model from the Mat\'ern correlation matrix. Here, a scale parameter proportional to the membership probability divided by the standard error was used to give more weight to data points with lower uncertainties in the measurements and with higher cluster membership probability.
  \item Afterwards, we applied R-INLA using a linear predictor structure with the SPDE model, which includes the effects of measurement errors and spatial correlations.
  \item Finally, we projected the resulting fields of $\mu_\alpha$ and $\mu_\delta$ on the positions of the original stars and on a regular grid.
\end{enumerate}

The application of this method results in a discretised field. At each position of this field, we had access to the inferred posteriors of the proper motion components, as represented by their means and standard deviations. So we could promptly reconstruct the most probable value of the proper motion vector at each position of the field and its error. 

In addition to the field reconstruction, we used this field to estimate smooth velocity curves as a function of the projected radius in the plane of the sky for all clusters. The curves were derived directly from the Gaia EDR3 catalogue data and from the INLA reconstructed fields. We adopted a smooth weighted local linear regression \citep{doi:10.1080/01621459.1979.10481038} through the fANCOVA package \citep{Wang2010}. Weights for each star were selected as $w_i=p_{memb,i}/\sigma_{\mu,i}^2$ and the smoothing length was selected based on generalised cross-validation \citep{doi:10.1080/00401706.1979.10489751}. The curves were estimated for the total proper motion $\mu_{tot}$, and for a polar decomposition into a radial component $\mu_\rho$ that indicates expansion and contraction, and into an angular component $\mu_{\theta}$ that indicates anticlockwise and clockwise rotation.


\section{Analysis} \label{Analysis}

We applied the method to the membership lists of 1275 clusters derived by \citet{upmask1, upmask2} and were able to reconstruct 1237 proper motion fields. To visualise these reconstructions, we represented each field with scatter plots of the cluster member positions and added the vectors representing the inferred smooth field direction and magnitude at the position of each member star. We also created the distributions of the polar field decomposition ($\mu_\theta$ and $\mu_\rho$) as functions of the distance to the cluster centre. We call these plots spatio-kinematic diagrams. All spatio-kinematic diagrams for the reconstructed fields were then visually inspected, and we looked for rotation-like as well as expansion- and contraction-like spatio-kinematic patterns.

We also created standard deviation maps from the reconstructed fields, and when the magnitude of the estimated standard deviation of the field was greater than the pattern that indicated the kinematic signal, we refrained from drawing conclusions. Additionally, in this work we concentrate on the most clear patterns, as the EDR3 data can have systematic errors that are spatially correlated due to non-astrophysical reasons on scales of $\lesssim0.5^\circ$ of the order of a few tens of $\mu$as/yr \citep[e.g.][]{2021A&A...649A...2L}.

\begin{figure*}[htb]
    \centering 
  \includegraphics[width=\columnwidth]{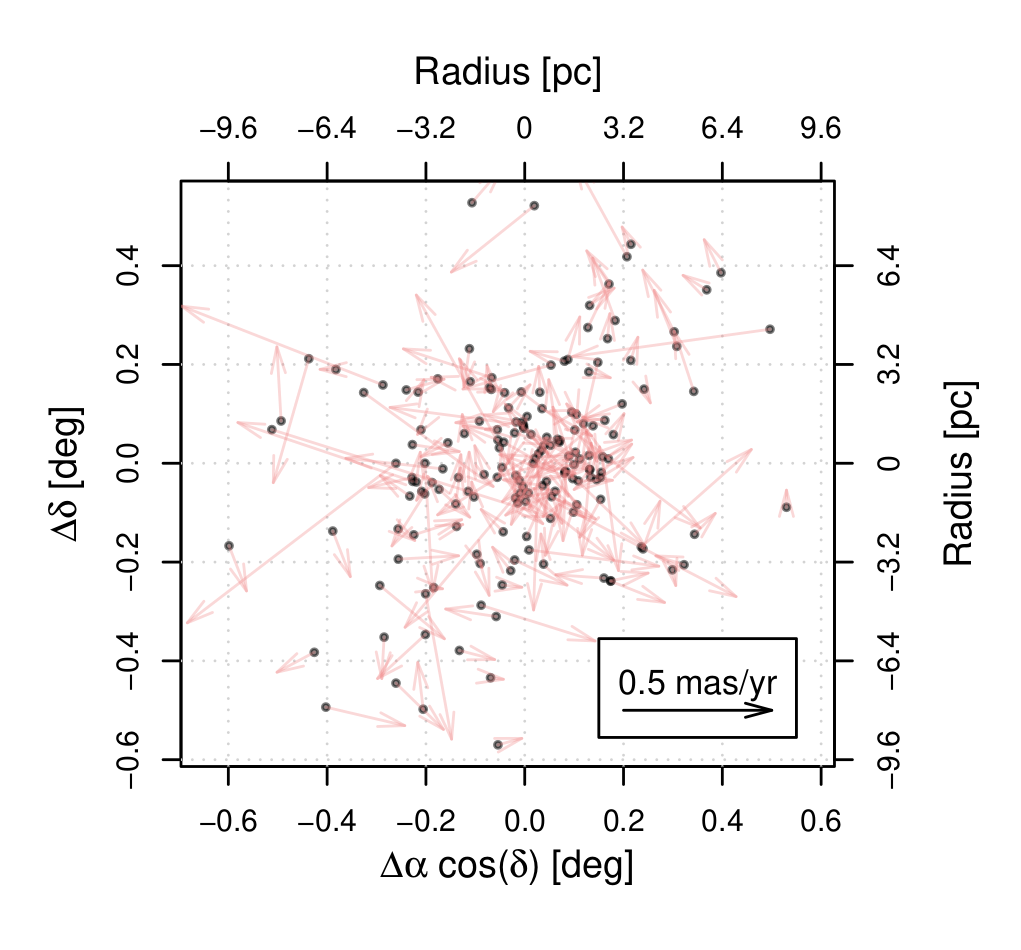}
  \includegraphics[width=\columnwidth]{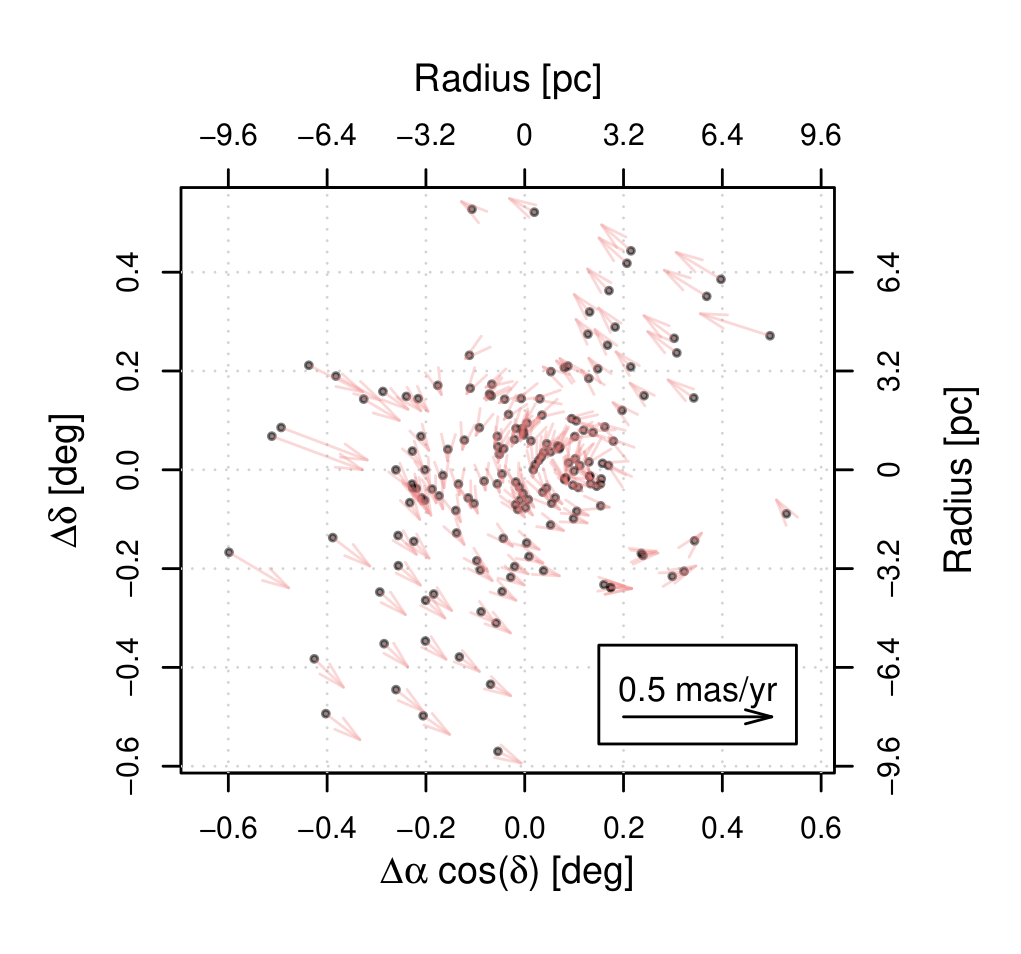}
\caption{Positions and proper motions of ASCC 114 members for the Gaia EDR3 data on the left and the inferred field on the right. The inferred proper motion field reaches a maximum angular component $\mu_\theta = 0.13\ \pm0.01\ mas/yr$ and a maximum radial component $\mu_\rho = -0.01\ \pm0.01\ mas/yr$, indicating anticlockwise rotation.}
\label{fig:teste2}
\end{figure*}

\begin{figure*}[htb]
    \centering 
  \includegraphics[width=\columnwidth]{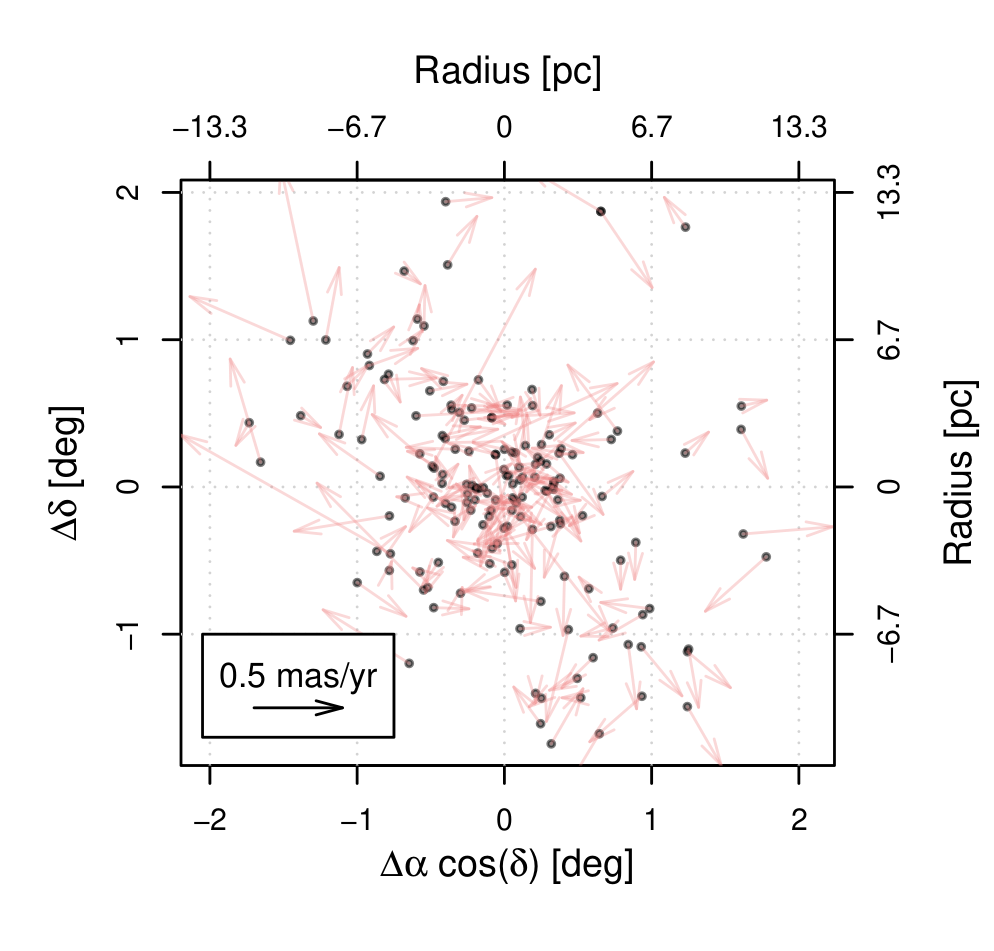}
  \includegraphics[width=\columnwidth]{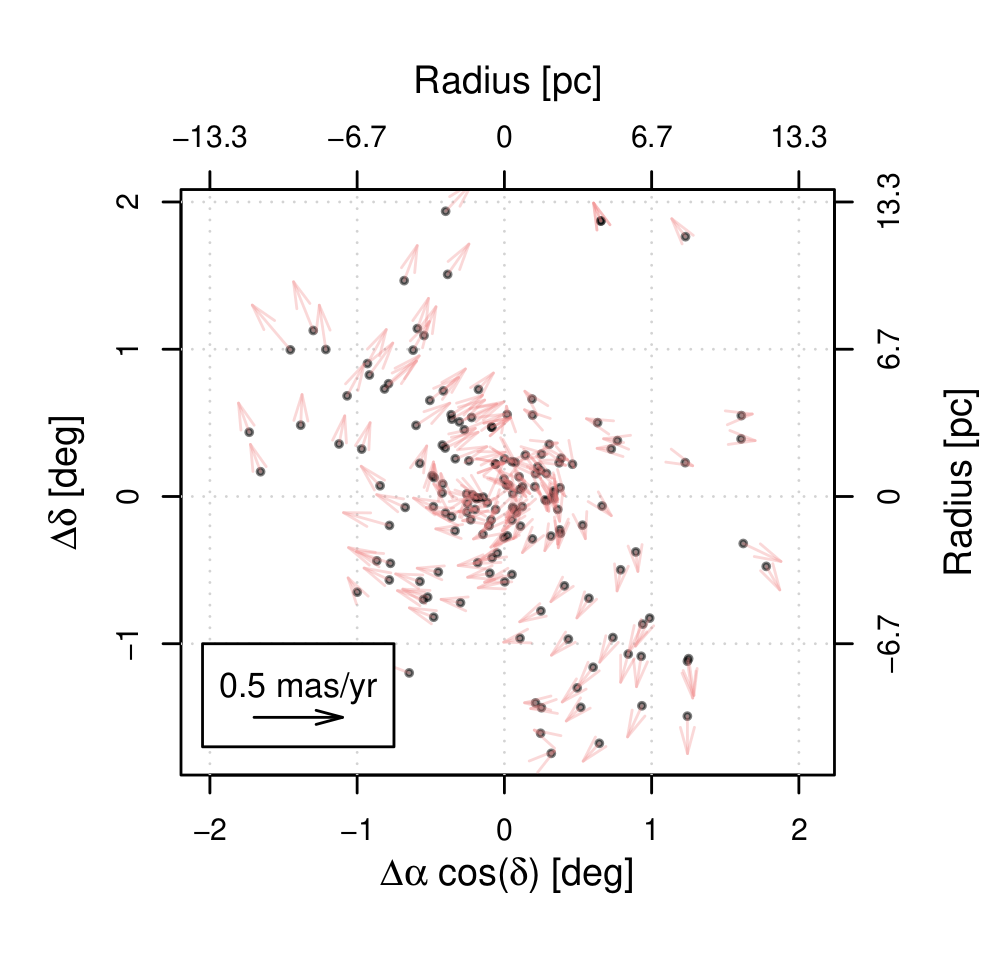}
\caption{Positions and proper motions of Collinder 140 members for the Gaia EDR3 data on the left and the inferred field on the right. The inferred proper motion field reaches a maximum angular component $\mu_\theta = -0.16\ \pm0.01\ mas/yr$ and a maximum radial component $\mu_\rho = 0.13\ \pm0.02\ mas/yr$, indicating clockwise rotation.}
\label{fig:teste1}
\end{figure*}

\begin{figure}
\centering
  \includegraphics[width=1\columnwidth]{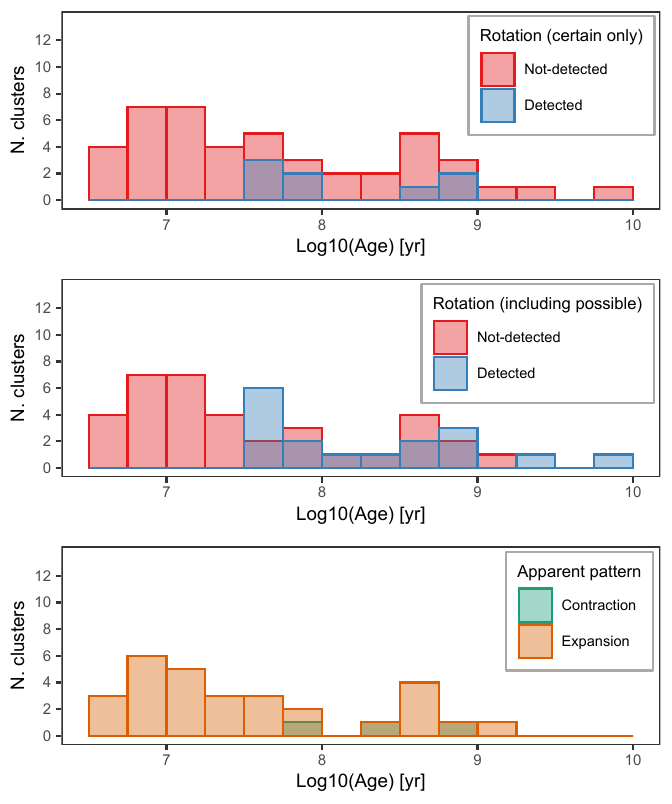}
  \caption{Upper plot: Age distribution of the clusters for which the detection of a rotation pattern was more certain. Middle plot: Same distribution, but including the clusters for which the rotation detection is much less certain -- with the inclusion of these possible rotation cases, the age distribution of the clusters with detected rotation just seems to raise, without significantly changing its shape. Bottom plot: Cluster age distribution for objects with some apparent pattern of contraction or expansion.}
  \label{fig:agespread}
\end{figure}

We show in Figures \ref{fig:teste2} and \ref{fig:teste1} the original Gaia EDR3 and the vector field reconstruction for the clusters ASCC 114 and Collinder 140. These figures indicate that spatio-kinematic patterns can be perceived even in a careful inspection of the original Gaia EDR3 data. By considering the errors in the proper motions and the assumption that members of a same cluster should share common overall motion, and thus that their proper motions should be physically correlated, the method described herein makes the spatio-kinematic patterns of these vector fields stand out more clearly, and reveals interesting rotational as well as expansion and contraction patterns in several of the analysed OCs. 

For the detection of kinematic patterns, we note that clusters with clear rotational as well as expansion and contraction patterns should have higher projected absolute velocities in the $\theta$ (rotation) and $\rho$ (radial expansion or contraction) components. Moreover, reliable signals should display smooth patterns throughout the cluster's radial direction. Taking these aspects into account, we adopted two types of indicators.
The first type consists of the components $\mu_\theta$ and $\mu_\rho$ of the INLA reconstructed field. The second type consists of the areas under the INLA reconstructed velocity curves. For determining the area under the data points, a locally estimated scatter-plot smoothing \citep[LOESS][]{Cleveland1979} regression was performed and the area under the fitted curve was calculated. We sorted the reconstructed fields using these indicators. Then by visual inspection, we identified those with clear kinematic patterns and we used them to set threshold values for the area under the INLA reconstructed velocity curves $A_\theta$ and $A_\rho$. Rotation candidates seem to follow both of the following criteria: $|\mu_\theta|\geq0.02\ mas/yr$ and $A_\theta \geq 0.74$. While expansion and contraction candidates follow both of the following criteria: $|\mu_\rho|\geq0.058\ mas/yr$ and $A_\rho\geq0.4$.

From the initial set of 1237 reconstructions, applying these criteria results in 98 candidates with detections of kinematic patterns. A final inspection of the candidates was performed to identify and remove spurious reconstructions, resulting into a list of $53$ clusters. In Appendix \ref{appendix:clusterprop} we summarise the detected patterns of these $53$ clusters. In Appendix \ref{appendix:atlas}, we show the spatio-kinematic diagrams for all these clusters.

Within the $53$ clusters, eight show clear rotation, with nine other objects presenting a less clear rotation signal. Expansion is clearly seen in $14$ objects, while two display contraction. An additional $15$ objects and another one show(s) less clear expansion and contraction patterns, respectively. Finally, there are $17$ objects for which the reconstruction showed spatio-kinematics patterns not compatible with any of the above behaviours (rotation, contraction, expansion). These unexpected patterns can appear due to factors such as remaining field contamination, errors in catalogued cluster centres, asymmetries introduced by variable extinction, external gravitational disturbances, or multiple populations with different kinematics.
Multiple populations may appear because of the alignment of more than one cluster along the same line of sight (e.g. Trumpler 22), or by substructure expected in younger groups. This last possibility seems to be the case for objects such as NGC~2244, NGC~6193, NGC~6531, NGC~6871, FSR~0904, Gulliver~9, IC~1396, van den Bergh~92, Stock~8, and Trumpler~16, all with age estimates $\lesssim20$ Myr \citep[e.g.][]{2019A&A...623A.108B, rv_source}. The age spread of clusters presenting unexpected reconstructions is however large, spanning almost the entire age interval of the sample considered in this work. We note that contraction and expansion patterns can appear as artefacts created by non-homogeneous distributions of the available samples of cluster members and/or under- or over-corrected perspective effects due to the bulk radial velocity of the cluster. This under- or over-correction can be seen as a source of concern for the detection of the effect in some objects, as this is driven by the cluster radial velocities which in many cases are ill-constrained, with errors at the level of several km/s. For the clusters for which we report the detection of a kinematic pattern, we used radial velocities from \citet{rv_source}, which were double-checked. Finally, we note that within the group of $53$ clusters for which we have found kinematic signals, we identify clear elongated spatial distributions, suggesting tidal tails, in five clusters: Platais~3, Platais~8, NGC~6991, Mamajek~4, and IC~4655.

The age distribution of the clusters with detected spatio-kinematical patterns is presented in Fig. \ref{fig:agespread}. The age determinations used therein are mostly from  \citet{2019A&A...623A.108B}, complemented by \citet{2019MNRAS.487.2385M} for Alessi 13 and by \citet{2005A&A...438.1163K} for Alessi 9. Although the spread is large, the age distribution of clusters presenting rotation patterns presents two groups, a younger one with $\log_{10}(\mathrm{age}) \sim 7.5-8$ and an older group at $\log_{10}(\mathrm{age}) \sim 8.5-9$. At the age ranges of these groups, there appears to be a marginal tendency to favour rotation for the younger ages if only certain detections are included in the analysis (upper panel of Fig.~3). One possible mechanism could be the enhanced destruction of clusters that rotate in the same direction of Galactic rotation \citep{2014BaltA..23..272O}. Thus, with time the disruption of clusters with unfavourable rotation would lead to a smaller fraction of older rotating clusters. However, if we consider the cases with possible rotation (middle panel of Fig.~3), this picture becomes blurred. At this moment, we consider it an observational suggestion for which a statistical or physical explanation requires further investigation.

\begin{figure*}
\centering
  \includegraphics[width=1.6\columnwidth,trim={1.4cm 2.5cm 1.4cm  2.5cm},clip]{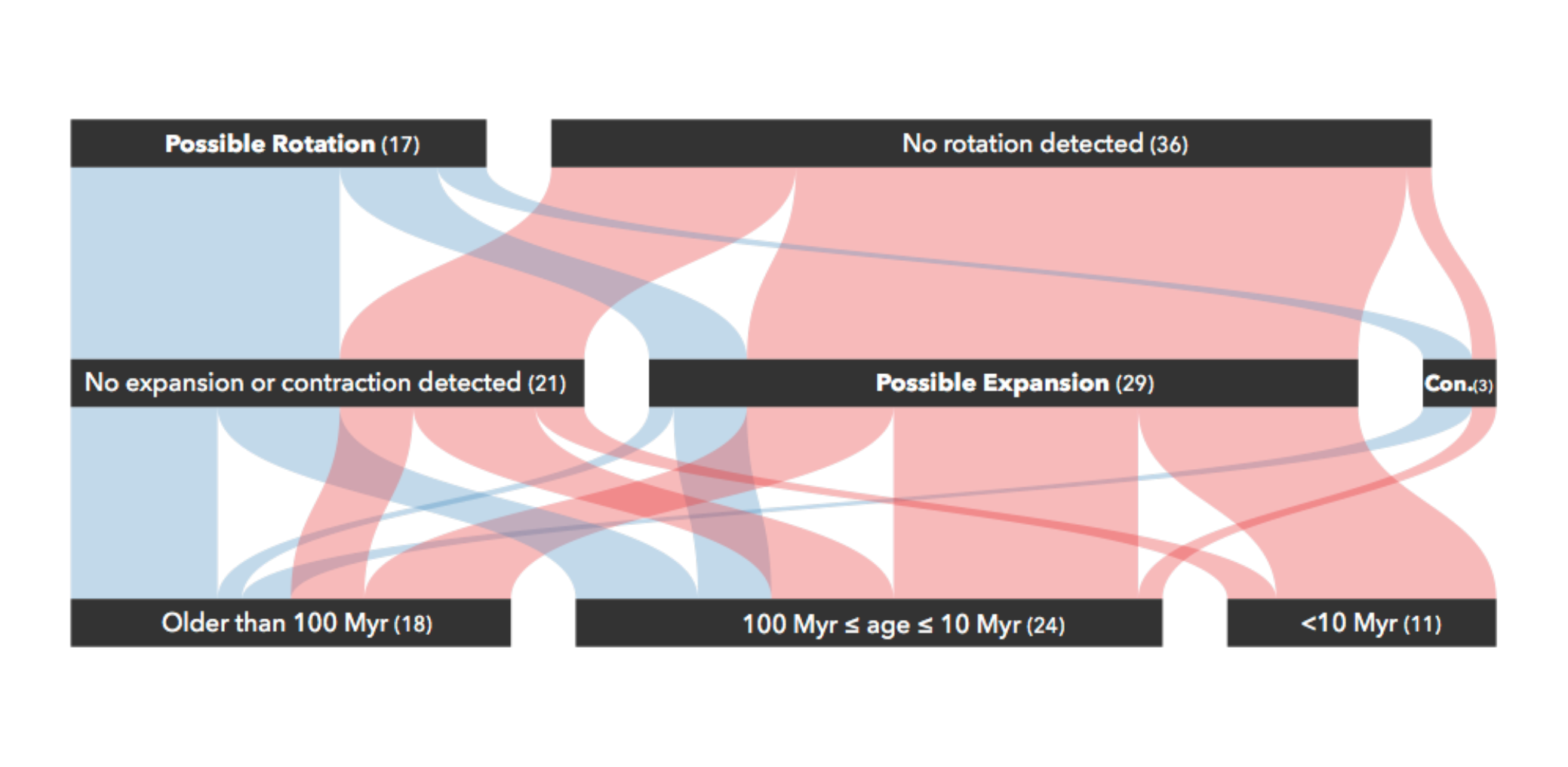}
  \caption{Correlations between cluster ages and two different classes of kinematic patterns: rotation and no-rotation as well as expansion, contraction, and others (represented as `no expansion or contraction detected') for the sample where some pattern could be detected in this work. The numbers in each box indicate the total number of objects for which the patterns were detected in this sample. The blue bands represent clusters with possible rotation patterns while the red bands the clusters where no rotation was detected. The bands are proportional to the number of clusters with each pattern (e.g. about half of the clusters older than 100 Myr had a detectable rotation in this sample), and they connect the different dimensions, revealing correlations in the patterns of the sample among the different dimensions (e.g. most clusters where a possible rotation and a possible expansion were detected have ages between 10 and 100 Myr).}
  \label{fig:agepattercorrelations}
\end{figure*}

The correlation of the ages and the possible kinematic patterns detected in this sample, including less uncertain pattern detection, is represented in Fig.~\ref{fig:agepattercorrelations}. This figure indicates that more than half of the clusters for which a possible rotation was detected presented no detectable expansion or contraction pattern from the reconstructions based on Gaia EDR3 data. It also indicates that half of the objects older than $100$ Myr in this sample are possibly rotating, and that most objects that are possibly rotating and at the same time might be expanding are younger than $\sim 100$ Myr. Finally, Fig.~\ref{fig:agepattercorrelations} also indicates that no rotation pattern was detected for the majority of objects with a possible detection of expansion, and that the large majority of these objects have ages $\lessapprox 100$ Myr. 

To validate the results, we also performed an additional test using a different method and implementation of the reconstruction. We created reconstructions using a simpler spatial Gaussian process with a Laplace approximation and an exponential spatial correlation structure with its length determined by generalised cross validation. The results obtained with this simpler method were similar to those resulting from INLA, in part due to the correlation structure and the posterior approximation being similar. Other methods that can be adopted for spatially correlated vector field reconstructions, as $\epsilon-$support vector regression with matrix valued kernels \citep[e.g.][]{IMPA2008}, can further provide interesting physical information as the stars acting as support vectors could be interpreted as naturally indicating boundaries that define distinct kinematic behaviours in the OCs. Such methods can also provide a faster estimation, perhaps enabling internal kinematics and dynamics to be considered in iterative cluster membership analysis such as UPMASK, in addition to a rigorous vector formulation providing the possibility to enforce curl and divergence properties as optional constraints. This is however at the expense of the posterior distribution inference as they are based on strict mathematical optimisation paradigms. Finally, methods such as NIFTy \citep{2013A&A...554A..26S, 2019ascl.soft03008A} may enable the Gaussian approximation for the posterior to be relaxed and also provide conditions to be placed in the power spectrum of the distribution and full three-dimensional inference, possibly including reconstructions of the internal positions of the objects within the cluster, however, in exchange for higher computational complexity. These avenues remain to be explored in future works making use of the upcoming Gaia Data Releases.

\section{Conclusions} \label{Conclusions}
We report the detection of rotation patterns in  eight OCs, with another nine possibly rotating, from Gaia EDR3 data. Additionally, we also detected expansion in $14$ OCs and contraction in two, with an additional $15$ objects possibly expanding and one possibly contracting. In addition to the kinematic patterns, we also identify clear elongated spatial distributions in five clusters, suggesting tidal tails. The signals reported here are above EDR3 systematic error levels, suggesting that there are many more interesting objects and effects having yet to be revealed with the increased astrometric accuracy and precision of the upcoming Gaia Data Releases.

To detect these patterns, we implemented a method to reconstruct OC proper motion vector fields using the INLA. The method reveals spatial correlations in vector fields, which in the case of astronomical objects such as OCs, are expected to exist for physical reasons due to the object kinematics and dynamics. We applied this method to astrometric data of OC members derived from Gaia data, resulting in the detection of objects with clear and interesting patterns in their internal proper motion fields, corresponding to the detection of systematic internal motions of stars within such a large set of OCs.

The vector-field reconstruction methods used in this work represent another step in the kinematic and dynamical study of star clusters. Application of these methodologies to precise astrometry from the upcoming Gaia Data Releases and proposed missions such as JASMINE \citep{Gouda:2011}, GaiaNIR \citep{2016arXiv160907325H}, and Theia \citep{2017arXiv170701348T}, open a path for future dynamical studies of astronomical systems, such as stellar clusters, streams, nearby dwarfs, the entire Milky Way, or even flows of larger-scale cosmological structures.

\begin{acknowledgements} We wish to thank the anonymous referee for constructive comments. 
    This work was partially supported by the Portuguese Funda\c c\~ao para a Ci\^encia e a Tecnologia (FCT) through the Portuguese Strategic Programme UIDB/FIS/00099/2020 for CENTRA. AKM additionally acknowledges the support from the Portuguese Funda\c c\~ao para a Ci\^encia e a Tecnologia (FCT) through grants SFRH/BPD/74697/2010, PTDC/FIS-AST/31546/2017, EXPL/FIS-AST/1368/2021, and from the Caltech Division of Physics, Mathematics and Astronomy for hosting research leaves during 2017-2018 and 2019, when some of the ideas and codes underlying this work were initially developed. This work has made use of results from the ESA space mission {\it Gaia}, the data from which were processed by the {\it Gaia} Data Processing and Analysis Consortium (DPAC). Funding for the DPAC has been provided by national institutions, in particular the institutions participating in the {\it Gaia} Multilateral Agreement. The {\it Gaia} mission website is:
http://www.cosmos.esa.int/gaia. Some of the authors are members of the {\it Gaia} Data Processing and Analysis Consortium (DPAC).

This research has made use of data obtained from the GES Data Archive, prepared and hosted by the Wide Field Astronomy Unit, Institute for Astronomy, University of Edinburgh, which is funded by the UK Science and Technology Facilities Council.

Funding for RAVE (www.rave-survey.org) has been provided by institutions of the RAVE participants and by their national funding agencies.

Guoshoujing Telescope (the Large Sky Area Multi-Object Fiber Spectroscopic Telescope LAMOST) is a National Major Scientific Project built by the Chinese Academy of Sciences. Funding for the project has been provided by the National Development and Reform Commission. LAMOST is operated and managed by the National Astronomical Observatories, Chinese Academy of Sciences.

This work has made use of APOGEE data. Funding for the Sloan Digital Sky Survey IV has been provided by the Alfred P. Sloan Foundation, the U.S. Department of Energy Office of Science, and the Participating Institutions. SDSS-IV acknowledges support and resources from the Center for High-Performance Computing at the University of Utah. The SDSS web site is www.sdss.org. SDSS-IV is managed by the Astrophysical Research Consortium for the
Participating Institutions of the SDSS Collaboration including the Brazilian
Participation Group, the Carnegie Institution for Science, Carnegie Mellon University, the Chilean Participation Group, the French Participation Group, Harvard-Smithsonian Center for Astrophysics, Instituto de Astrofísica de
Canarias, The Johns Hopkins University, Kavli Institute for the Physics and Mathematics of the Universe (IPMU) / University of Tokyo, the Korean
Participation Group, Lawrence Berkeley National Laboratory, Leibniz Institut für Astrophysik Potsdam (AIP), Max-Planck-Institut für Astronomie (MPIA Heidelberg), Max-Planck-Institut für Astrophysik (MPA Garching), MaxPlanck-Institut für Extraterrestrische Physik (MPE), National Astronomical Observatories of China, New Mexico State University, New York University, University of Notre Dame, Observatário Nacional / MCTI, The Ohio State University, Pennsylvania State University, Shanghai Astronomical Observatory,
United Kingdom Participation Group, Universidad Nacional Autónoma de México, University of Arizona, University of Colorado Boulder, University of Oxford, University of Portsmouth, University of Utah, University of Virginia,
University of Washington, University of Wisconsin, Vanderbilt University, and Yale University. This work has made use of GALAH data, based on data acquired through the Australian Astronomical Observatory, under programmes:
A/2013B/13 (The GALAH pilot survey); A/2014A/25, A/2015A/19, and A2017A/18 (The GALAH survey). We acknowledge the traditional owners of the land on which the AAT stands, the Gamilaraay people, and pay our respects to elders past and present.
\end{acknowledgements}

\bibliographystyle{aa}
\bibliography{bibliography} 

\clearpage
\onecolumn
\begin{appendix}
\section{Detected kinematic patterns}

\label{appendix:clusterprop}
\begin{longtable}{|c|c|c|c|c|} 
\caption{Kinematic patterns detected.} 
\label{table:tab1} \\
\hline
Cluster       & Rotation     & Expansion    & Contraction & Other \\ \hline
Alessi 3      & $\CIRCLE$    &              &             &                      \\ \hline
Alessi 6      & $\ocircle$   &              &             &                      \\ \hline
Alessi 9      &              & $\ocircle$   &             &                      \\ \hline
Alessi 13     & $\CIRCLE$    &   $\CIRCLE$  &             &                      \\ \hline
Alessi 19     &              &              &  $\CIRCLE$  &                      \\ \hline
Alessi 37     &              & $\ocircle$   &             &                      \\ \hline
Alessi 43     &              & $\CIRCLE$    &             &                      \\ \hline
Alessi 44     &              &              &             & $\CIRCLE$            \\ \hline
ASCC 13       &              &  $\ocircle$  &             &                      \\ \hline
ASCC 16       &              & $\CIRCLE$    &             &                      \\ \hline
ASCC 19       &              & $\CIRCLE$    &             &                      \\ \hline
ASCC 58       & $\CIRCLE$    &              &             &                      \\ \hline
ASCC 71       & $\ocircle$   &              &             &                      \\ \hline
ASCC 73       & $\ocircle$   &              &             &                      \\ \hline
ASCC 114      &  $\ocircle$  &              &             &                      \\ \hline
ASCC 127      &              & $\ocircle$   &             &                      \\ \hline
Aveni Hunter 1 &             & $\CIRCLE$    &             &                      \\ \hline
BDSB96        &              & $\ocircle$   &             &                      \\ \hline
BH 99         &              & $\ocircle$   &             &                      \\ \hline
BH 164        & $\ocircle$   & $\CIRCLE$    &             &                      \\ \hline
Collinder 69  &              & $\CIRCLE$    &             &                      \\ \hline
Collinder 132 &              & $\CIRCLE$    &             &                      \\ \hline
Collinder 140 &  $\CIRCLE$   &              &             &                      \\ \hline
Collinder 197 &              & $\ocircle$   &             &                      \\ \hline
Collinder 359 & $\CIRCLE$    & $\CIRCLE$    &             &                      \\ \hline
FSR 0904      &              &              &             & $\CIRCLE$   \\ \hline
Gulliver 9    &              & $\CIRCLE$    &             & $\ocircle$           \\ \hline
IC 1396       &              & $\CIRCLE$    &             & $\ocircle$           \\ \hline
IC 1805       &              & $\ocircle$   &             &                      \\ \hline
IC 2602       &  $\ocircle$  &              &             &                       \\ \hline
IC 4665       & $\CIRCLE$    &              &             &                       \\ \hline
Mamajek 4     &              & $\ocircle$   &             &                       \\ \hline
NGC 188       & $\ocircle$   &              &             &                       \\ \hline
NGC 2194      &              &              &             & $\CIRCLE$             \\ \hline
NGC 2244      &              &              &             & $\CIRCLE$             \\ \hline
NGC 6193      &              &              &             & $\CIRCLE$             \\ \hline
NGC 6531      &              &              &             & $\CIRCLE$             \\ \hline
NGC 6871      &              &  $\CIRCLE$   &             & $\CIRCLE$   \\ \hline
NGC 6991      &              &  $\ocircle$  &             &                       \\ \hline
NGC 7380      &              &  $\ocircle$  &             &                       \\ \hline

Platais 3     &  $\ocircle$  &              & $\ocircle$  & $\CIRCLE$             \\ \hline
Platais 8     &              &              &             & $\ocircle$            \\ \hline
Roslund 2     &              &  $\ocircle$  &             &                       \\ \hline
Ruprecht 41   &              &              &             & $\CIRCLE$             \\ \hline
Ruprecht 98   &              &  $\CIRCLE$   &             &                       \\ \hline
Ruprecht 147  & $\ocircle$  & &        &                       \\ \hline
Ruprecht 161  & $\CIRCLE$    &  $\ocircle$  &             &                       \\ \hline
Stock 1       &              & $\CIRCLE$    &             &                       \\ \hline
Stock 2       & $\CIRCLE$    &              & $\CIRCLE$   & $\ocircle$            \\ \hline
Stock 8       &              &              &             & $\CIRCLE$             \\ \hline
Trumpler 16   &              & $\ocircle$   &             & $\ocircle$   \\ \hline
Trumpler 22   &              &              &             & $\CIRCLE$             \\ \hline
vdBergh 92    &              & $\ocircle$   &             & $\ocircle$            \\ \hline
\end{longtable}
The symbol $\CIRCLE$ indicates a detection, while the symbol $\ocircle$ indicates possible detection from the reconstructions based on Gaia EDR3. The column labelled `Other' marks clusters displaying unexpected velocity field patterns. For these, filled circles mark strong signals, mostly due to multiple kinematic groups in the field, and open circles indicate less clear signals of an unidentified nature.

\section{Proper motion and radial velocity corrections}
\label{appendix:corrections}
We followed \citet{2009A&A...497..209V} for the corrections to the proper motions and radial velocities due to the projections effects. The corrections that were applied for this work are the following:

\begin{equation}
\Delta \mu_{\alpha *, i} \approx \Delta \alpha_{i}\left(\mu_{\delta, 0} \sin \delta_{0}-\frac{V_{\mathrm{rad}, 0} \varpi_{0}}{\kappa} \cos \delta_{0}\right)
\end{equation}
\begin{equation}
\begin{aligned}
\Delta \mu_{\delta, i} \approx-\Delta \alpha_{i} \mu_{\alpha *, 0} \sin \delta_{0}-\Delta \delta_{i} \frac{V_{\mathrm{rad}, 0} \varpi_{0}}{\kappa}
\end{aligned}
\end{equation}
\begin{equation}
\begin{aligned}
\Delta V_{\mathrm{rad}, i} \approx \Delta \alpha_{i} \frac{\kappa \mu_{\alpha *, 0}}{\varpi_{0}} \cos \delta_{0}+\Delta \delta_{i} \frac{\kappa \mu_{\delta, 0}}{\varpi_{0}} 
\end{aligned}
,\end{equation}

The subscript '$i$' refers to each individual cluster member star, and the subscript '$0$' to the cluster itself;
 $\Delta \alpha_{i} = \alpha_i - \alpha_0$ and $\Delta \delta_{i} = \delta_i - \delta_0$;
 $\mu_{\alpha *} \equiv \mu_{\alpha} \cos \delta$;
 $\kappa = 4.740470446$ is the factor that converts $mas/yr$ to $km/s$ at $1 kpc$;
 $V_{\mathrm{rad}, 0}$ is the cluster's radial velocity; and
 $\varpi_{0}$ is the cluster's parallax.

\begin{figure*}[htb]
\includegraphics[width=\linewidth]{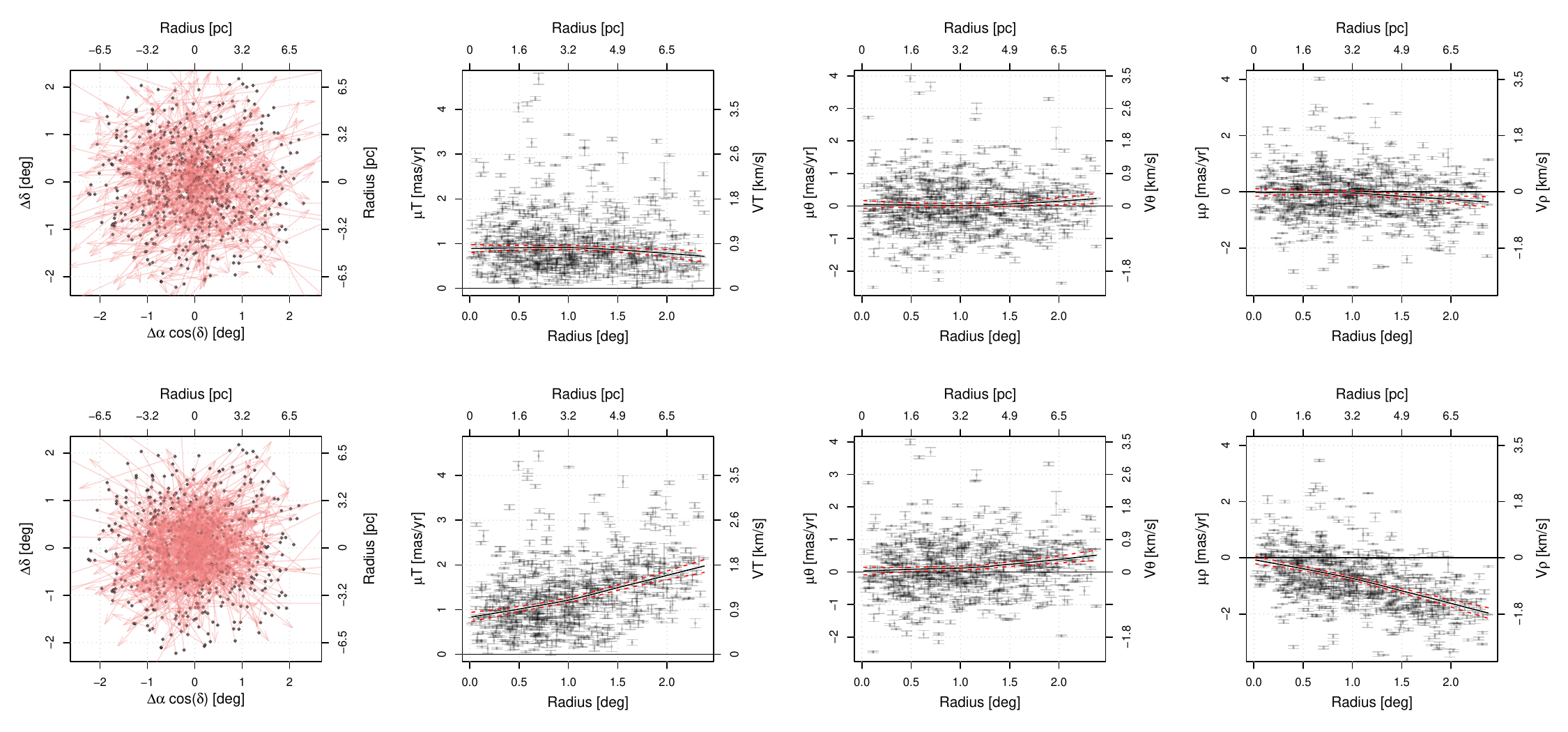}
\caption{Comparison between the corrected (top row) and uncorrected (bottom row) Gaia EDR3 data of the member of NGC~2632. The first column shows the positions of the stars with proper motions represented as vectors. The second column shows the total proper motion in function of the cluster's radius. The third column shows the component in the angular direction in function of the distance to the cluster's centre. This third column conveys if the cluster is rotating clockwise or anticlockwise if the component is negative or positive, respectively. The fourth column shows the radial component in function of the cluster's centre. This fourth column conveys if the cluster is contracting or expanding if the component is negative or positive, respectively.
}
\label{fig:corr}
\end{figure*}

To exemplify the importance of these corrections on proper motions, we can compare the uncorrected data with the corrected data in the case of the Praesepe cluster (NGC~2632). This is shown in Fig. \ref{fig:corr}, where it is possible to see that apparent patterns of rotation and contraction can appear due to the lack of the correction of these projection effects.

\onecolumn
\section{Spatio-kinematic diagrams}
\label{appendix:atlas}

\begin{figure*}[htb]
\includegraphics[width=\linewidth]{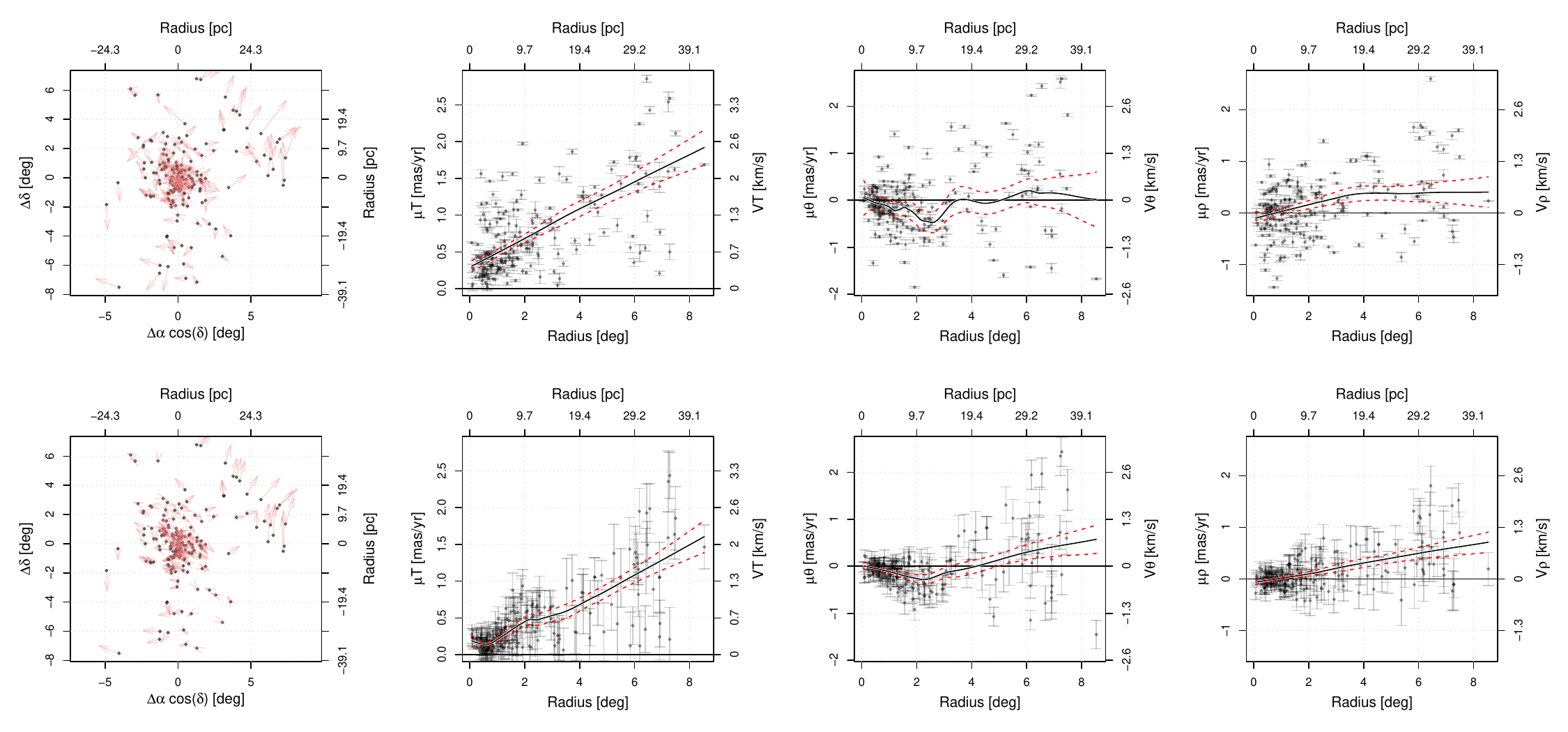}
\caption{Spatio-kinematic diagrams for Alessi~3. The real data from Gaia EDR3 is plotted in the first row. The second row shows the result of our method. In the first column, we represent the star positions and their proper motion vectors.
In the second column we show the total proper motion in function of the radius to the cluster's centre.
The third and fourth columns show a polar decomposition of the proper motion field. In the third column we show the component in the angular direction in function of the distance to the cluster centre, which indicates the rotational component of the proper motion: it is positive for anticlockwise rotations and negative for clockwise ones. In the fourth column, we show the radial component in function of the distance to the cluster centre, which can indicate if the cluster is contracting if it is negative, or expanding if it is positive.}
\label{fig:vcurve0}
\end{figure*}
\begin{figure*}[htb]
\includegraphics[width=\linewidth]{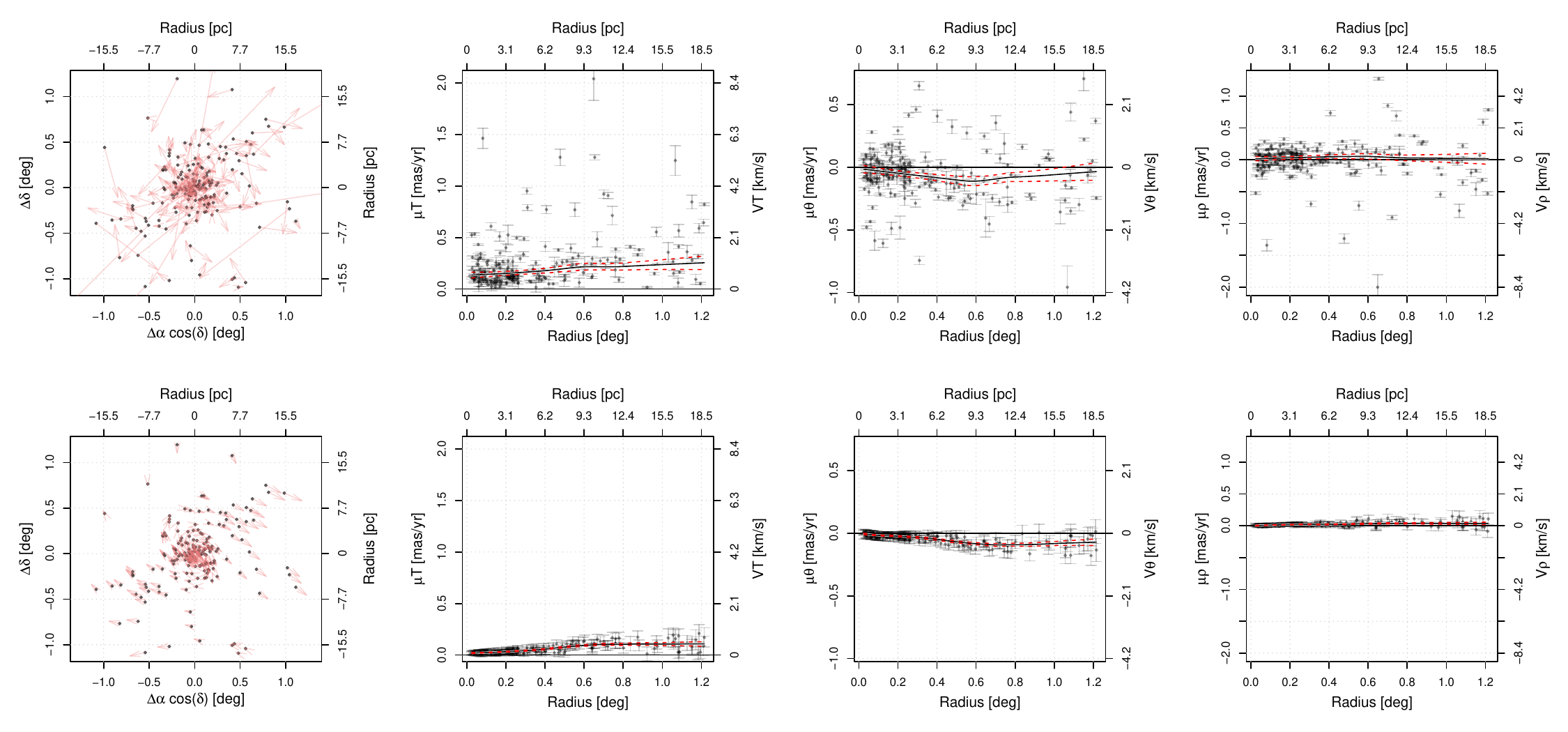}
\caption{Same as Fig.\ \ref{fig:vcurve0}, but for the Alessi 6 cluster.}
\label{fig:vcurve7}
\end{figure*}
\begin{figure*}[htb]
\includegraphics[width=\linewidth]{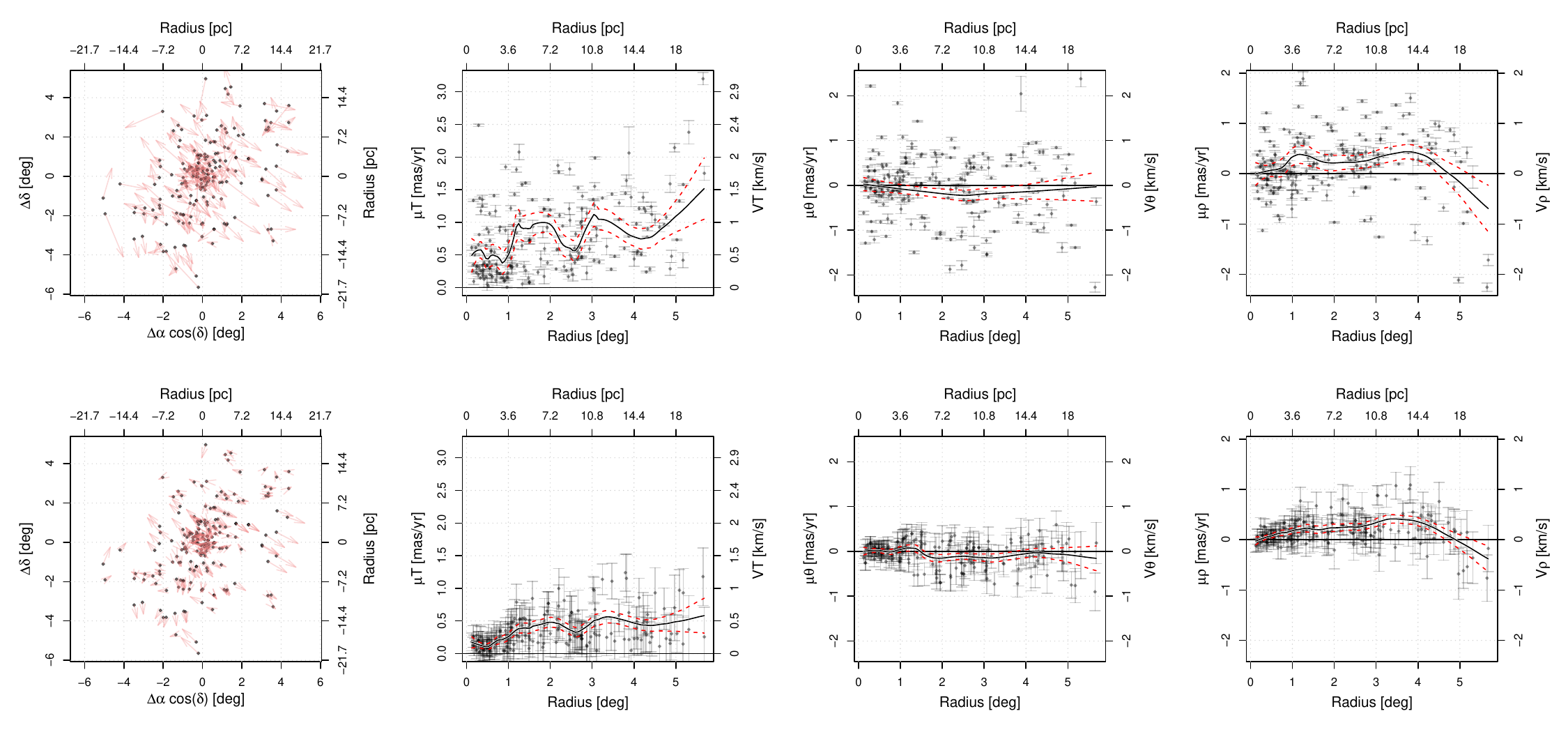}
\caption{Same as Fig.\ \ref{fig:vcurve0}, but for the Alessi 9 cluster.}
\label{fig:vcurve8}
\end{figure*}
\begin{figure*}[htb]
\includegraphics[width=\linewidth]{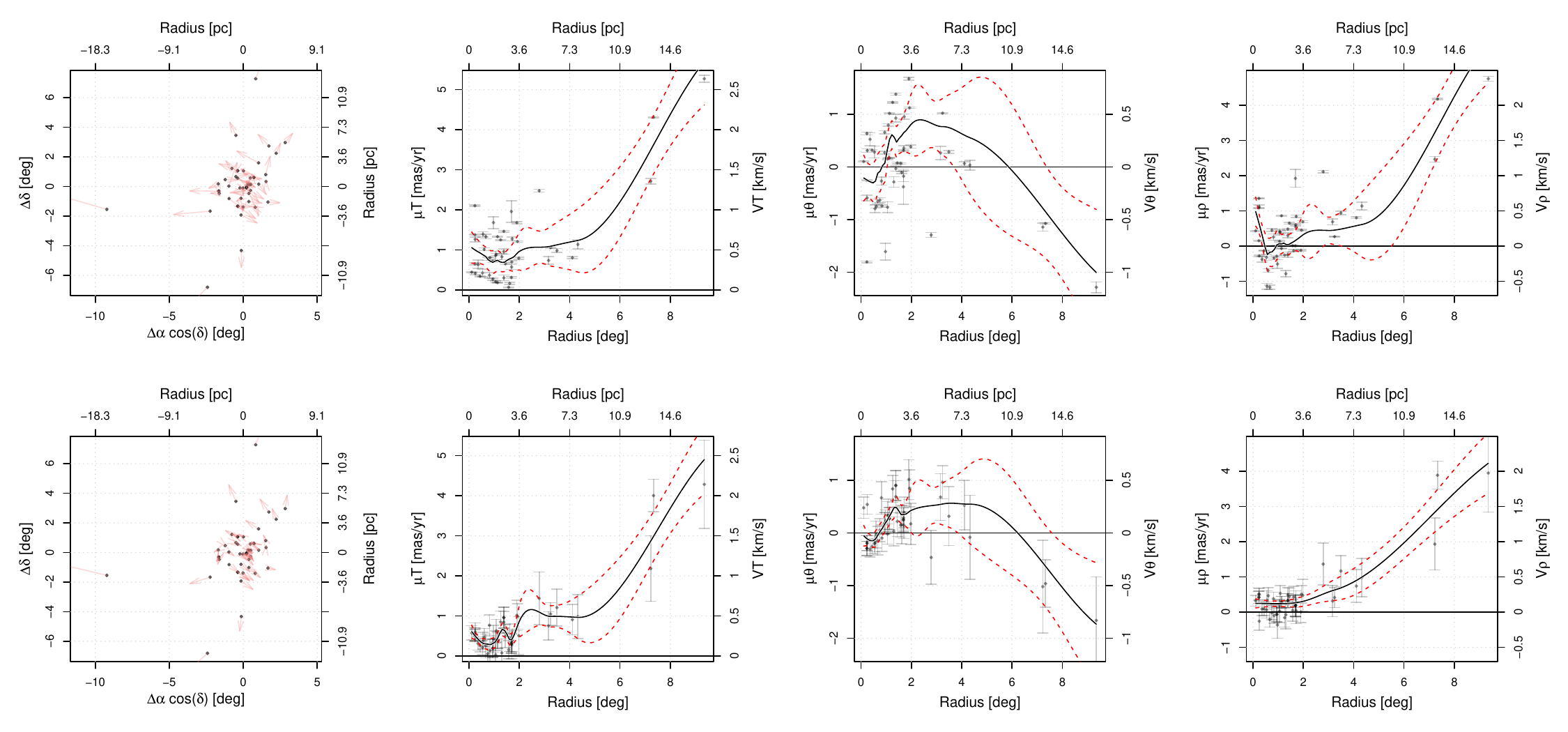}
\caption{Same as Fig.\ \ref{fig:vcurve0}, but for the Alessi 13 cluster. }
\label{fig:vcurve3}
\end{figure*}
\begin{figure*}[htb]
\includegraphics[width=\linewidth]{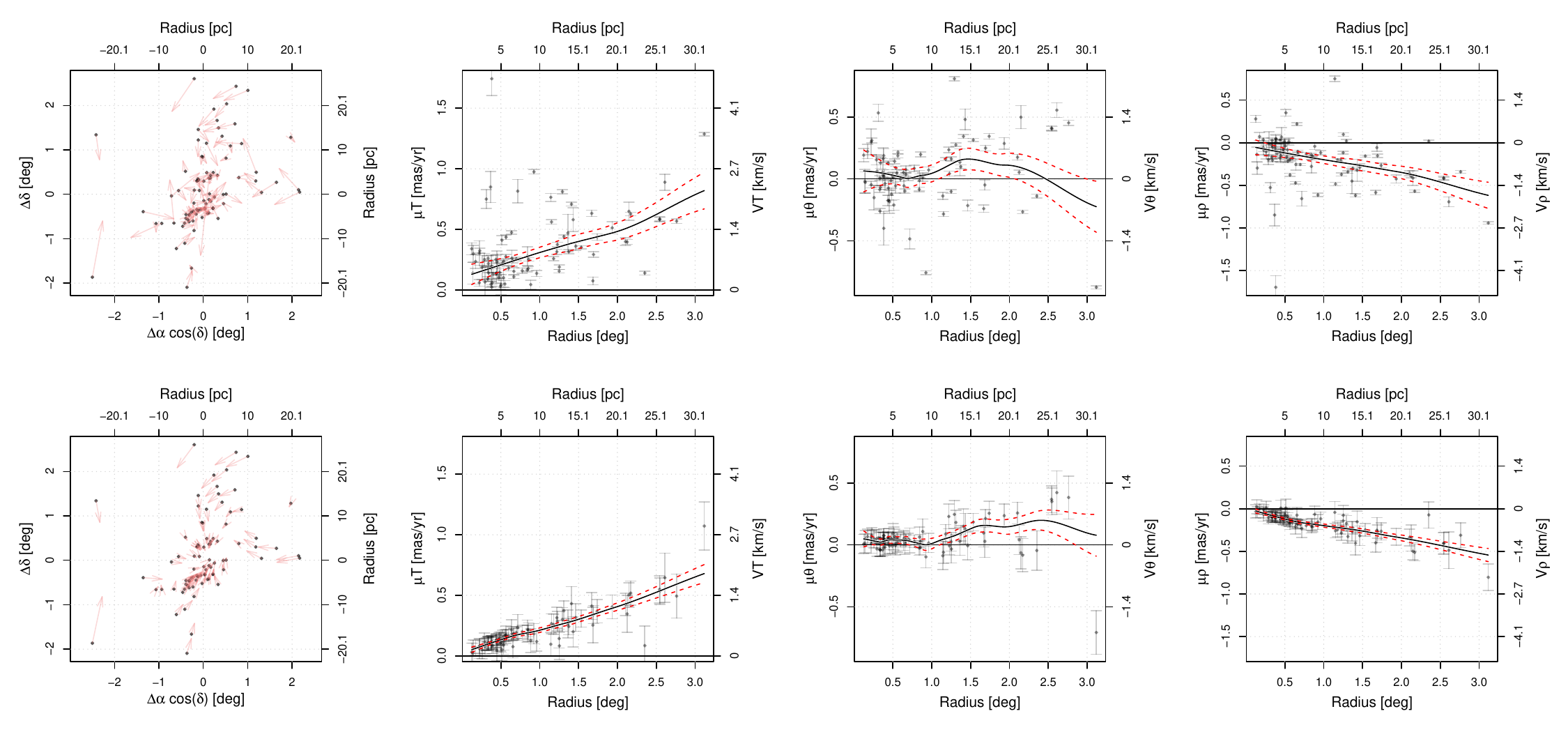}
\caption{Same as Fig.\ \ref{fig:vcurve0}, but for the Alessi 19 cluster. }
\label{fig:vcurve2}
\end{figure*}
\begin{figure*}[htb]
\includegraphics[width=\linewidth]{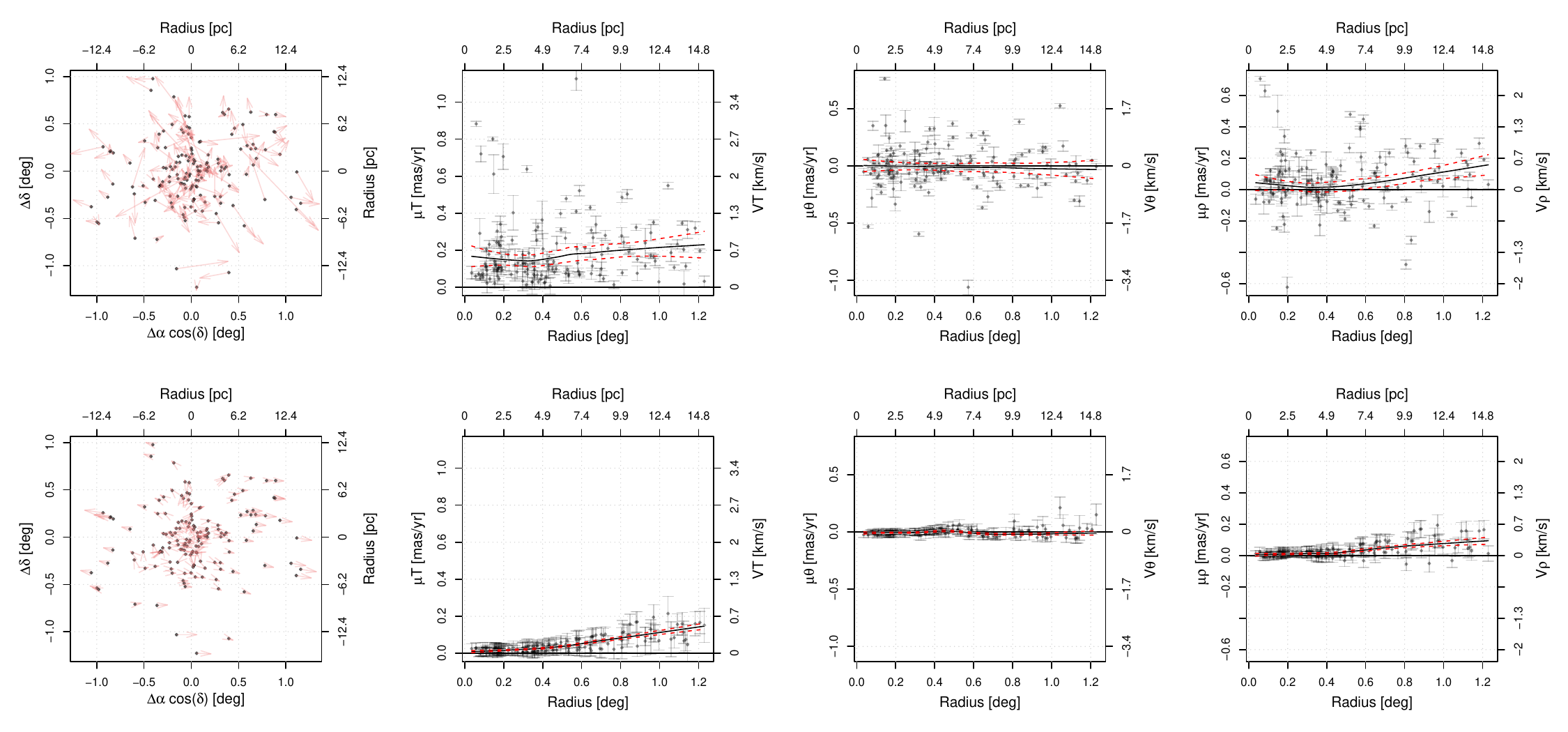}
\caption{Same as Fig.\ \ref{fig:vcurve0}, but for the Alessi 37 cluster.}
\label{fig:vcurve4}
\end{figure*}
\begin{figure*}[htb]
\includegraphics[width=\linewidth]{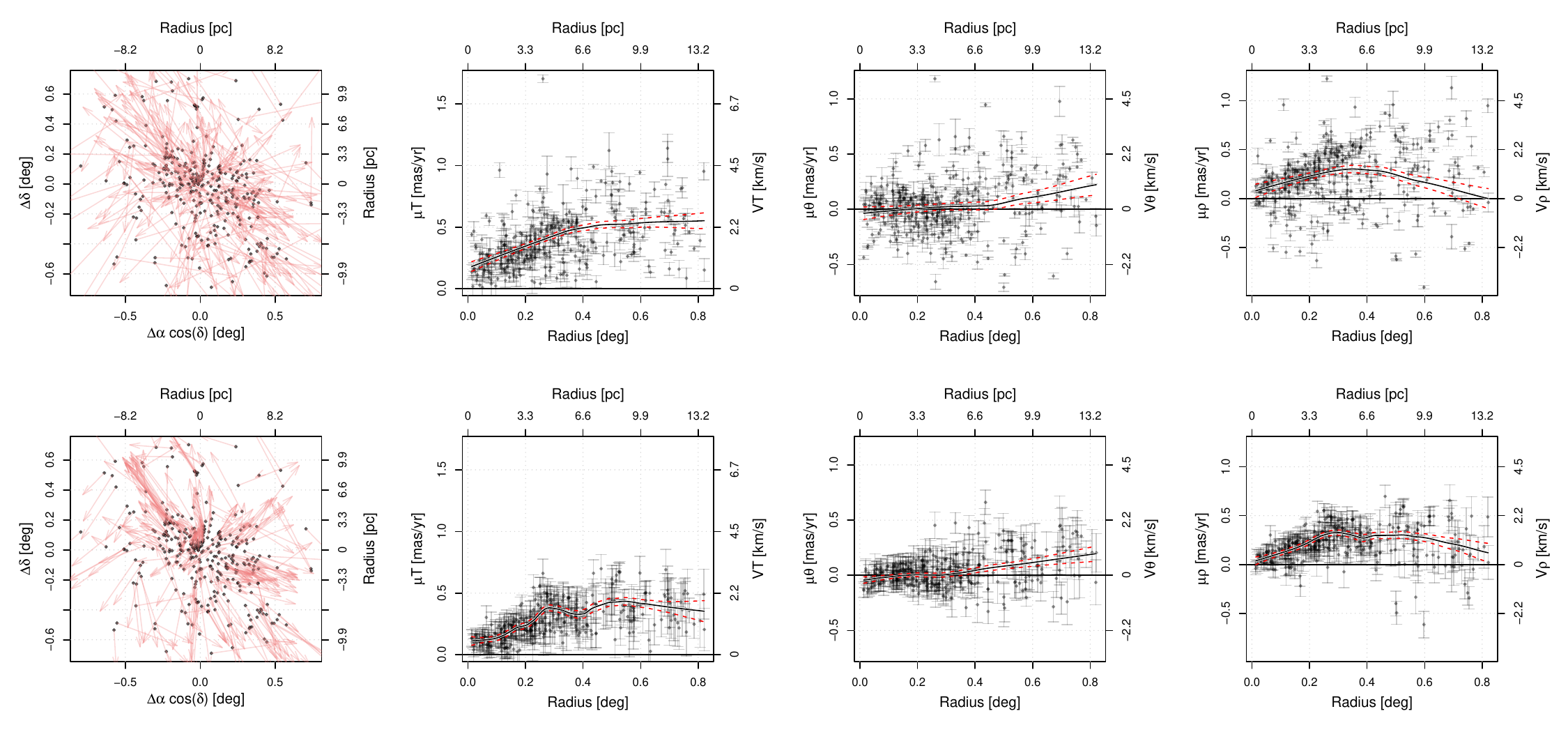}
\caption{Same as Fig.\ \ref{fig:vcurve0}, but for the Alessi 43 cluster. }
\label{fig:vcurve5}
\end{figure*}
\begin{figure*}[htb]
\includegraphics[width=\linewidth]{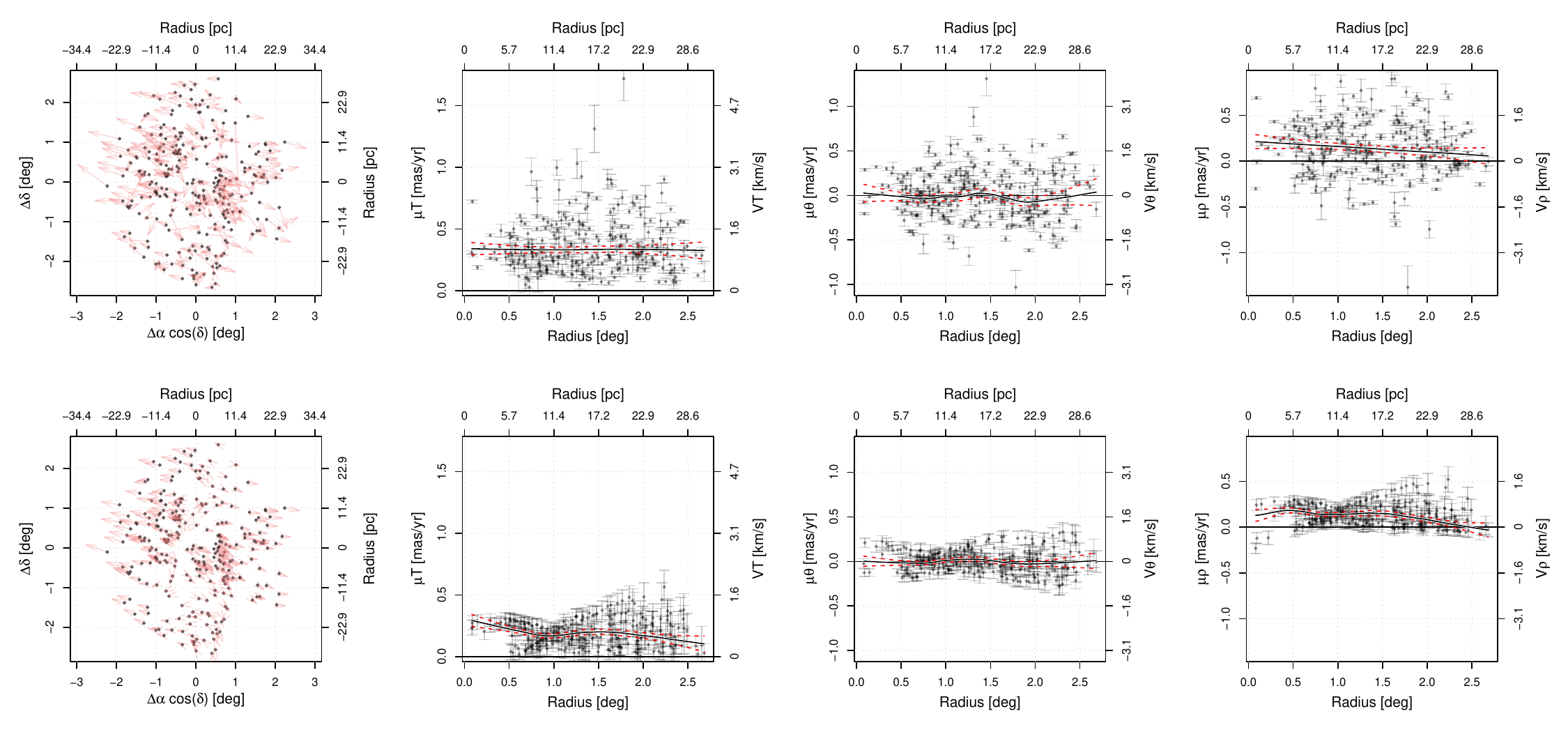}
\caption{Same as Fig.\ \ref{fig:vcurve0}, but for the Alessi 44 cluster. }
\label{fig:vcurve6}
\end{figure*}
\begin{figure*}[htb]
\includegraphics[width=\linewidth]{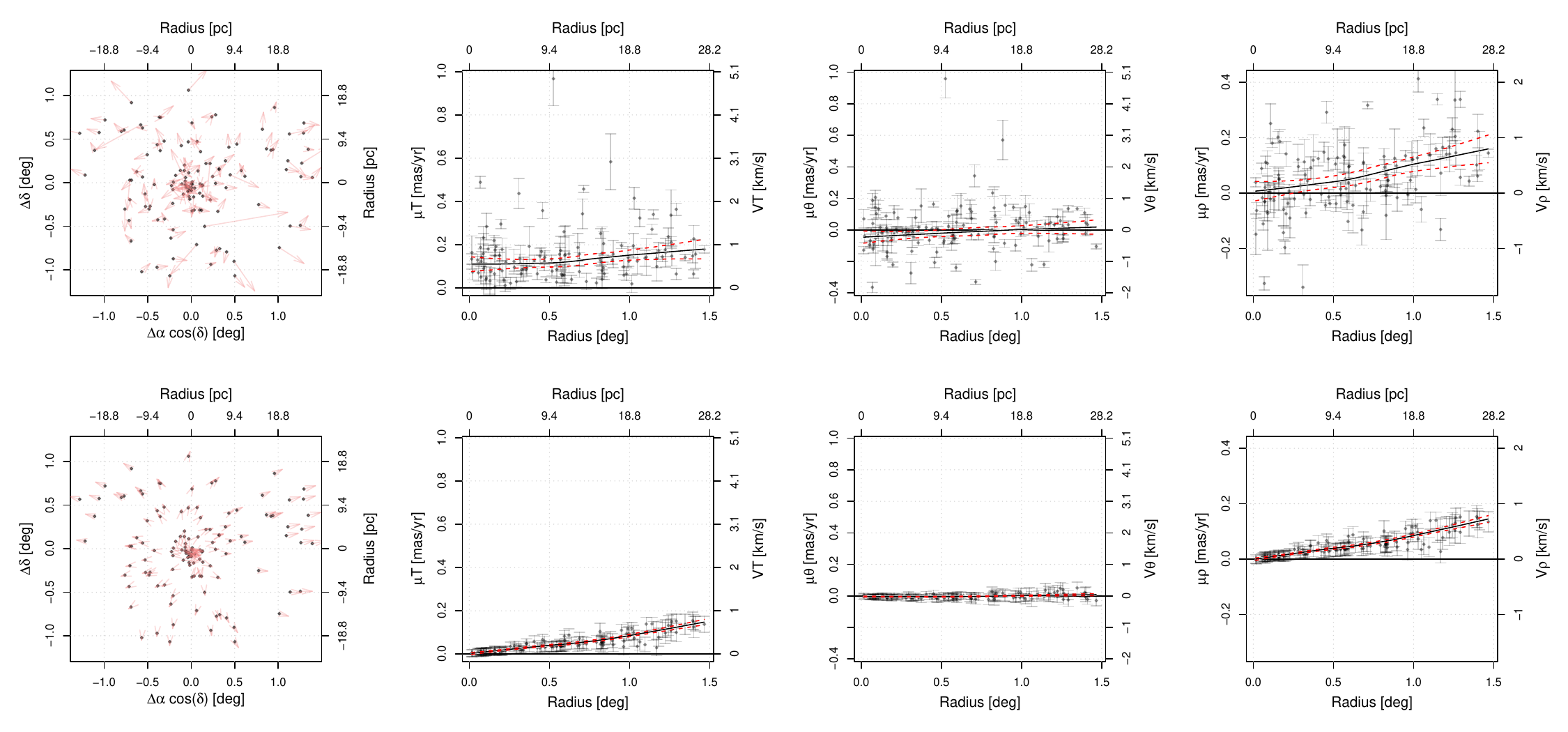}
\caption{Same as Fig.\ \ref{fig:vcurve0}, but for the ASCC 13 cluster.}
\label{fig:vcurve11}
\end{figure*}
\begin{figure*}[htb]
\includegraphics[width=\linewidth]{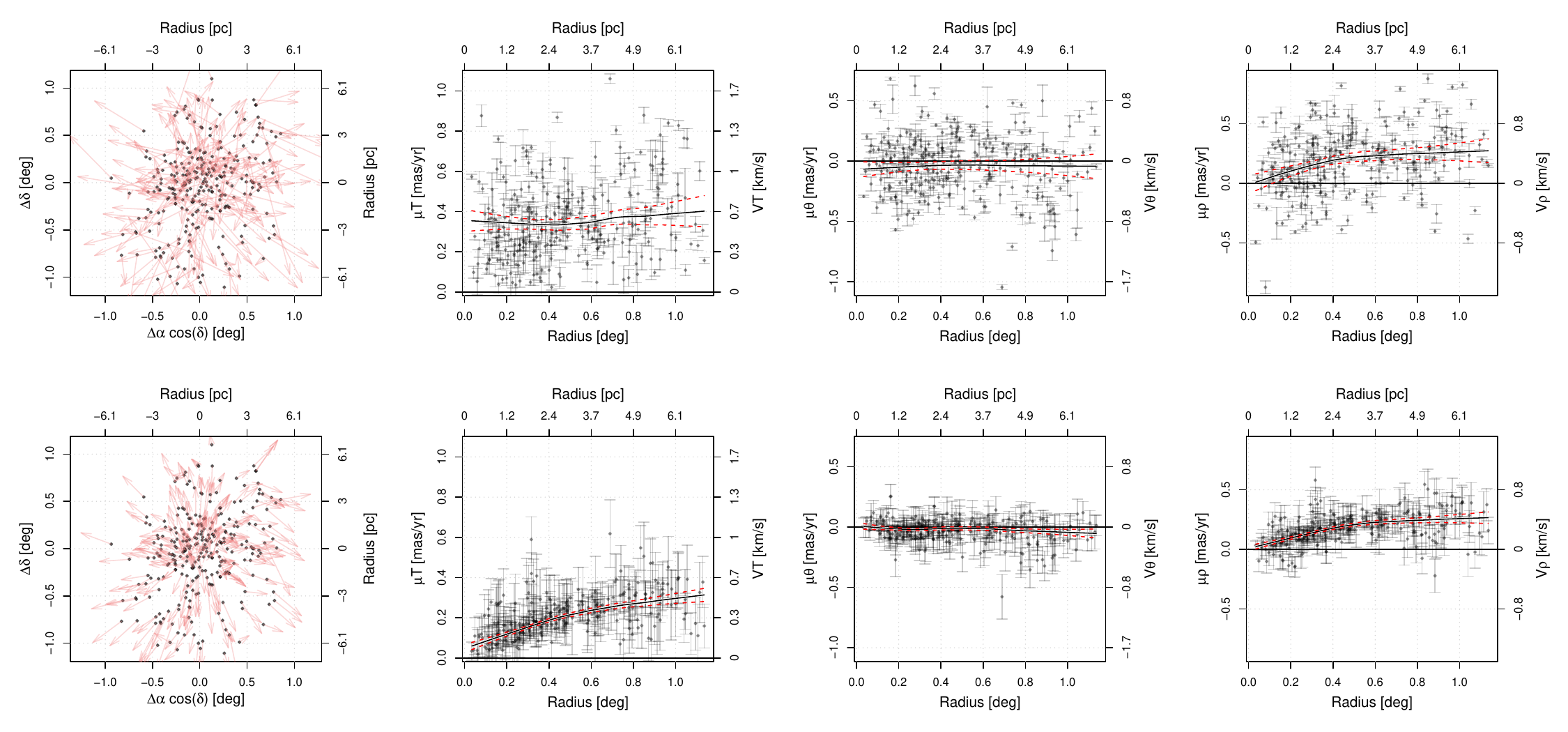}
\caption{Same as Fig.\ \ref{fig:vcurve0}, but for the ASCC 16 cluster. }
\label{fig:vcurve12}
\end{figure*}
\begin{figure*}[htb]
\includegraphics[width=\linewidth]{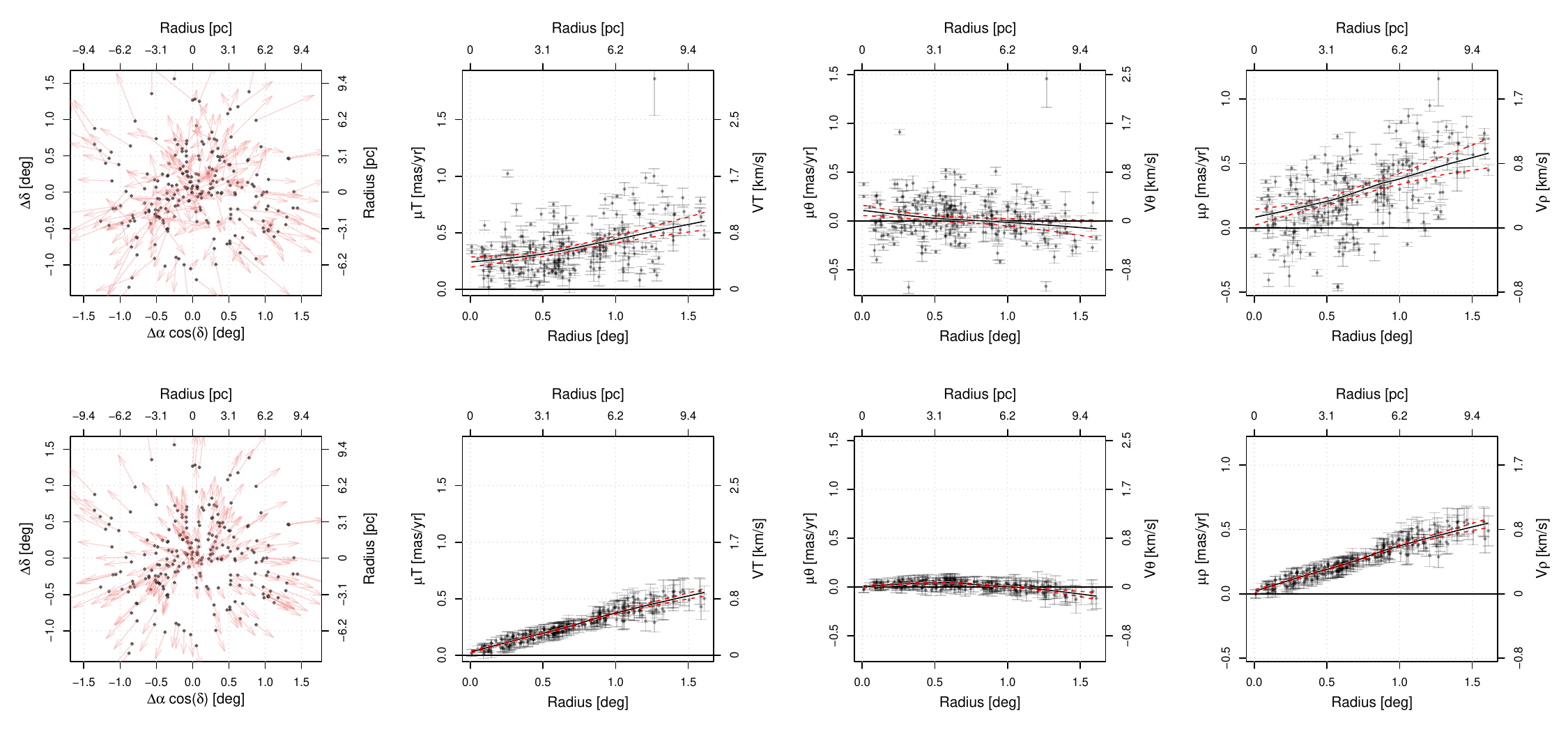}
\caption{Same as Fig.\ \ref{fig:vcurve0}, but for the ASCC 19 cluster. }
\label{fig:vcurve13}
\end{figure*}
\begin{figure*}[htb]
\includegraphics[width=\linewidth]{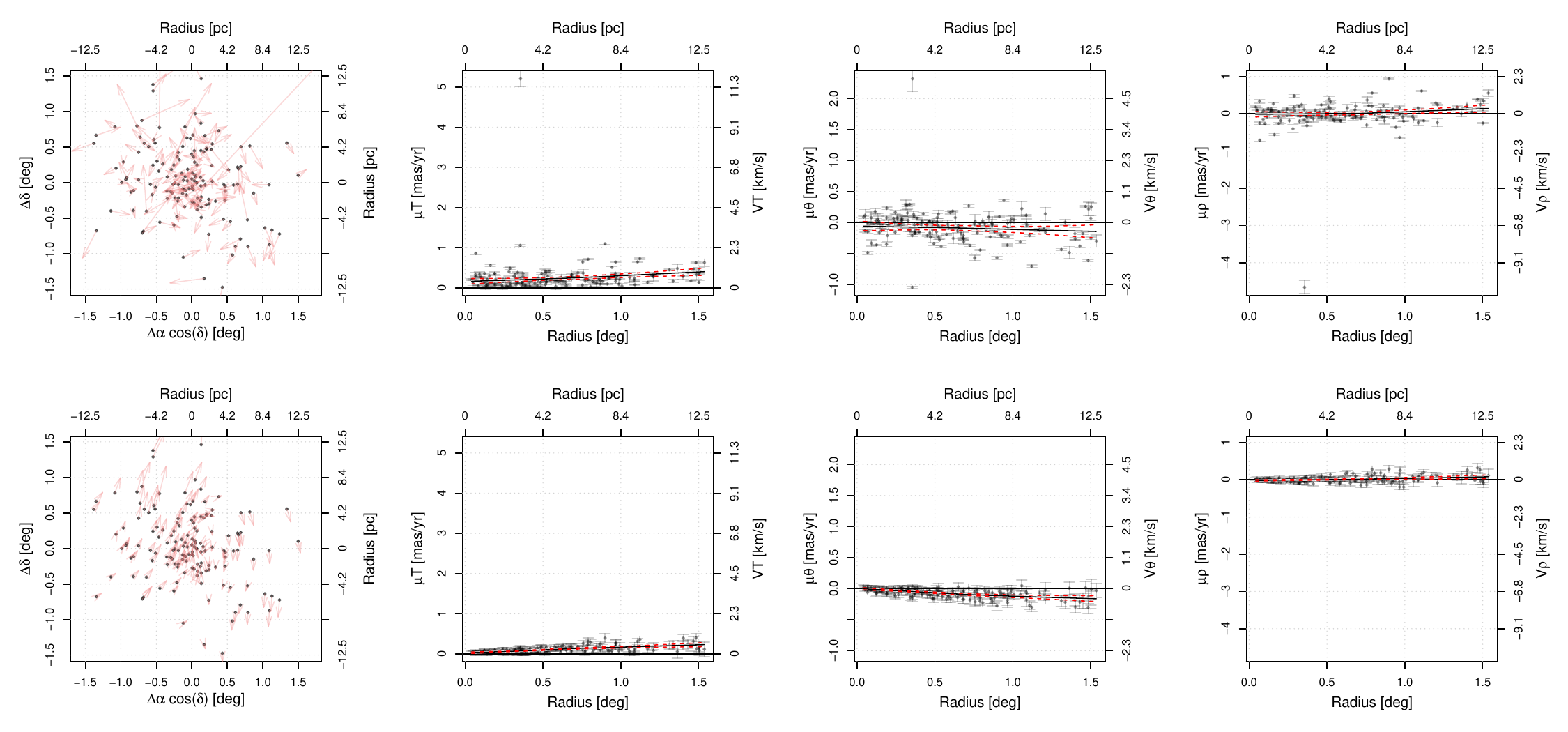}
\caption{Same as Fig.\ \ref{fig:vcurve0}, but for the ASCC 58 cluster. }
\label{fig:vcurve14}
\end{figure*}
\begin{figure*}[htb]
\includegraphics[width=\linewidth]{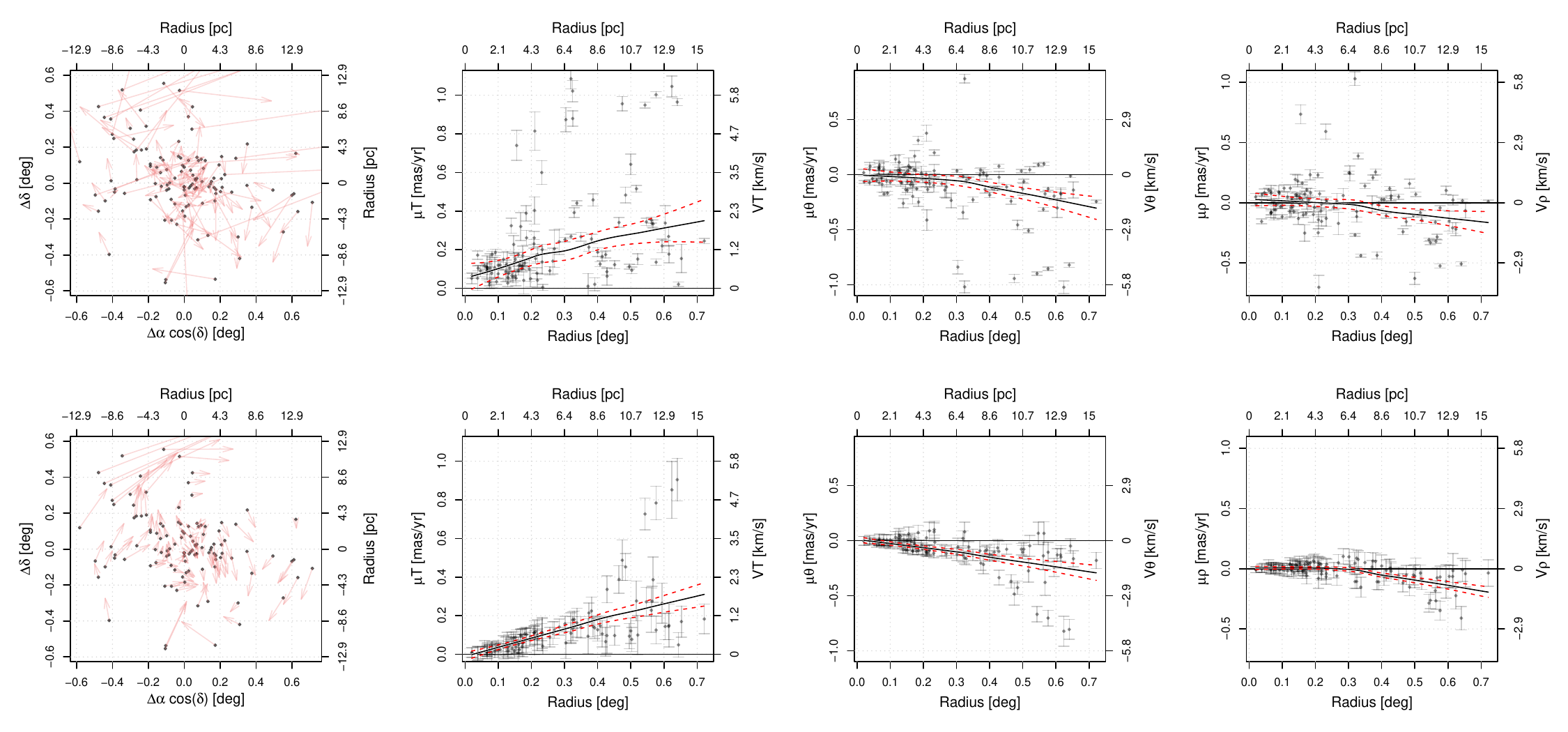}
\caption{Same as Fig.\ \ref{fig:vcurve0}, but for the ASCC 71 cluster. }
\label{fig:vcurve15}
\end{figure*}
\begin{figure*}[htb]
\includegraphics[width=\linewidth]{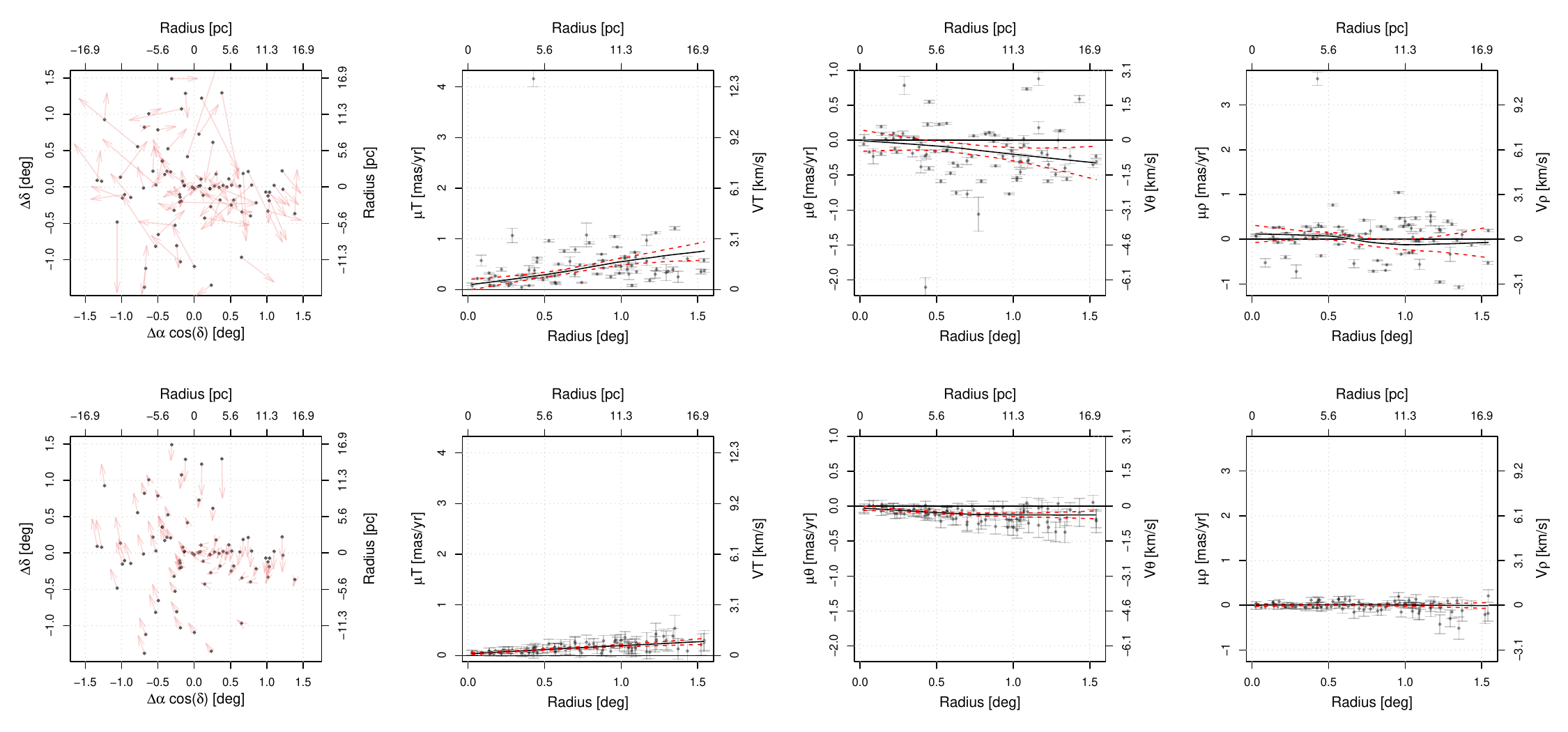}
\caption{Same as Fig.\ \ref{fig:vcurve0}, but for the ASCC 73 cluster. }
\label{fig:vcurve16}
\end{figure*}
\begin{figure*}[htb]
\includegraphics[width=\linewidth]{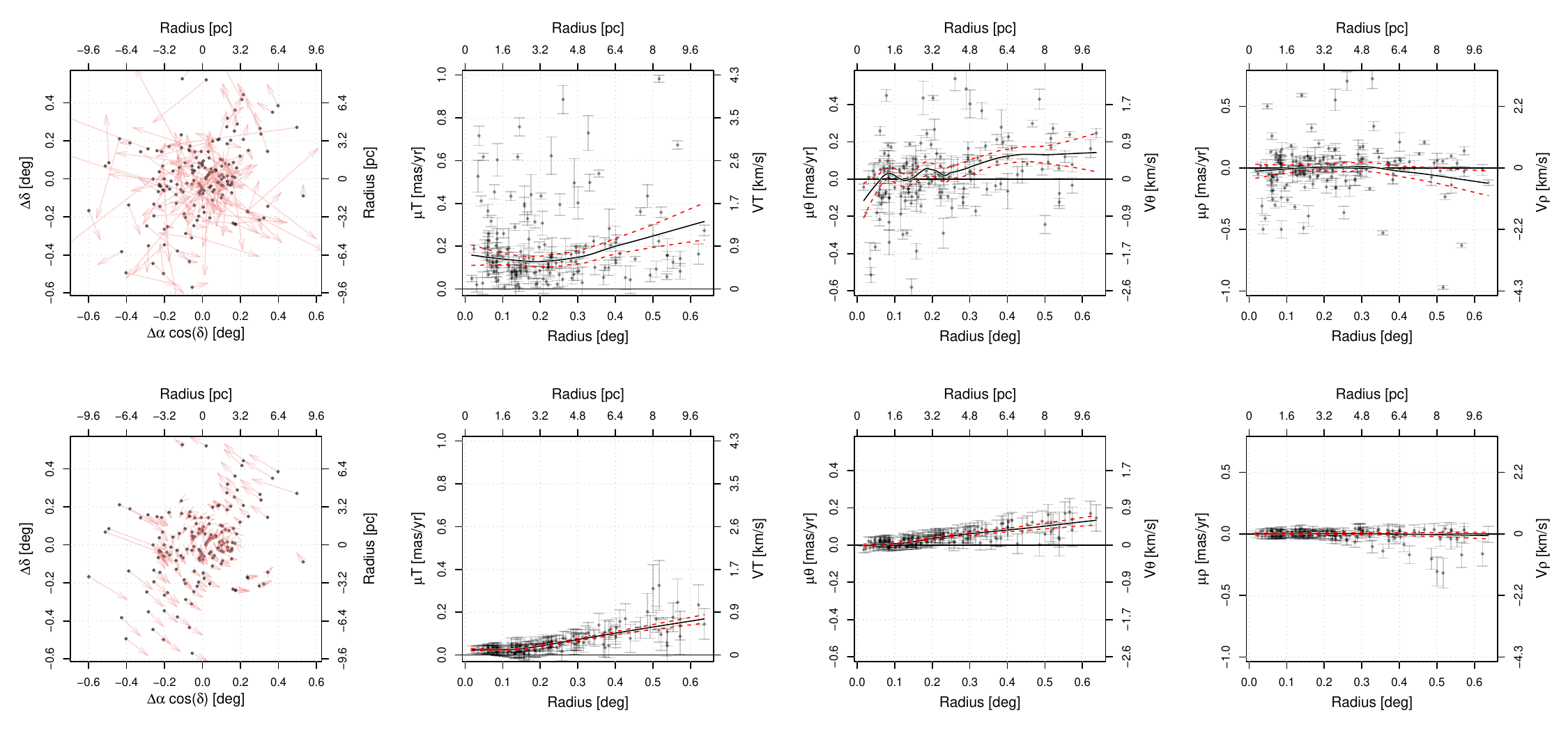}
\caption{Same as Fig.\ \ref{fig:vcurve0}, but for the ASCC 114 cluster.}
\label{fig:vcurve9}
\end{figure*}
\begin{figure*}[htb]
\includegraphics[width=\linewidth]{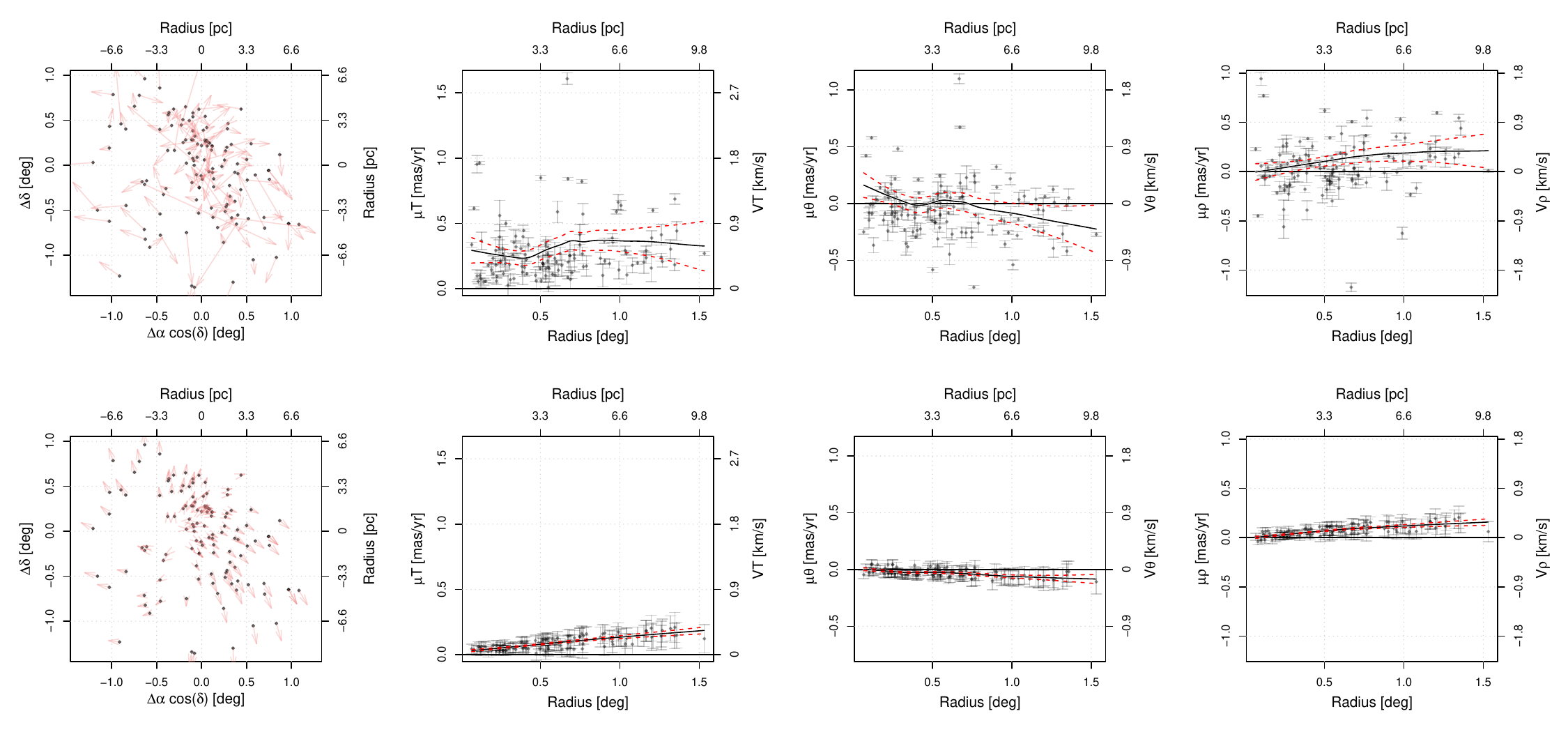}
\caption{Same as Fig.\ \ref{fig:vcurve0}, but for the ASCC 127 cluster. }
\label{fig:vcurve10}
\end{figure*}
\begin{figure*}[htb]
\includegraphics[width=\linewidth]{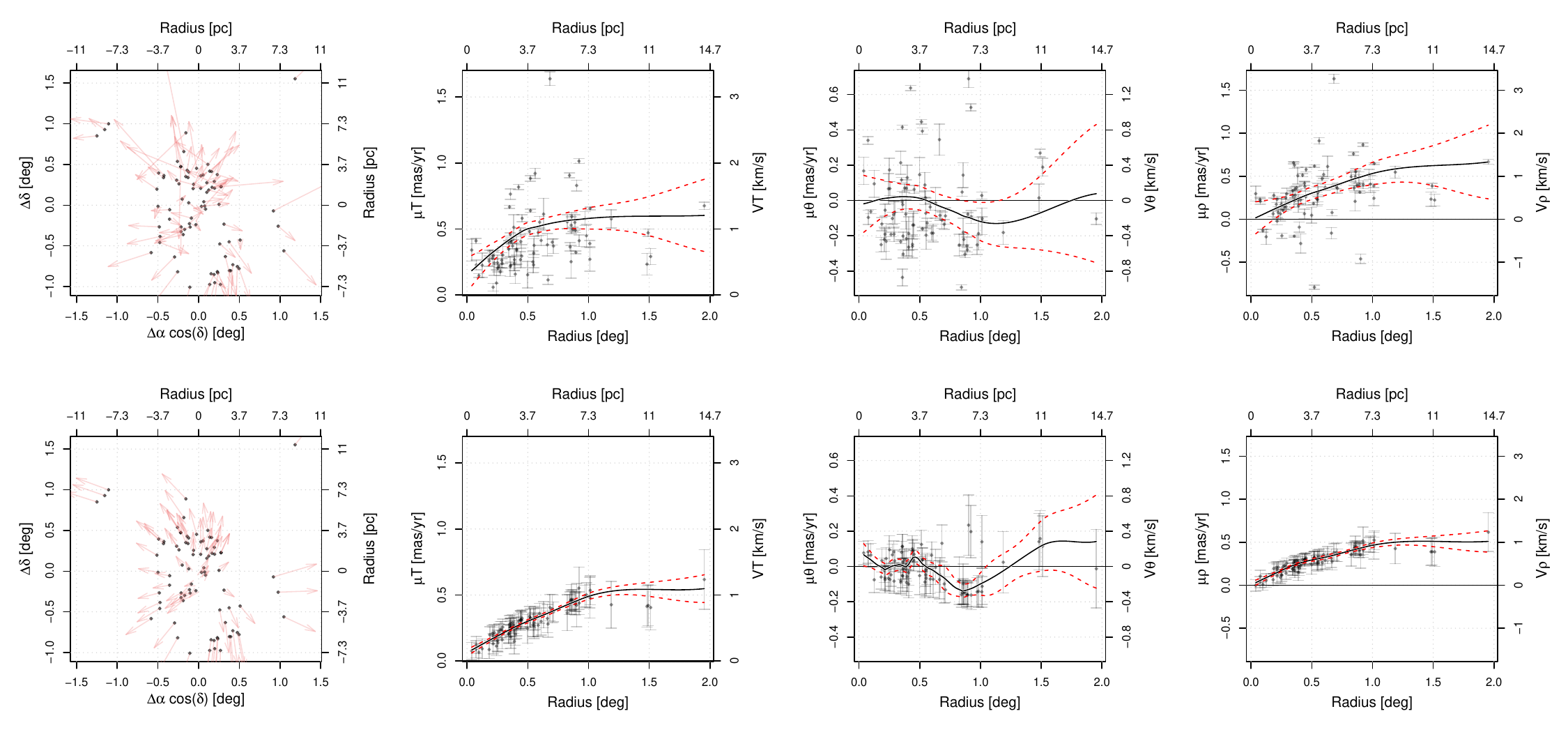}
\caption{Same as Fig.\ \ref{fig:vcurve0}, but for the Aveni Hunter 1 cluster. }
\label{fig:vcurve17}
\end{figure*}
\begin{figure*}[htb]
\includegraphics[width=\linewidth]{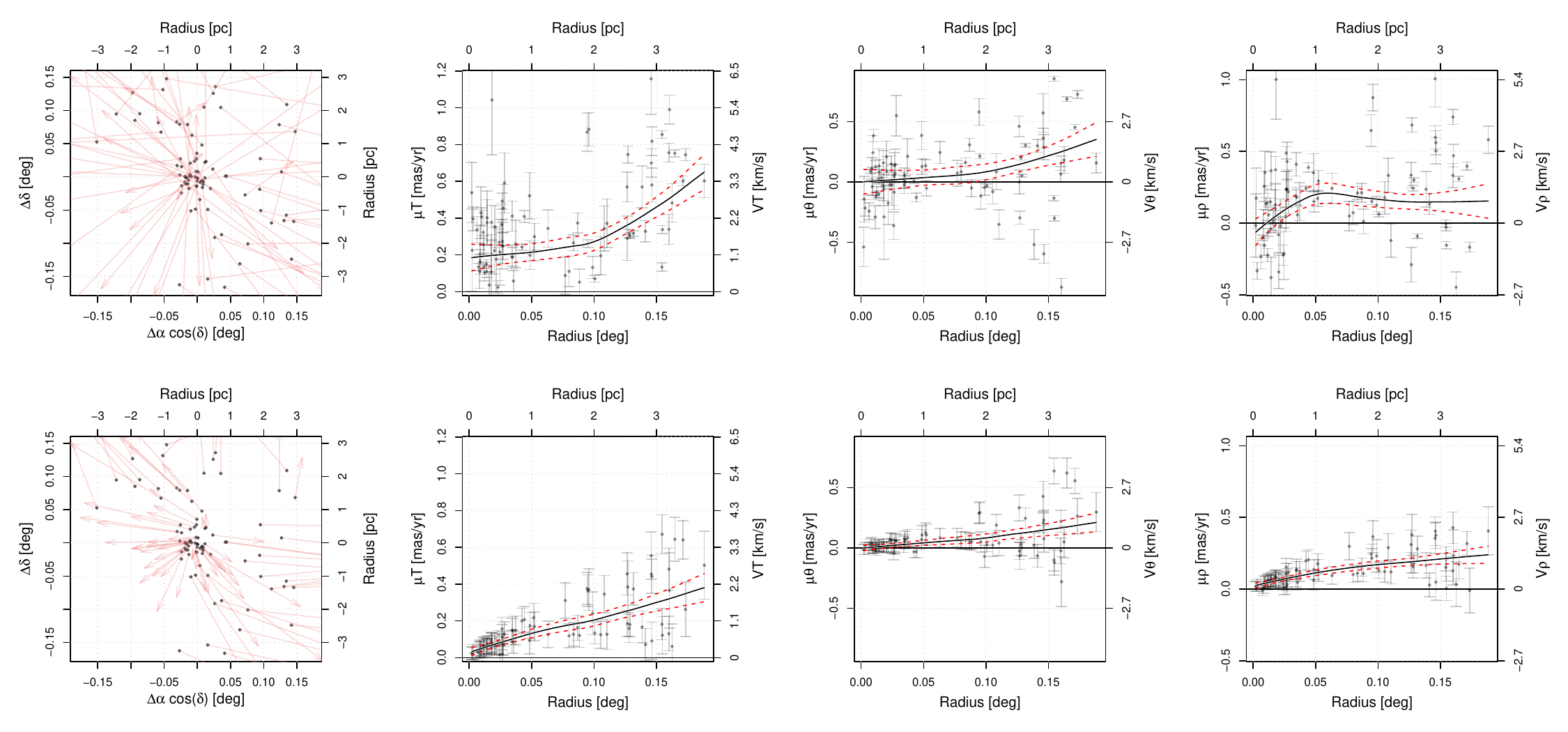}
\caption{Same as Fig.\ \ref{fig:vcurve0}, but for the BDSB96 cluster. }
\label{fig:vcurve18}
\end{figure*}
\begin{figure*}[htb]
\includegraphics[width=\linewidth]{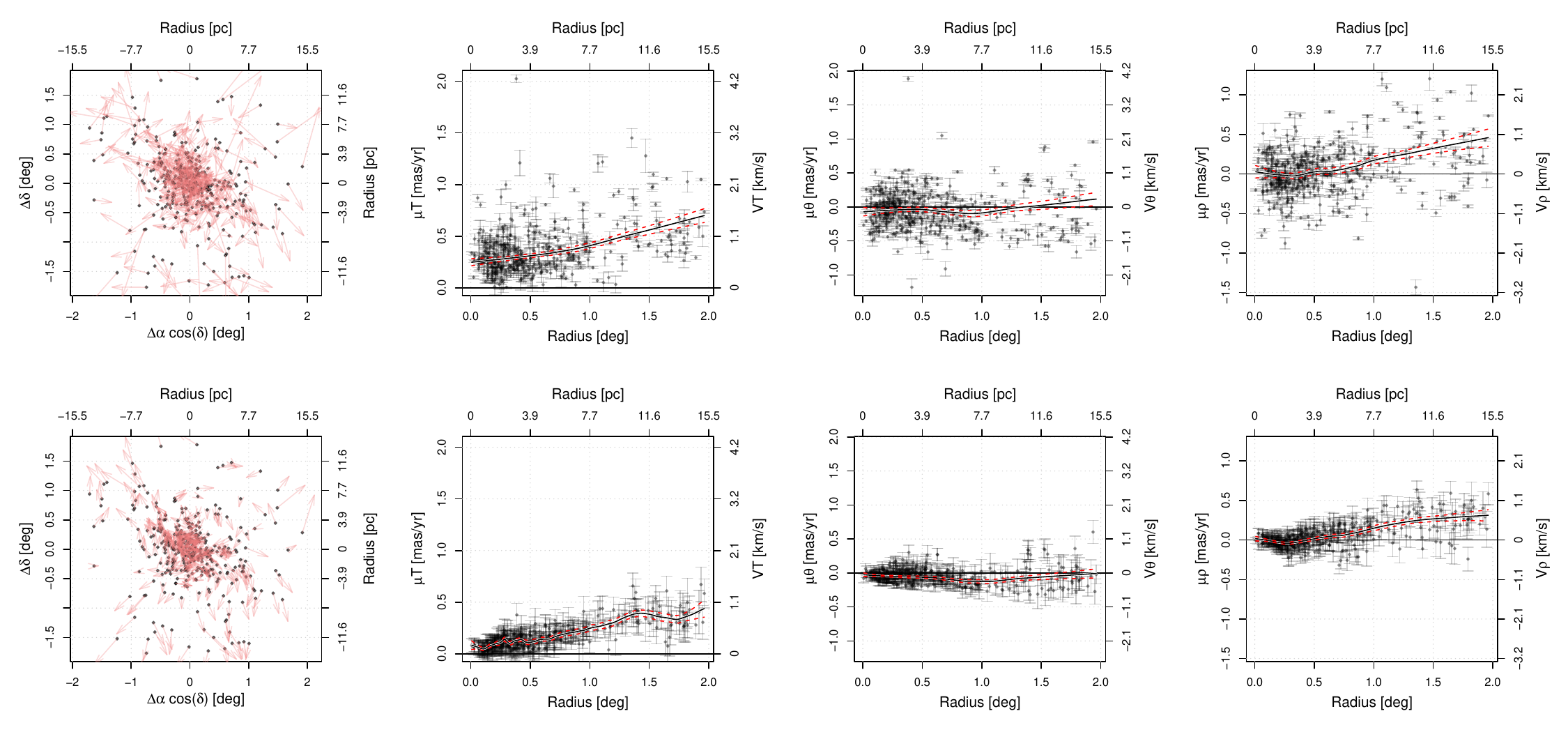}
\caption{Same as Fig.\ \ref{fig:vcurve0}, but for the BH 99 cluster. }
\label{fig:vcurve20}
\end{figure*}
\begin{figure*}[htb]
\includegraphics[width=\linewidth]{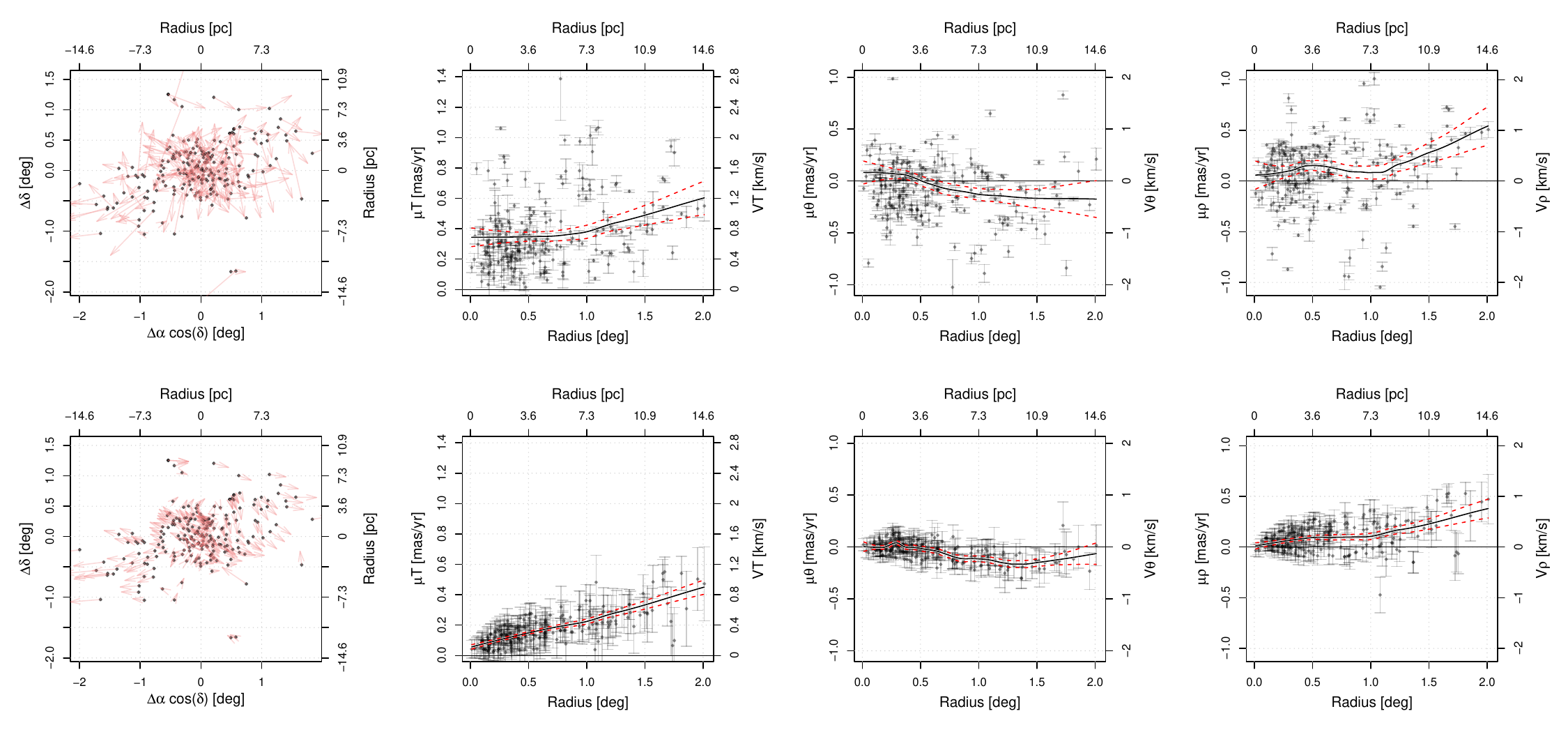}
\caption{Same as Fig.\ \ref{fig:vcurve0}, but for the BH 164 cluster. }
\label{fig:vcurve19}
\end{figure*}
\begin{figure*}[htb]
\includegraphics[width=\linewidth]{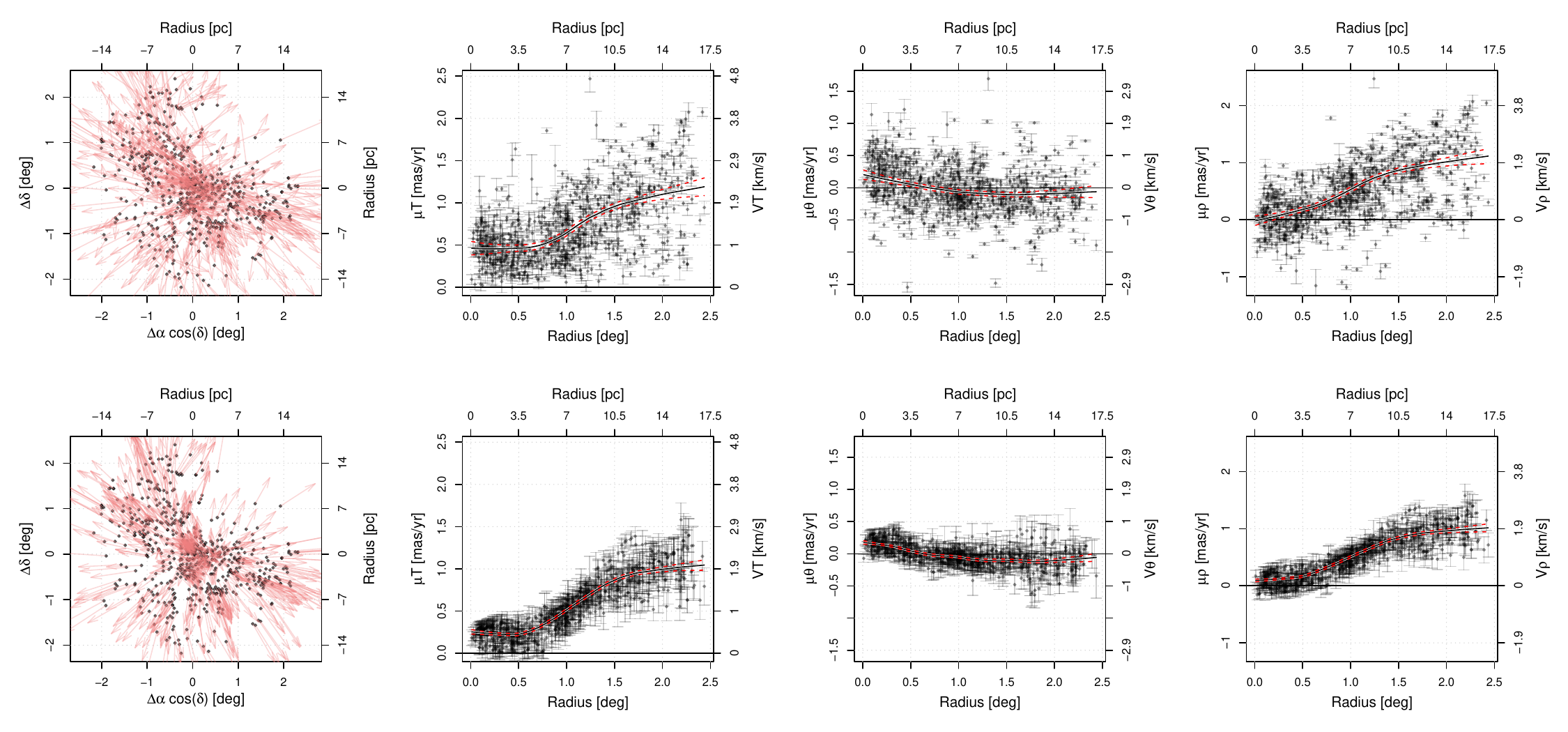}
\caption{Same as Fig.\ \ref{fig:vcurve0}, but for the Collinder 69 cluster. }
\label{fig:vcurve25}
\end{figure*}
\begin{figure*}[htb]
\includegraphics[width=\linewidth]{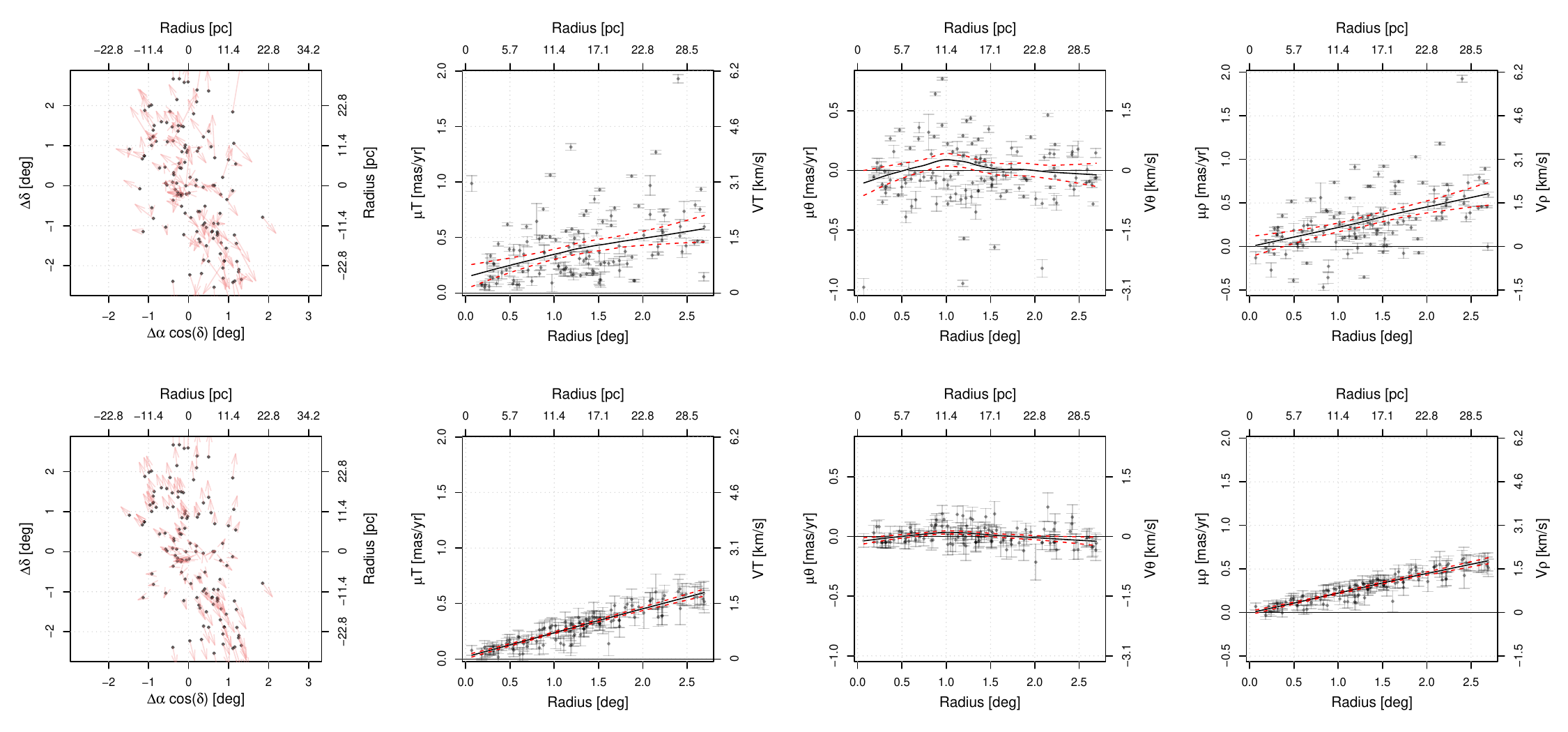}
\caption{Same as Fig.\ \ref{fig:vcurve0}, but for the Collinder 132 cluster. }
\label{fig:vcurve21}
\end{figure*}
\begin{figure*}[htb]
\includegraphics[width=\linewidth]{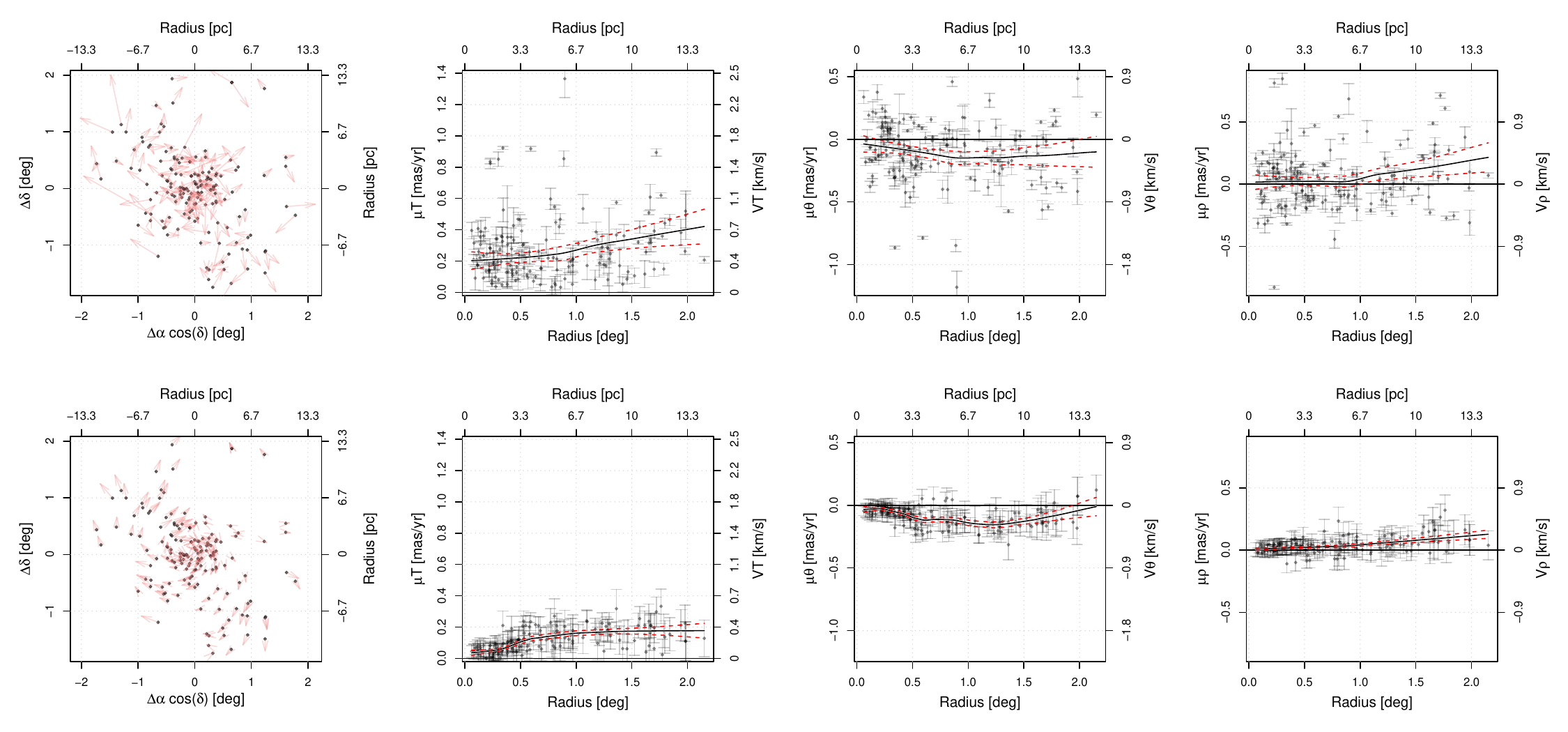}
\caption{Same as Fig.\ \ref{fig:vcurve0}, but for the Collinder 140 cluster. }
\label{fig:vcurve22}
\end{figure*}
\begin{figure*}[htb]
\includegraphics[width=\linewidth]{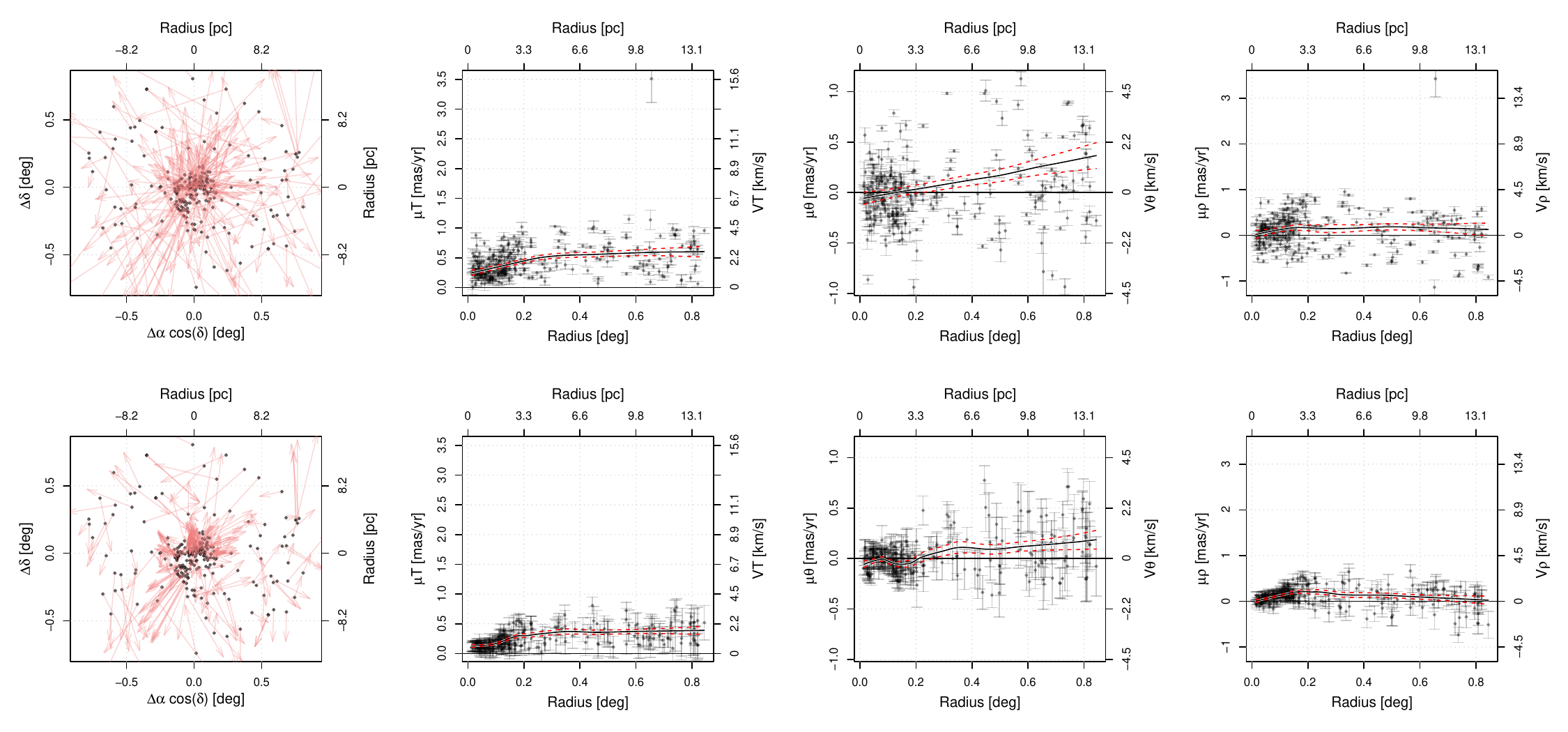}
\caption{Same as Fig.\ \ref{fig:vcurve0}, but for the Collinder 197 cluster.}
\label{fig:vcurve23}
\end{figure*}
\begin{figure*}[htb]
\includegraphics[width=\linewidth]{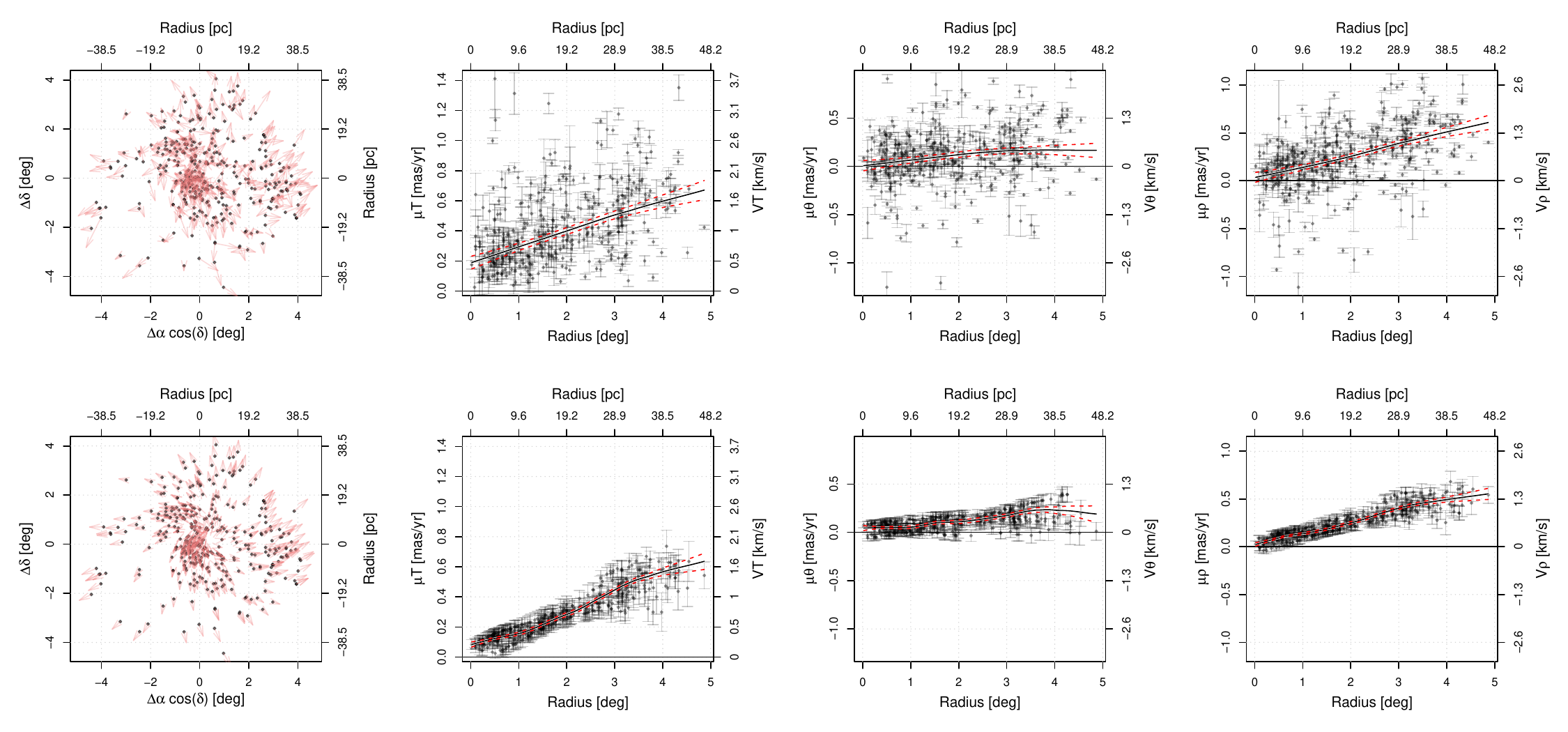}
\caption{Same as Fig.\ \ref{fig:vcurve0}, but for the Collinder 359 cluster. }
\label{fig:vcurve24}
\end{figure*}
\begin{figure*}[htb]
\includegraphics[width=\linewidth]{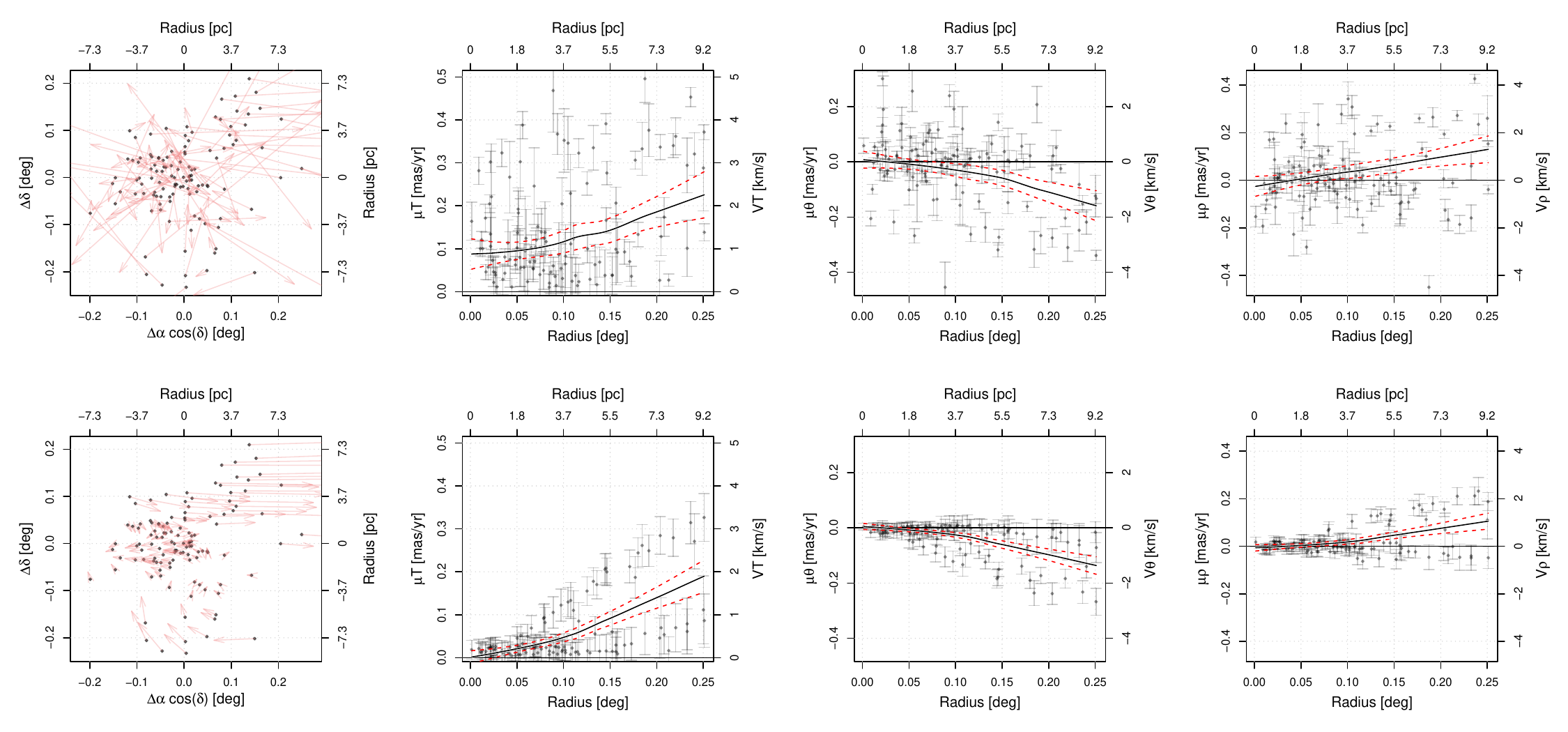}
\caption{Same as Fig.\ \ref{fig:vcurve0}, but for the FSR 0904 cluster. }
\label{fig:vcurve26}
\end{figure*}
\begin{figure*}[htb]
\includegraphics[width=\linewidth]{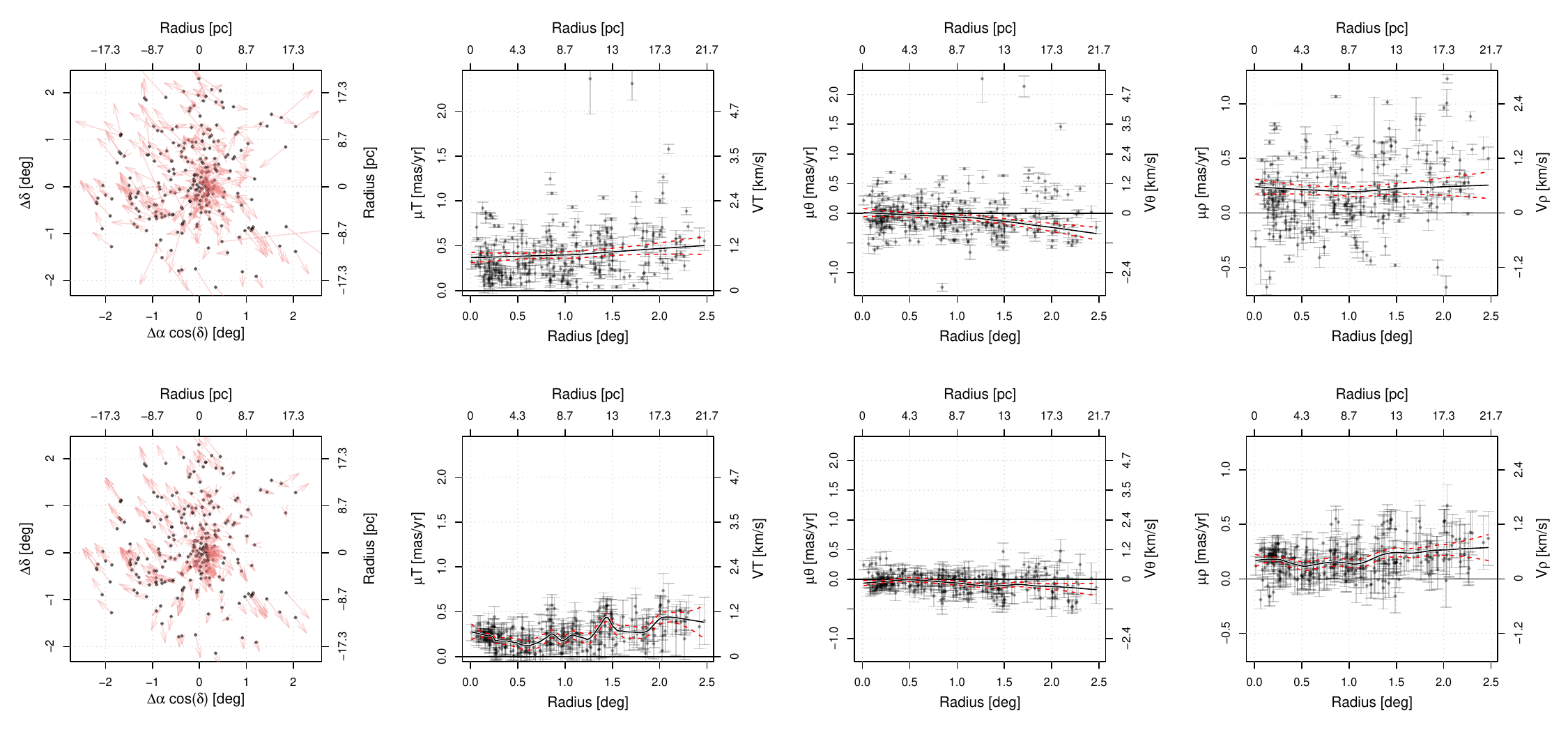}
\caption{Same as Fig.\ \ref{fig:vcurve0}, but for the Gulliver 9 cluster. }
\label{fig:vcurve27}
\end{figure*}
\begin{figure*}[htb]
\includegraphics[width=\linewidth]{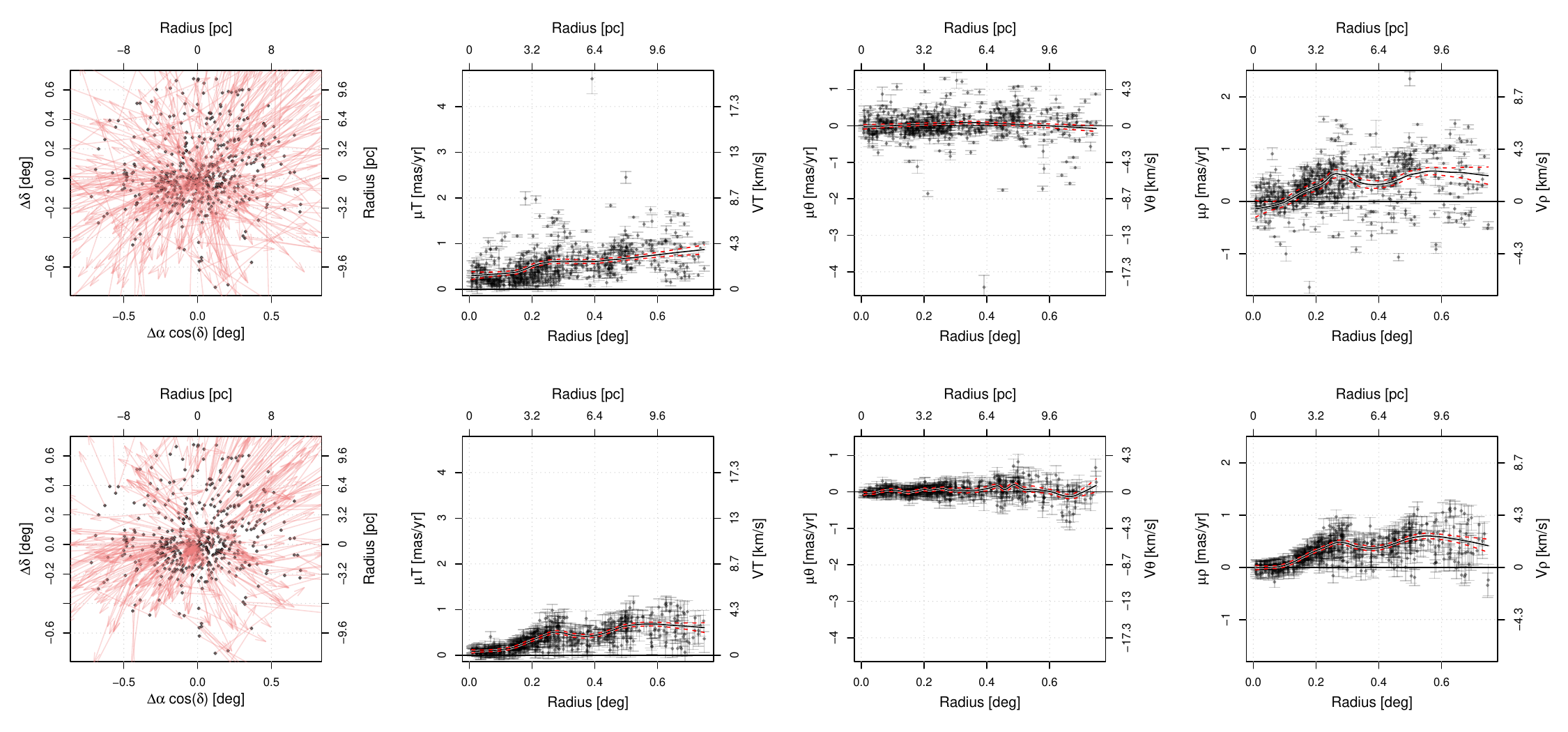}
\caption{Same as Fig.\ \ref{fig:vcurve0}, but for the IC 1396 cluster.}
\label{fig:vcurve28}
\end{figure*}
\begin{figure*}[htb]
\includegraphics[width=\linewidth]{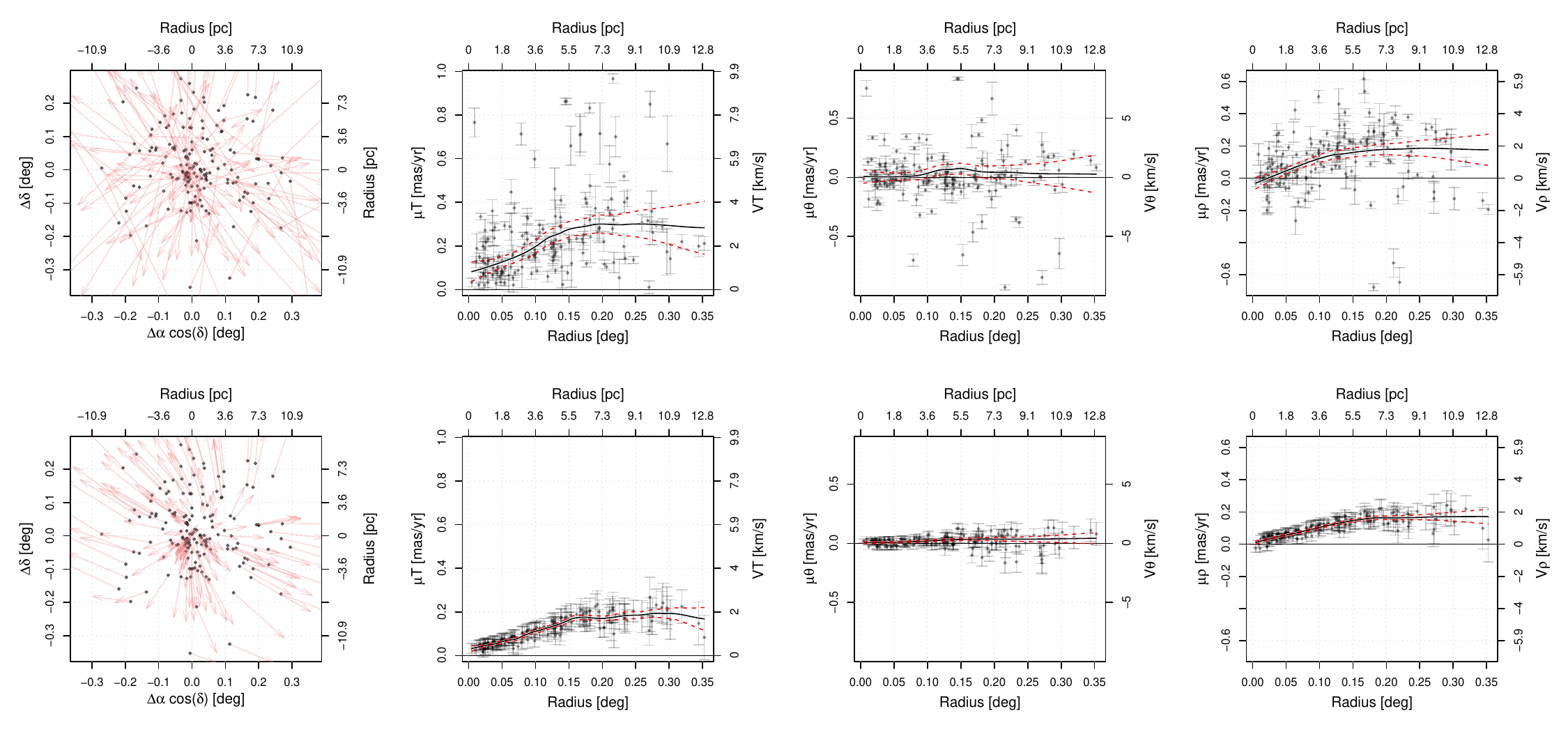}
\caption{Same as Fig.\ \ref{fig:vcurve0}, but for the IC 1805 cluster. }
\label{fig:vcurve29}
\end{figure*}
\begin{figure*}[htb]
\includegraphics[width=\linewidth]{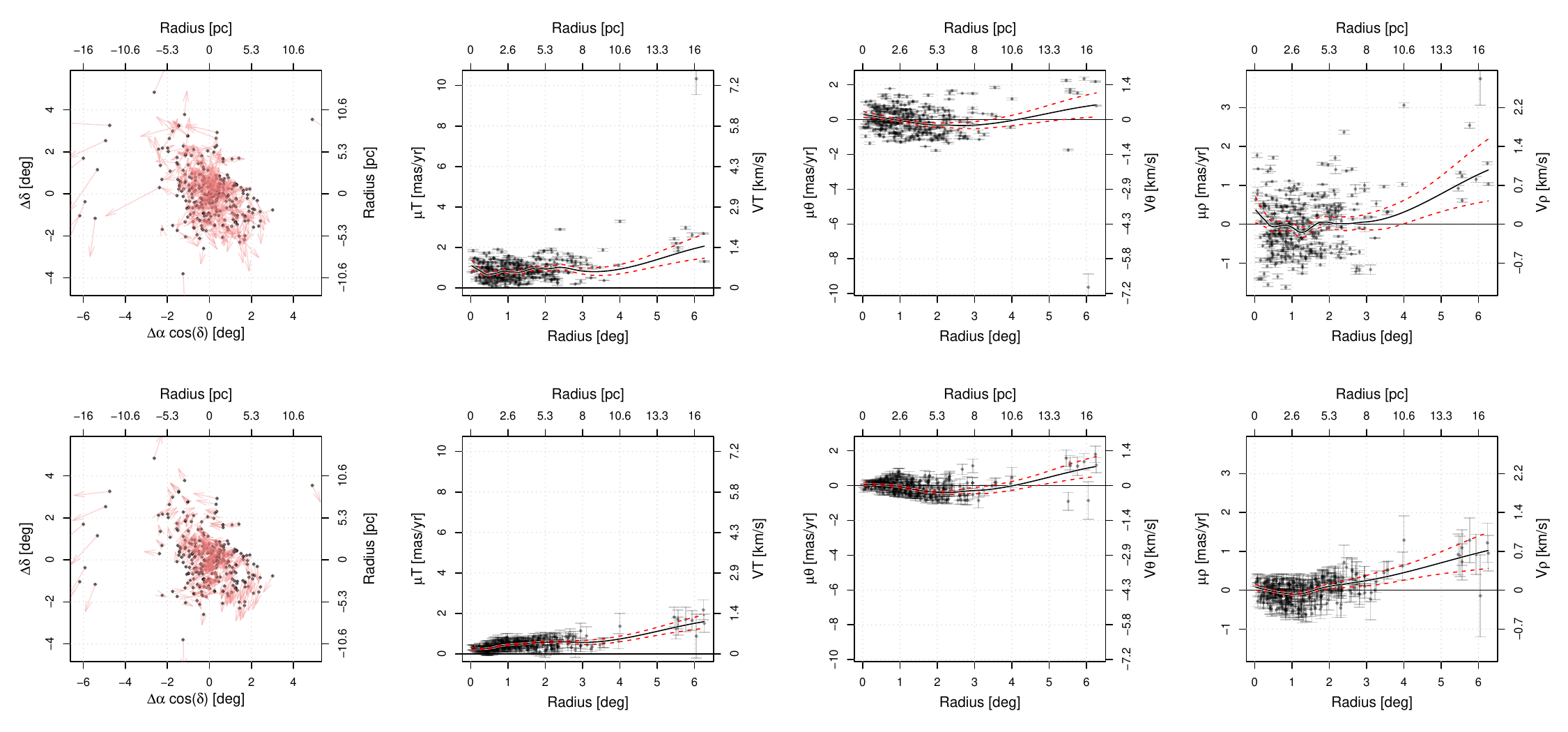}
\caption{Same as Fig.\ \ref{fig:vcurve0}, but for the IC 2602 cluster. }
\label{fig:vcurve30}
\end{figure*}
\begin{figure*}[htb]
\includegraphics[width=\linewidth]{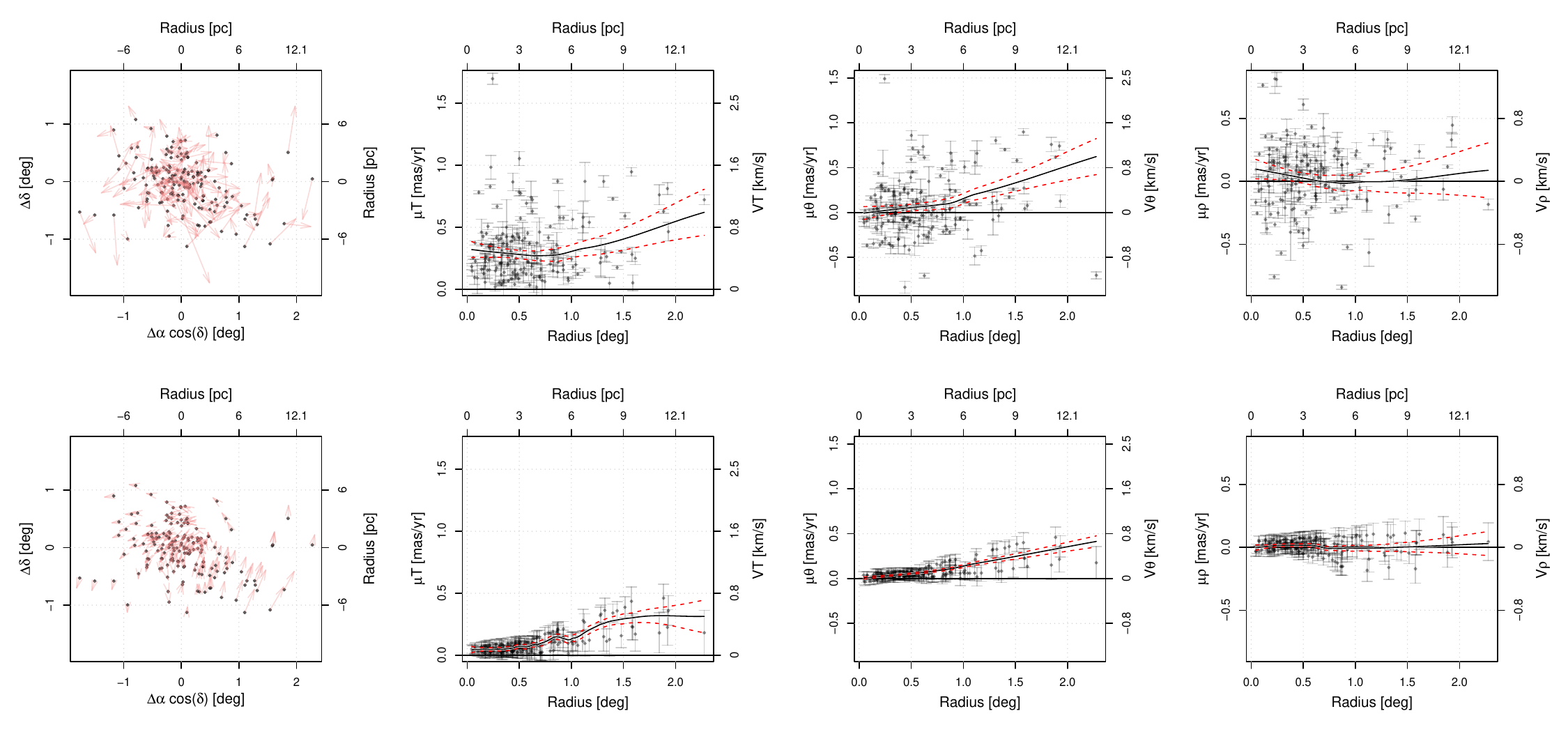}
\caption{Same as Fig.\ \ref{fig:vcurve0}, but for the IC 4665 cluster. }
\label{fig:vcurve31}
\end{figure*}
\begin{figure*}[htb]
\includegraphics[width=\linewidth]{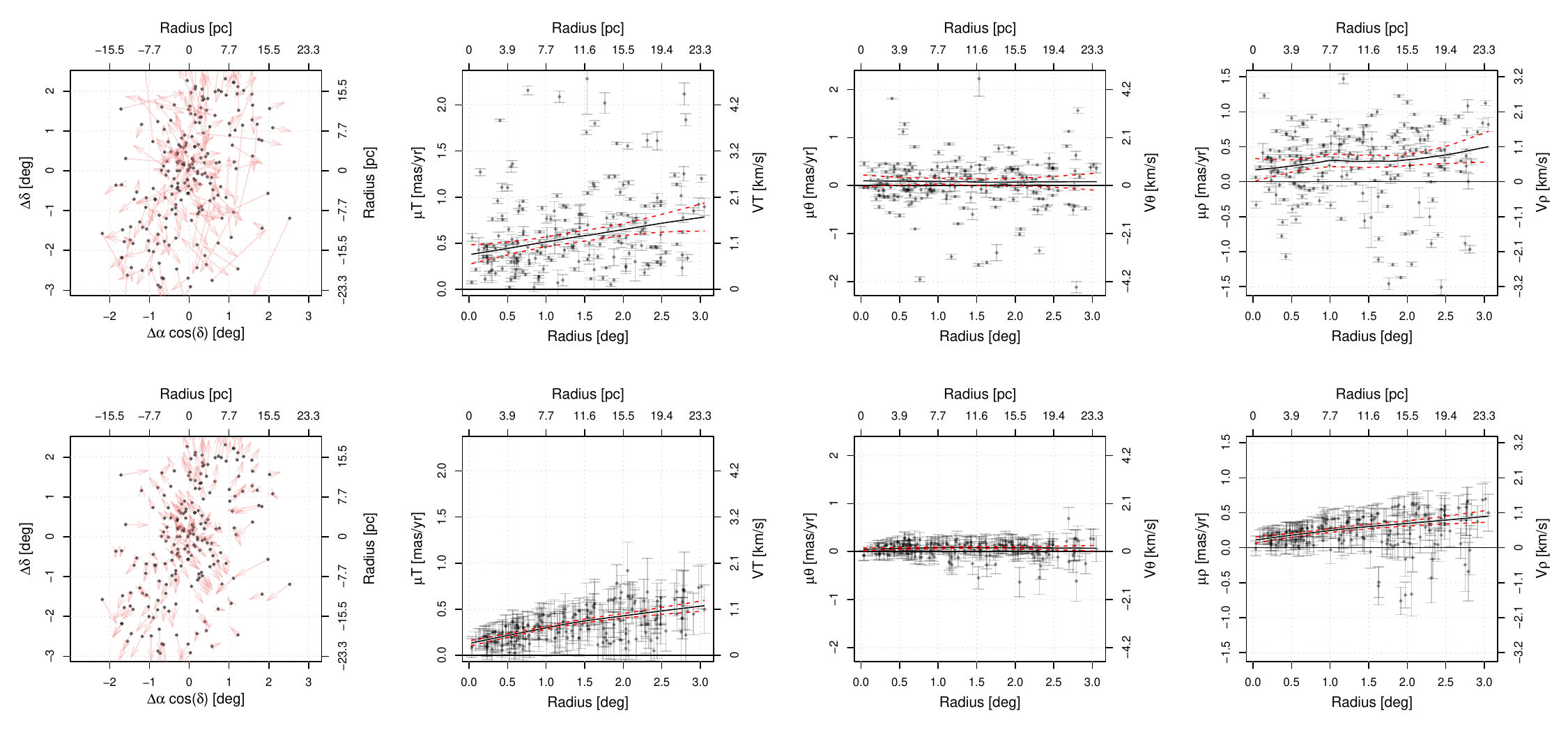}
\caption{ Same as Fig.\ \ref{fig:vcurve0}, but for the Mamajek 4 cluster. }
\label{fig:vcurve33}
\end{figure*}
\begin{figure*}[htb]
\includegraphics[width=\linewidth]{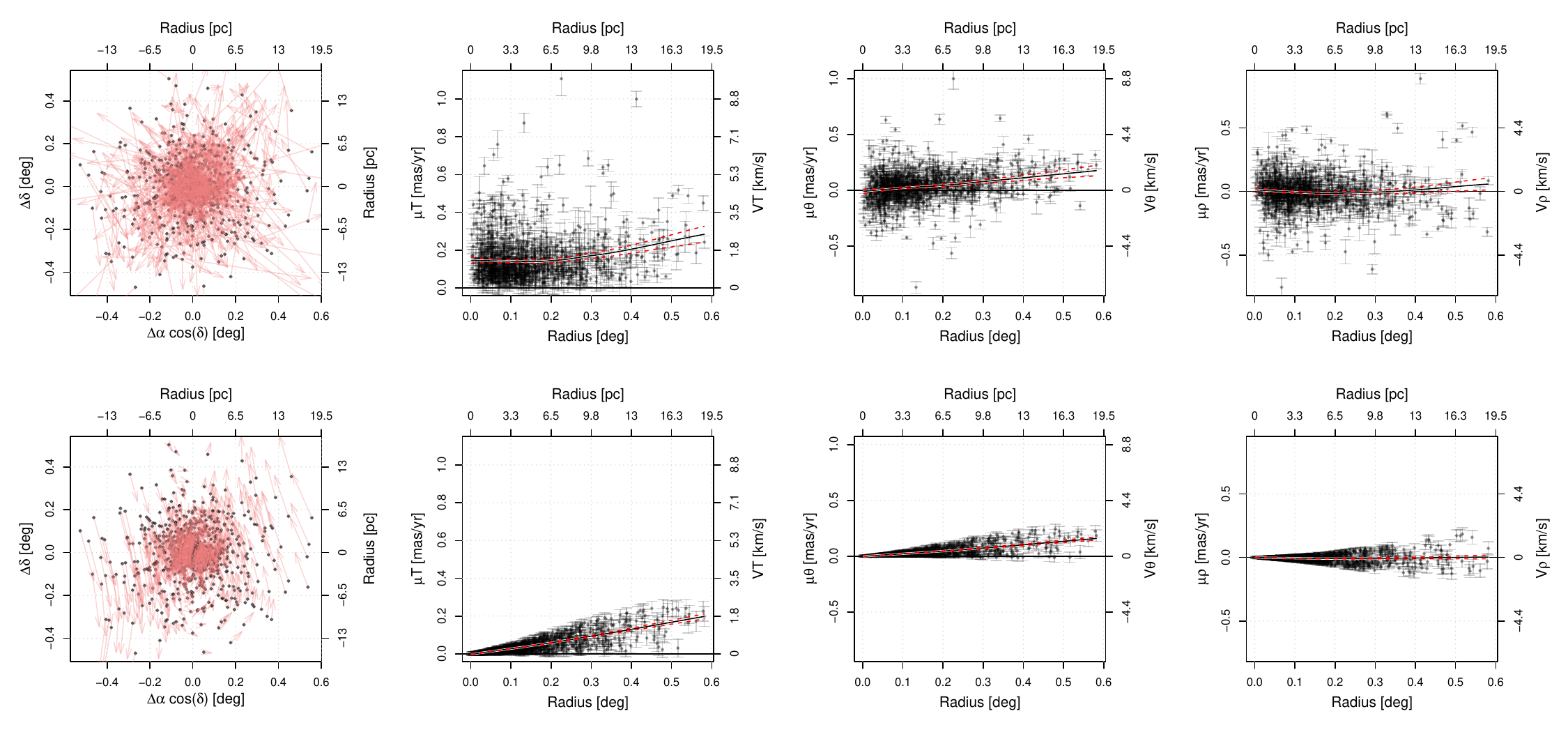}
\caption{Same as Fig.\ \ref{fig:vcurve0}, but for the NGC 188 cluster. }
\label{fig:vcurve34}
\end{figure*}
\begin{figure*}[htb]
\includegraphics[width=\linewidth]{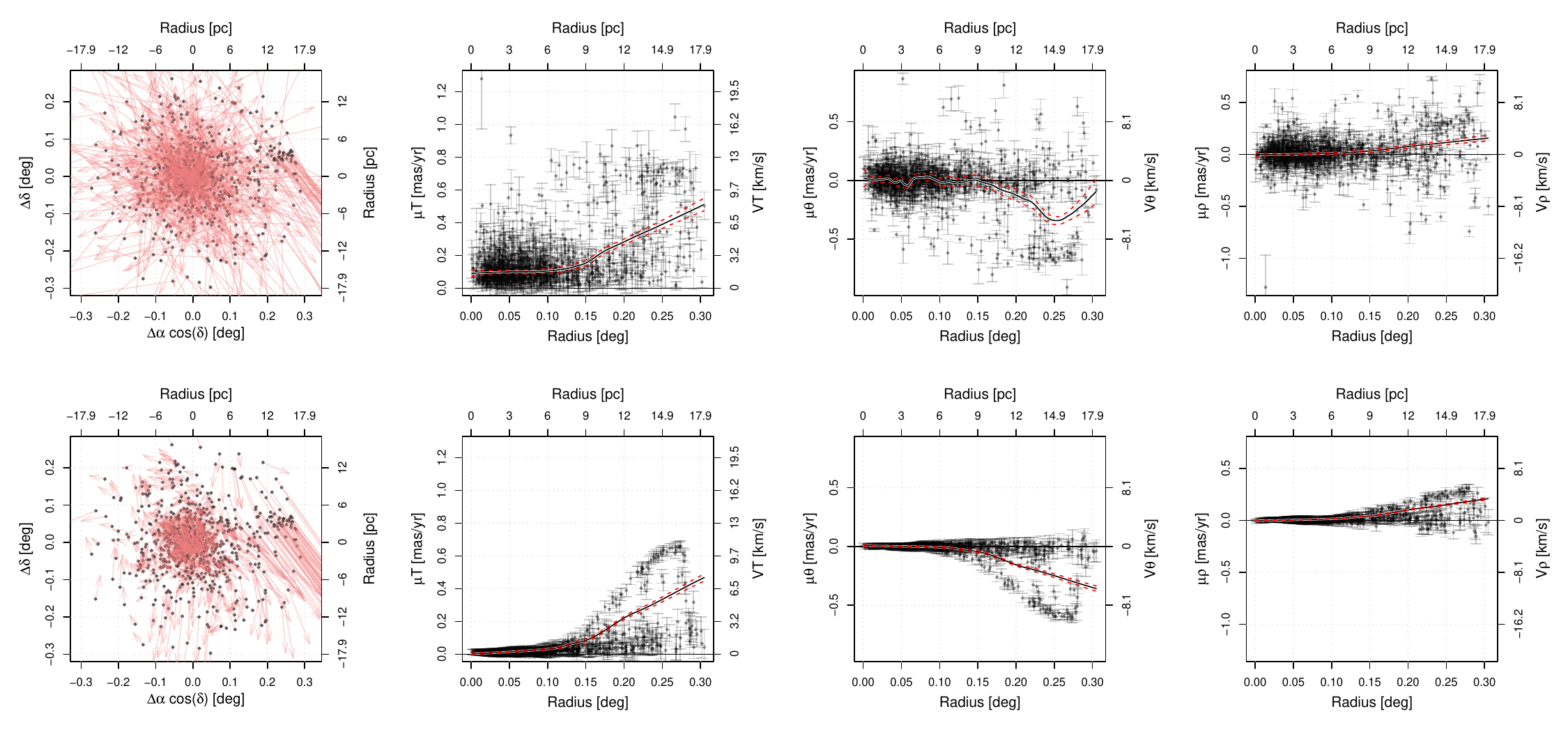}
\caption{Same as Fig.\ \ref{fig:vcurve0}, but for the NGC 2194 cluster.}
\label{fig:vcurve35}
\end{figure*}
\begin{figure*}[htb]
\includegraphics[width=\linewidth]{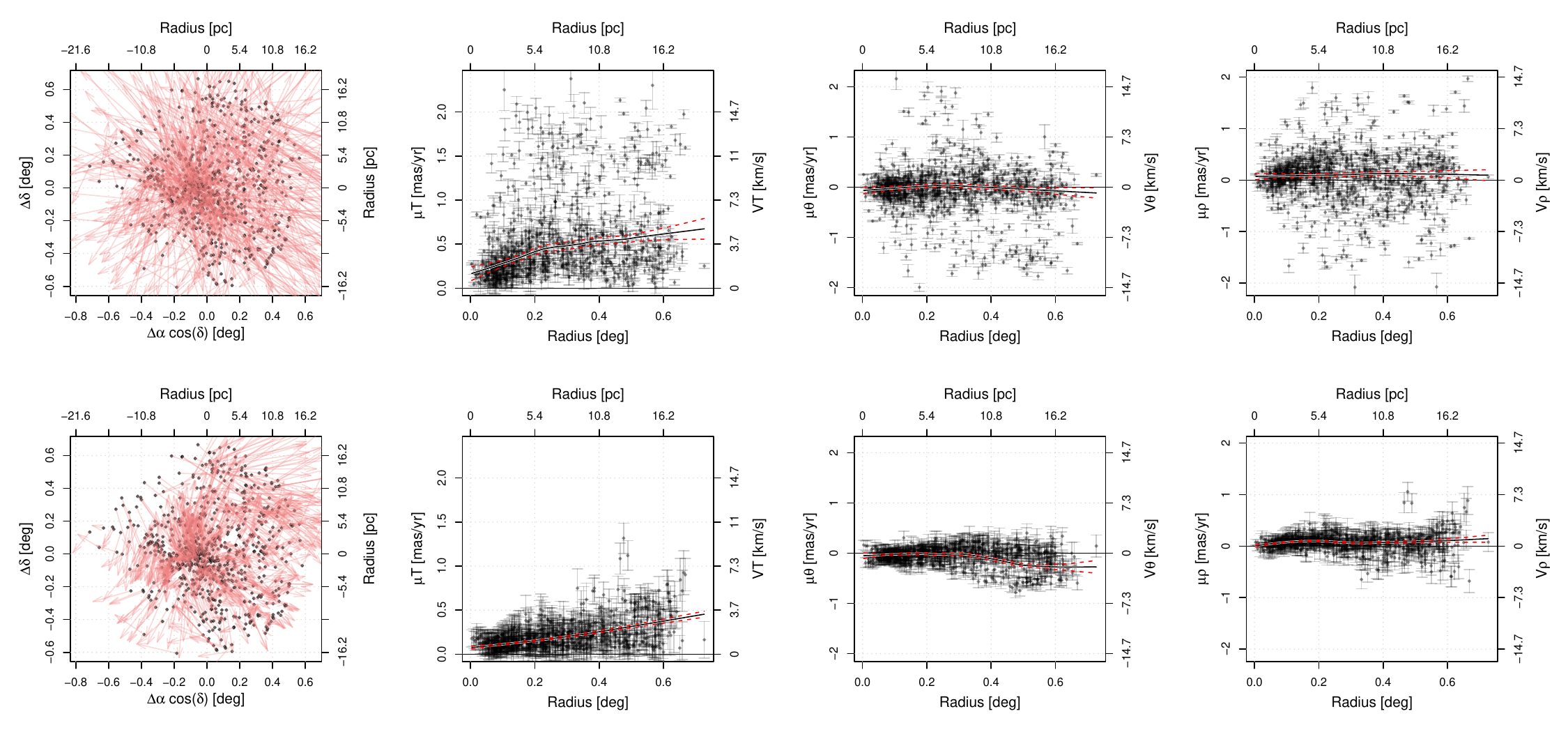}
\caption{Same as Fig.\ \ref{fig:vcurve0}, but for the NGC 2244 cluster. }
\label{fig:vcurve36}
\end{figure*}
\begin{figure*}[htb]
\includegraphics[width=\linewidth]{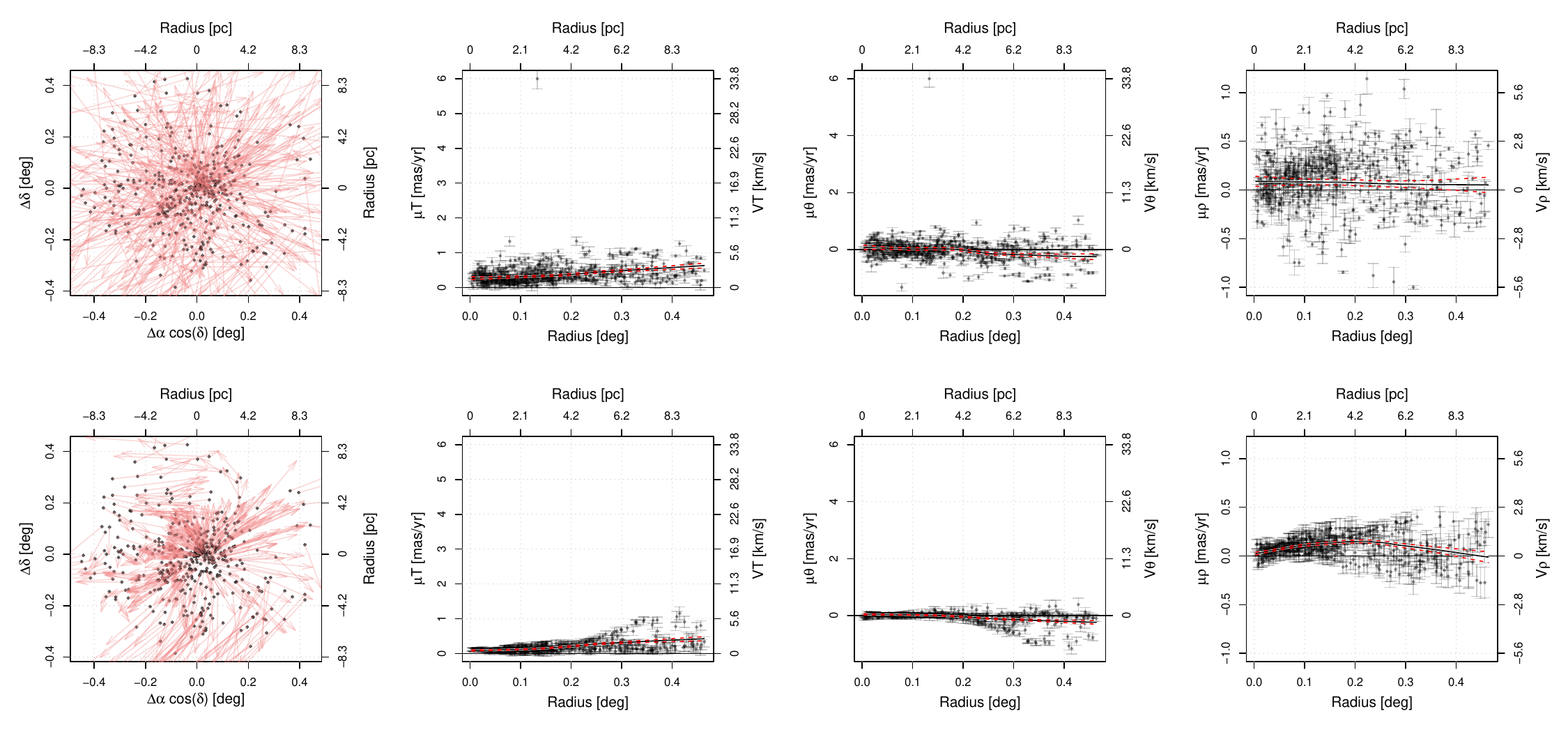}
\caption{Same as Fig.\ \ref{fig:vcurve0}, but for the NGC 6193 cluster. }
\label{fig:vcurve37}
\end{figure*}
\begin{figure*}[htb]
\includegraphics[width=\linewidth]{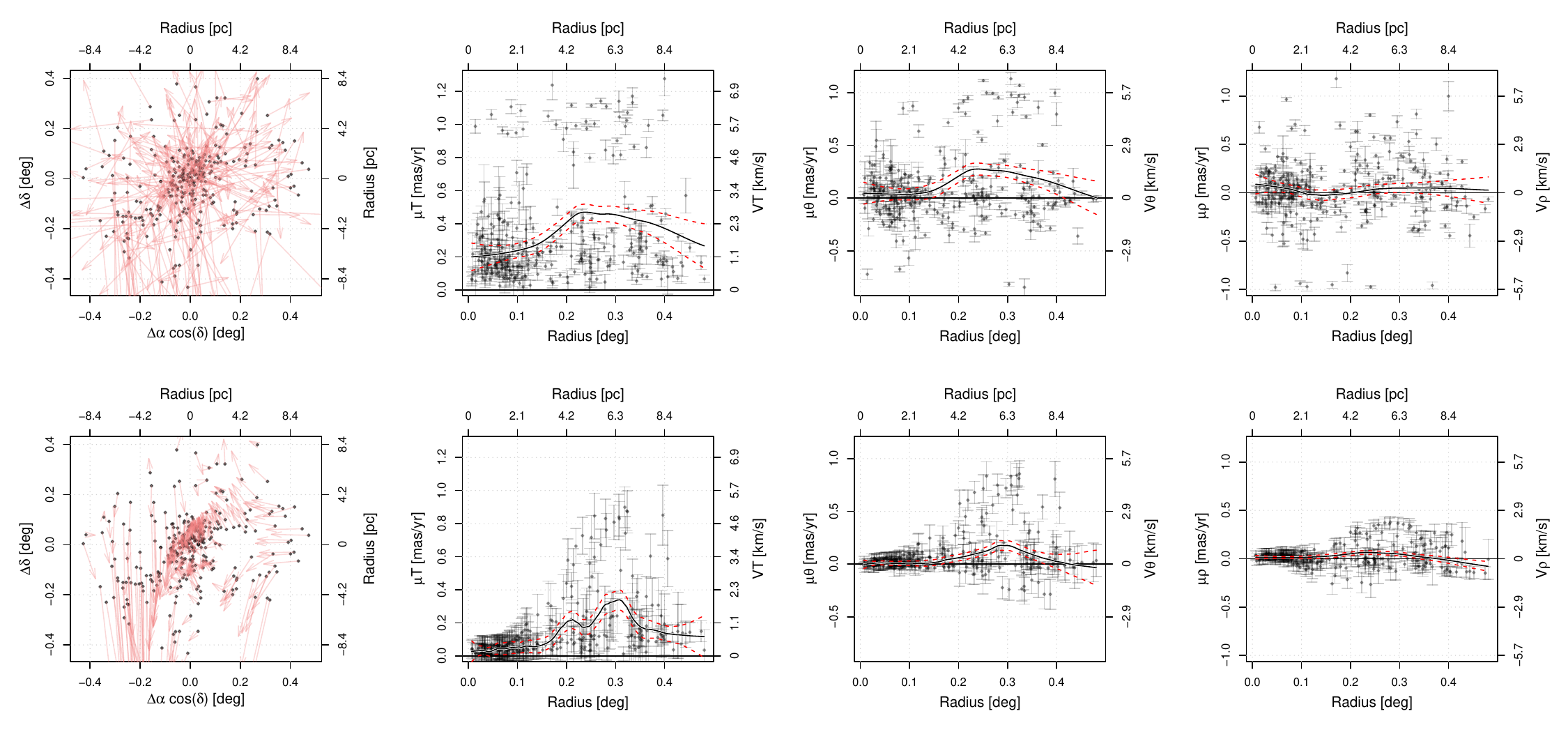}
\caption{Same as Fig.\ \ref{fig:vcurve0}, but for the NGC 6531 cluster. }
\label{fig:vcurve38}
\end{figure*}
\begin{figure*}[htb]
\includegraphics[width=\linewidth]{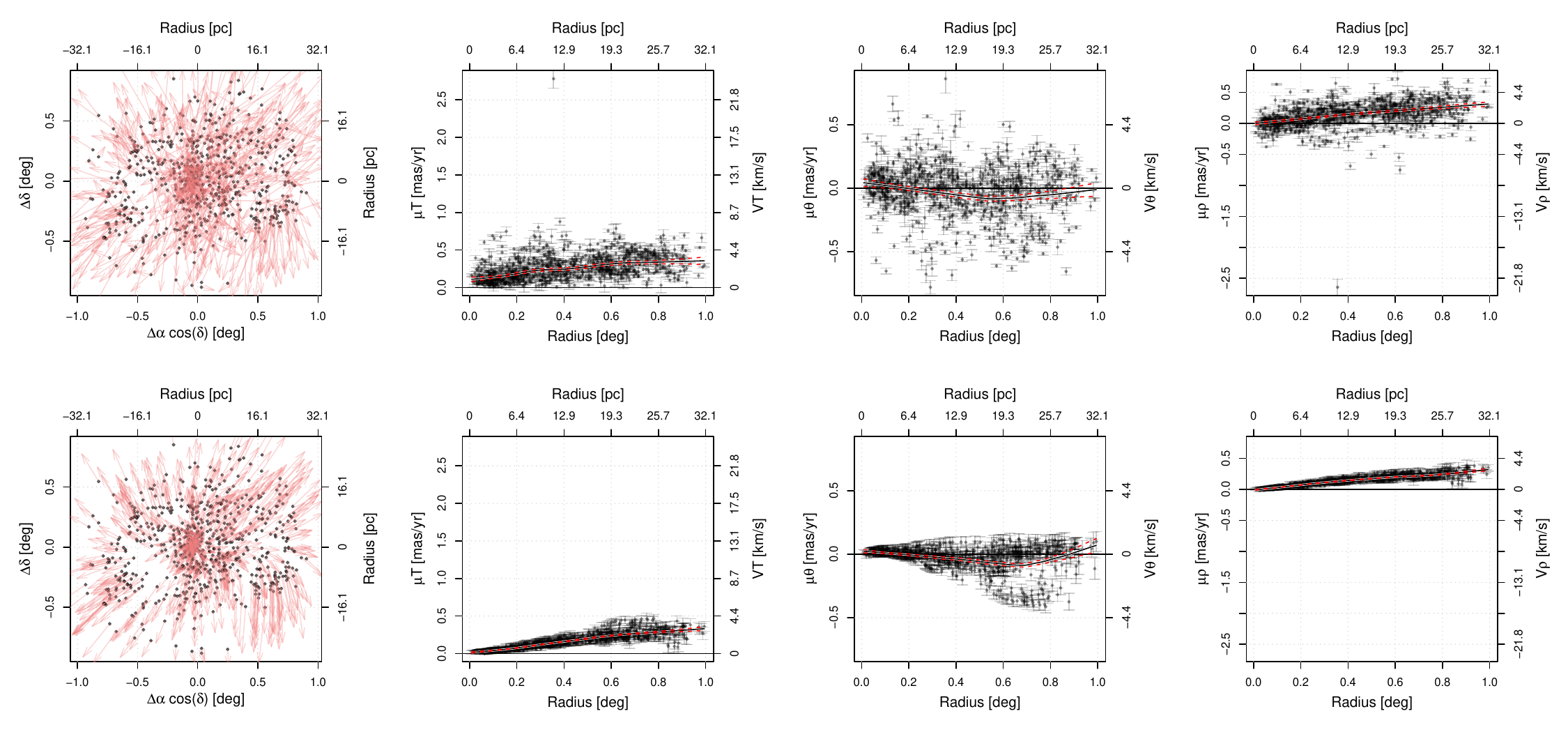}
\caption{Same as Fig.\ \ref{fig:vcurve0}, but for the NGC 6871 cluster. }
\label{fig:vcurve39}
\end{figure*}
\begin{figure*}[htb]
\includegraphics[width=\linewidth]{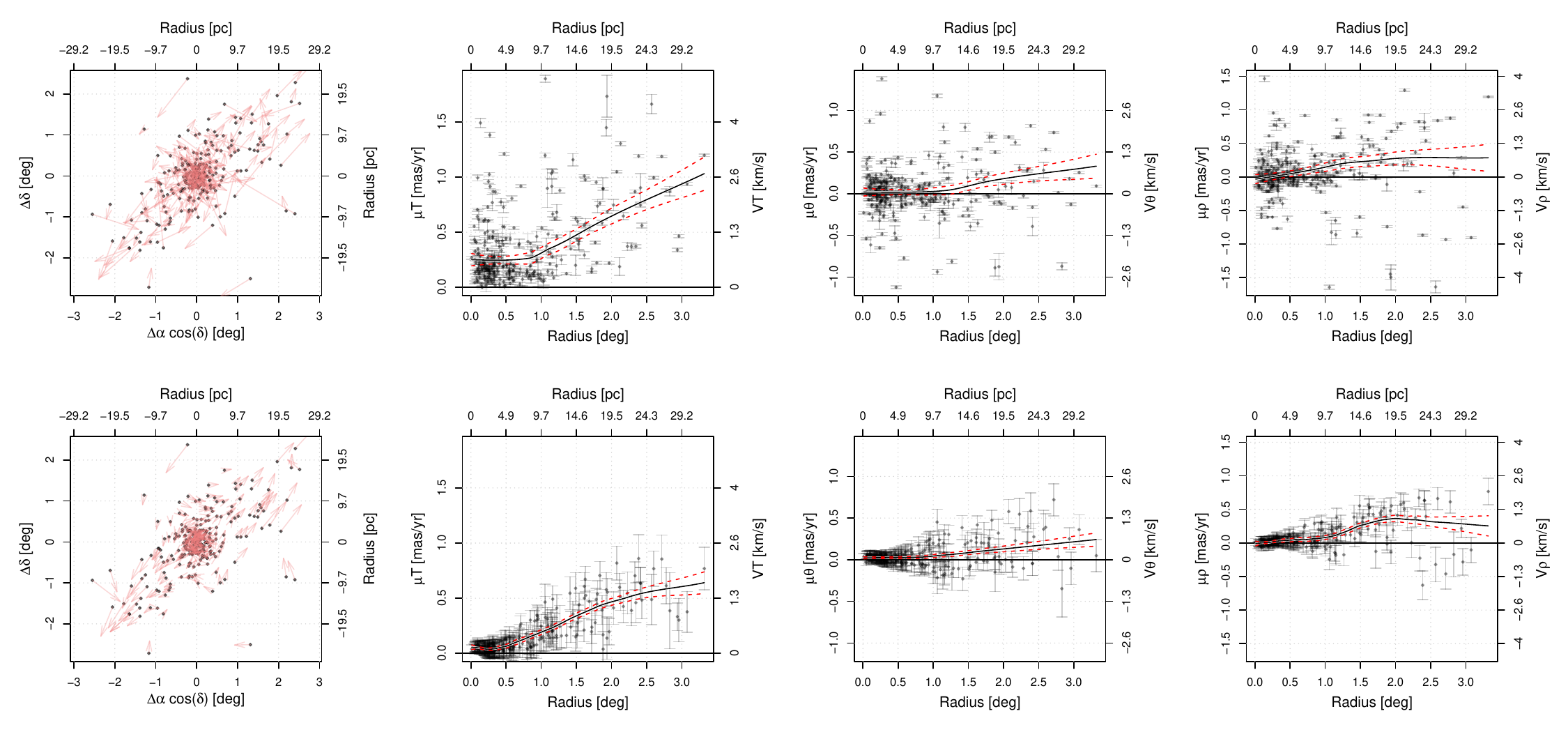}
\caption{Same as Fig.\ \ref{fig:vcurve0}, but for the NGC 6991 cluster. }
\label{fig:vcurve40}
\end{figure*}
\begin{figure*}[htb]
\includegraphics[width=\linewidth]{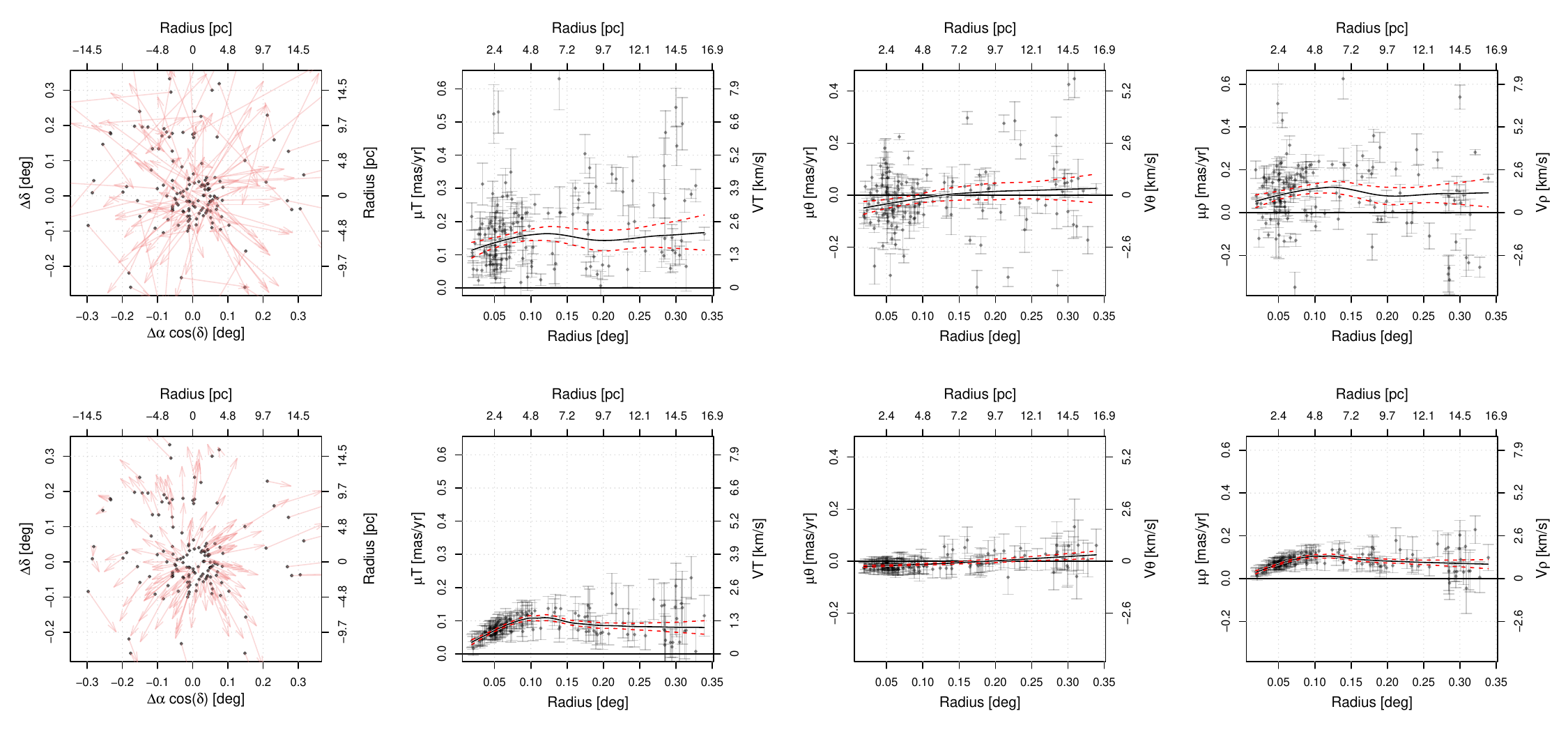}
\caption{Same as Fig.\ \ref{fig:vcurve0}, but for the NGC 7380 cluster.}
\label{fig:vcurve41}
\end{figure*}
\begin{figure*}[htb]
\includegraphics[width=\linewidth]{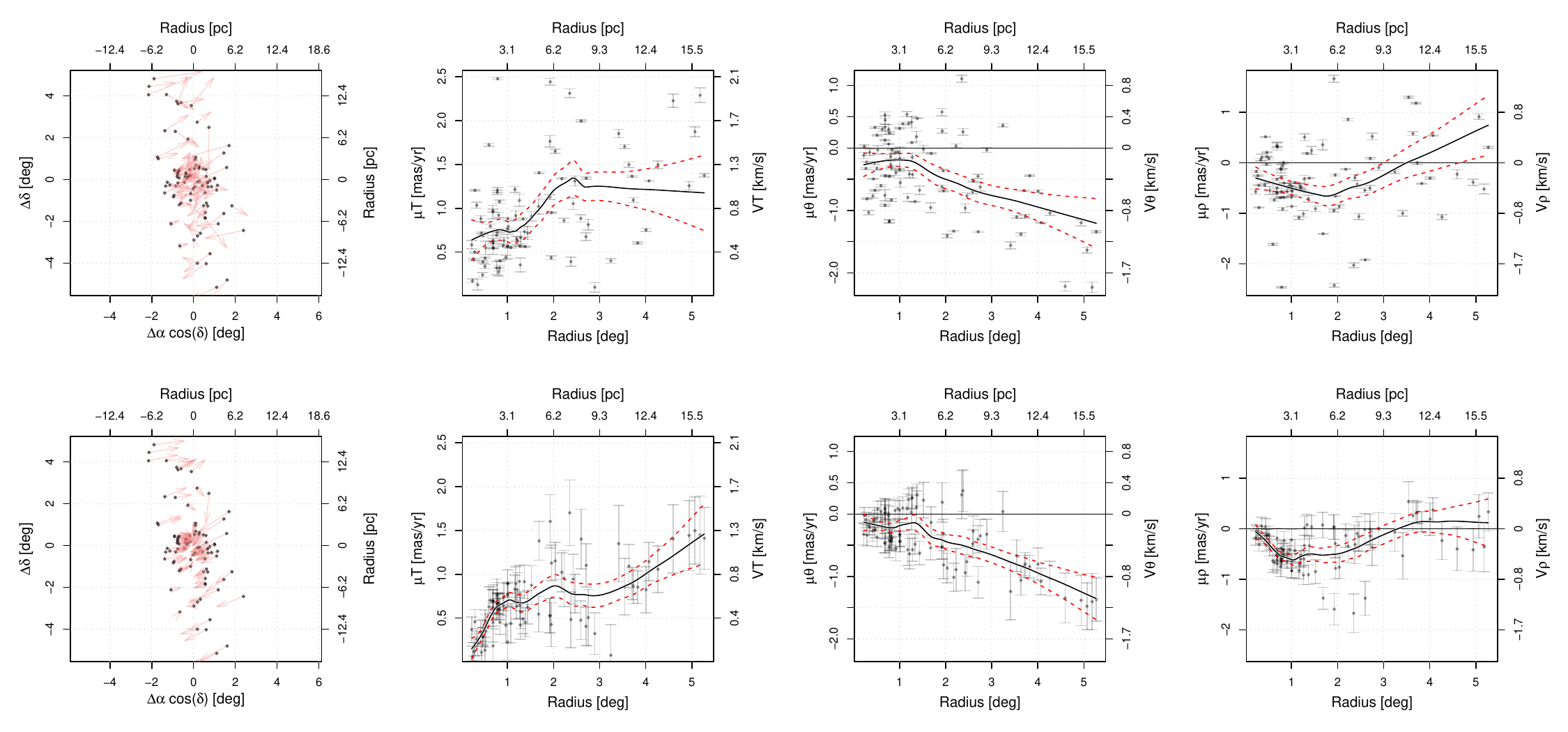}
\caption{Same as Fig.\ \ref{fig:vcurve0}, but for the Platais 3 cluster. }
\label{fig:vcurve42}
\end{figure*}
\begin{figure*}[htb]
\includegraphics[width=\linewidth]{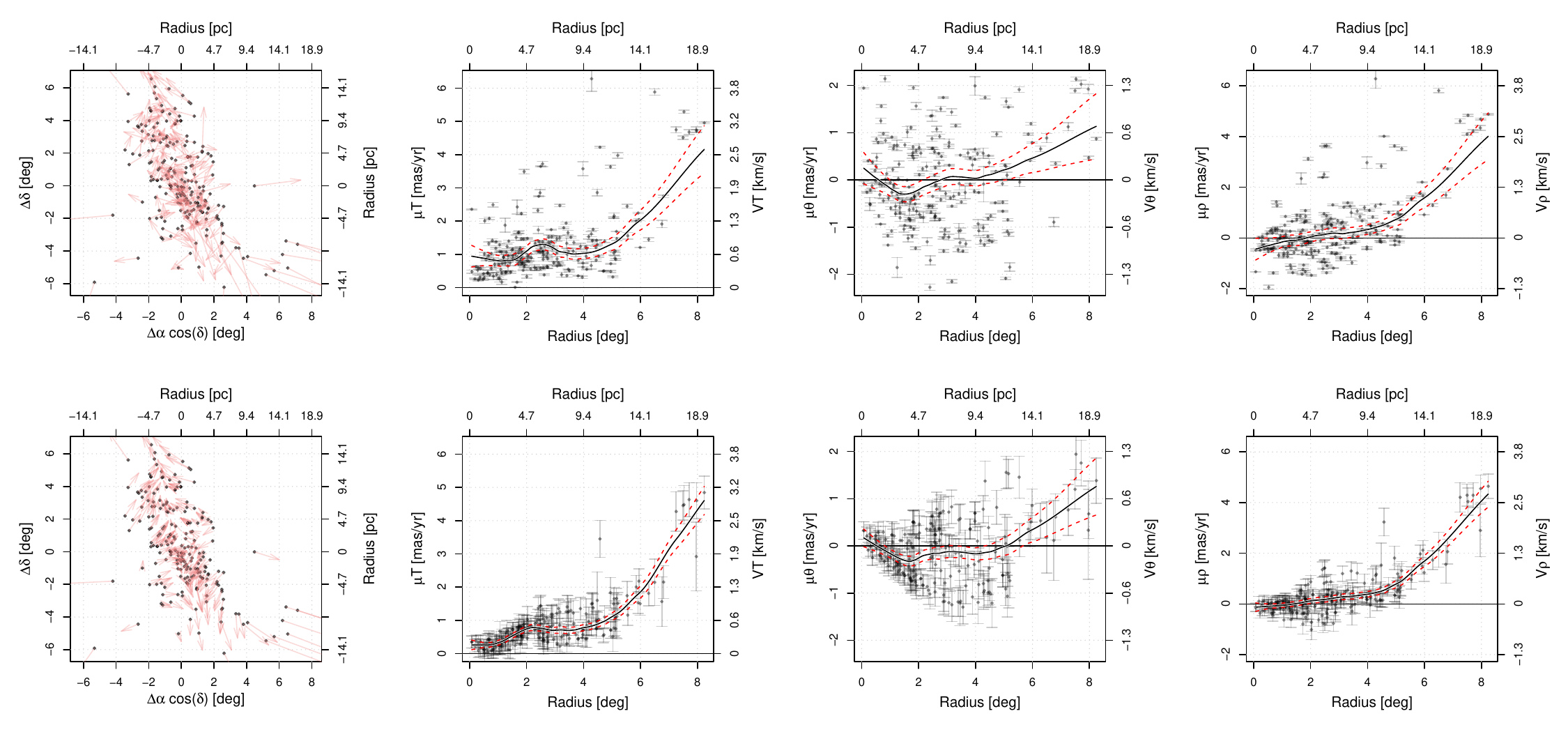}
\caption{Same as Fig.\ \ref{fig:vcurve0}, but for the Platais 8 cluster. }
\label{fig:vcurve43}
\end{figure*}
\begin{figure*}[htb]
\includegraphics[width=\linewidth]{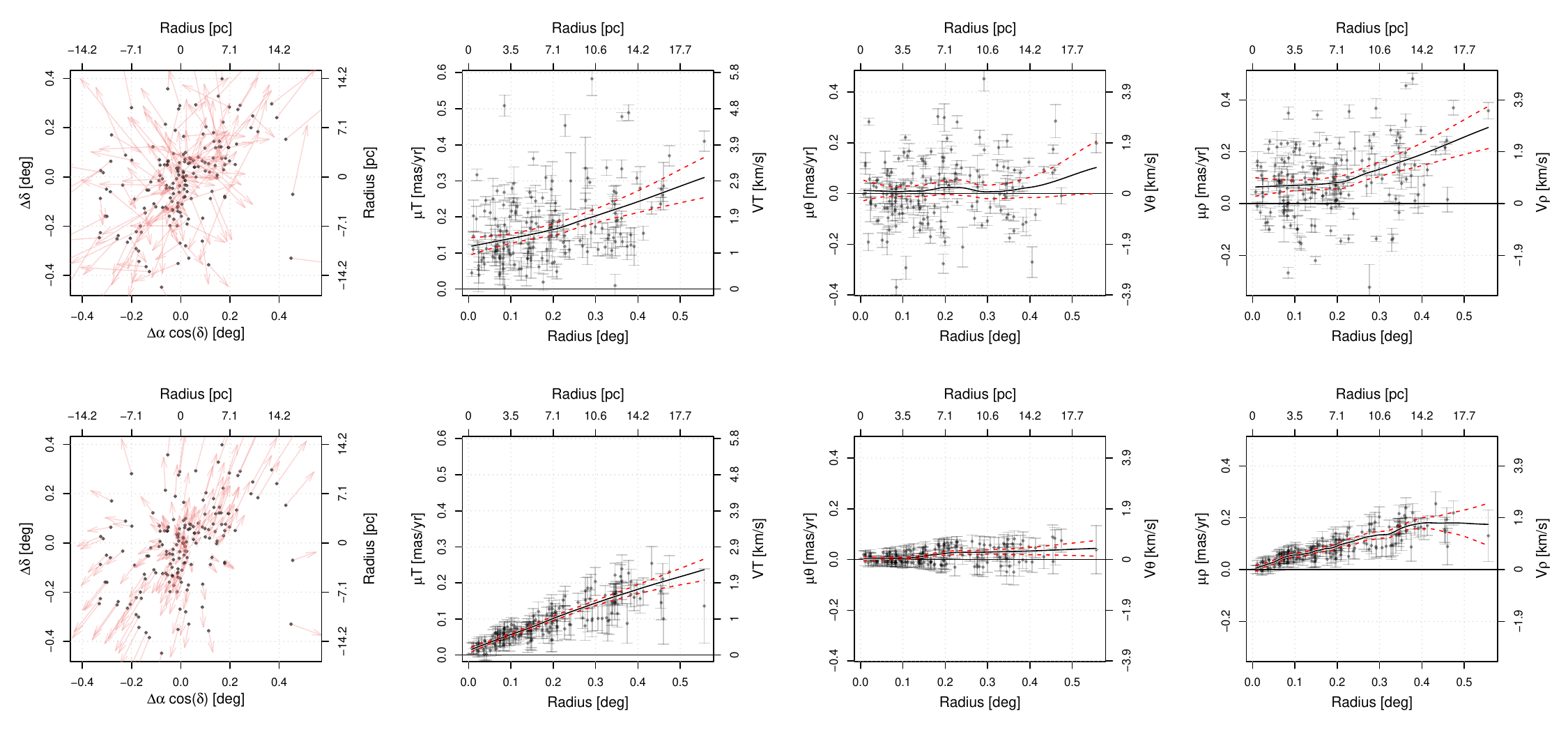}
\caption{Same as Fig.\ \ref{fig:vcurve0}, but for the Roslund 2 cluster.}
\label{fig:vcurve44}
\end{figure*}
\begin{figure*}[htb]
\includegraphics[width=\linewidth]{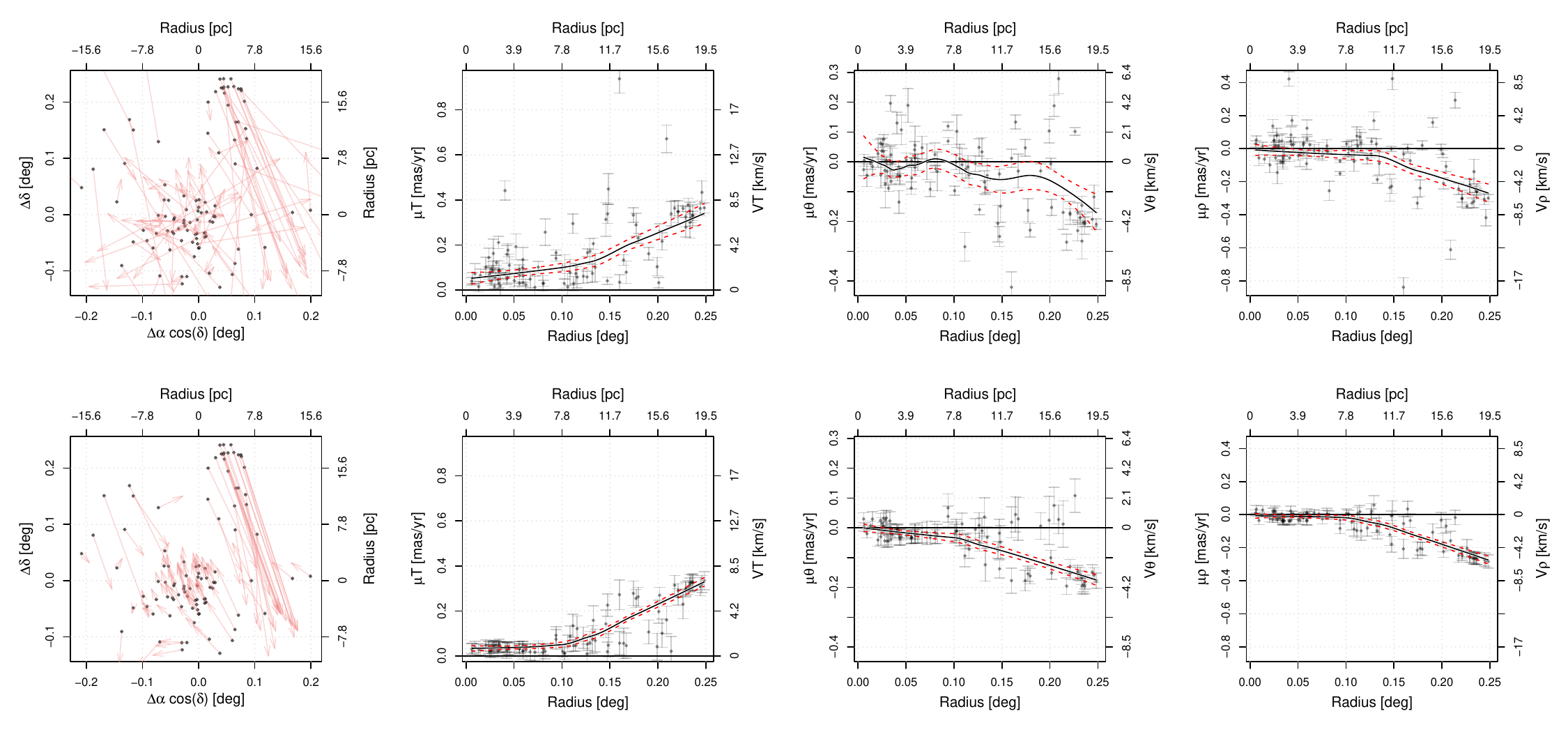}
\caption{Same as Fig.\ \ref{fig:vcurve0}, but for the Ruprecht 41 cluster. }
\label{fig:vcurve47}
\end{figure*}
\begin{figure*}[htb]
\includegraphics[width=\linewidth]{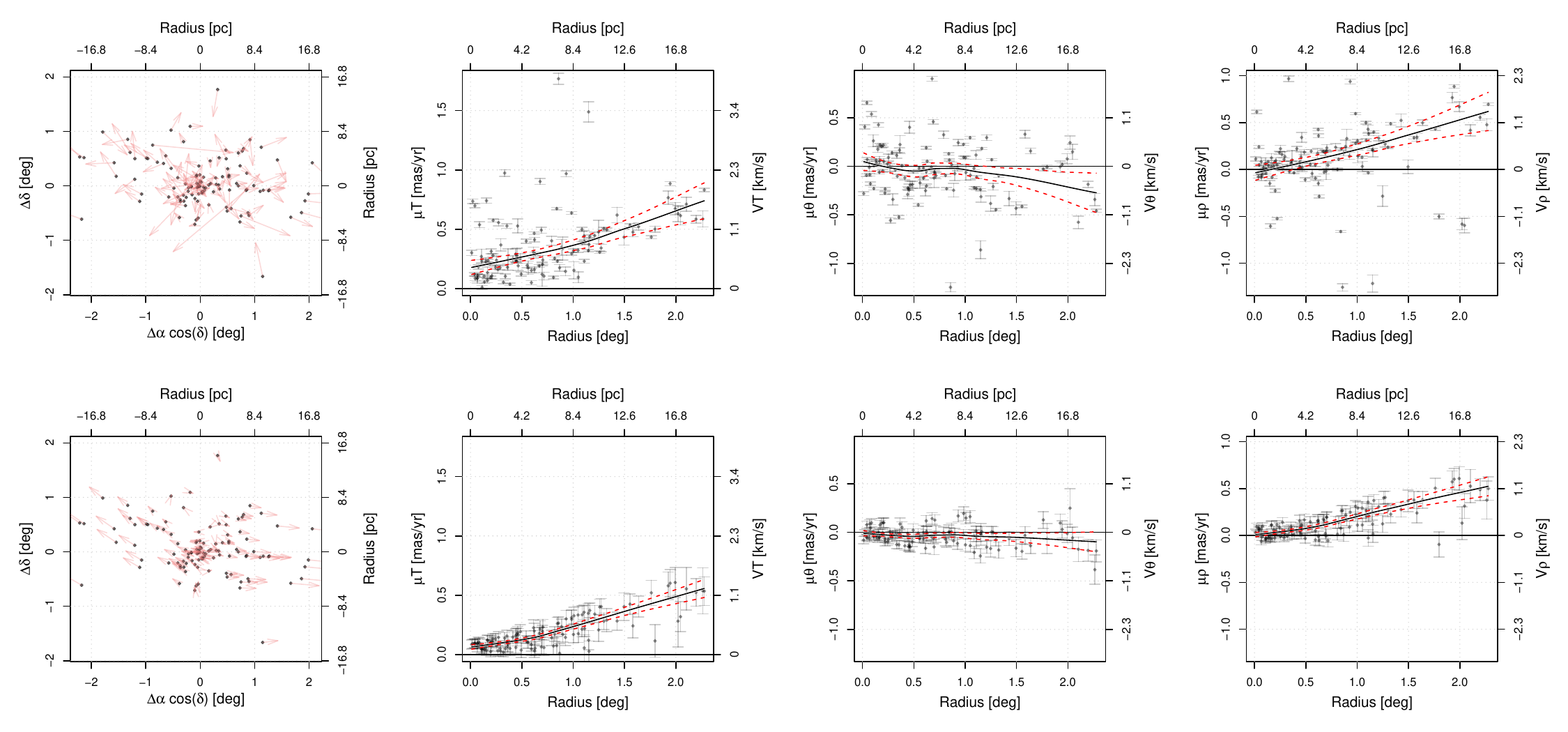}
\caption{Same as Fig.\ \ref{fig:vcurve0}, but for the Ruprecht 98 cluster. }
\label{fig:vcurve48}
\end{figure*}
\begin{figure*}[htb]
\includegraphics[width=\linewidth]{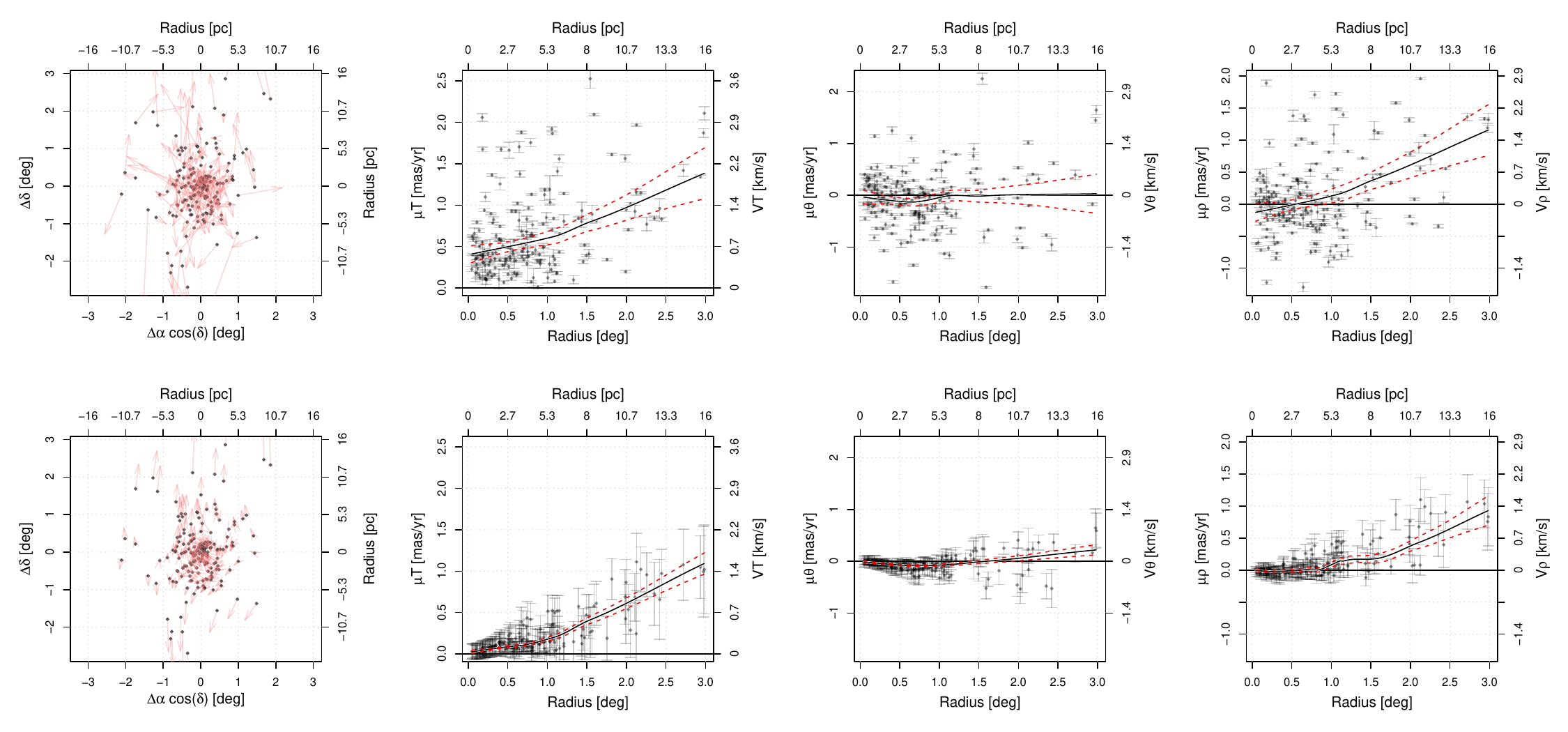}
\caption{Same as Fig.\ \ref{fig:vcurve0}, but for the Ruprecht 147 cluster. }
\label{fig:vcurve45}
\end{figure*}
\begin{figure*}[htb]
\includegraphics[width=\linewidth]{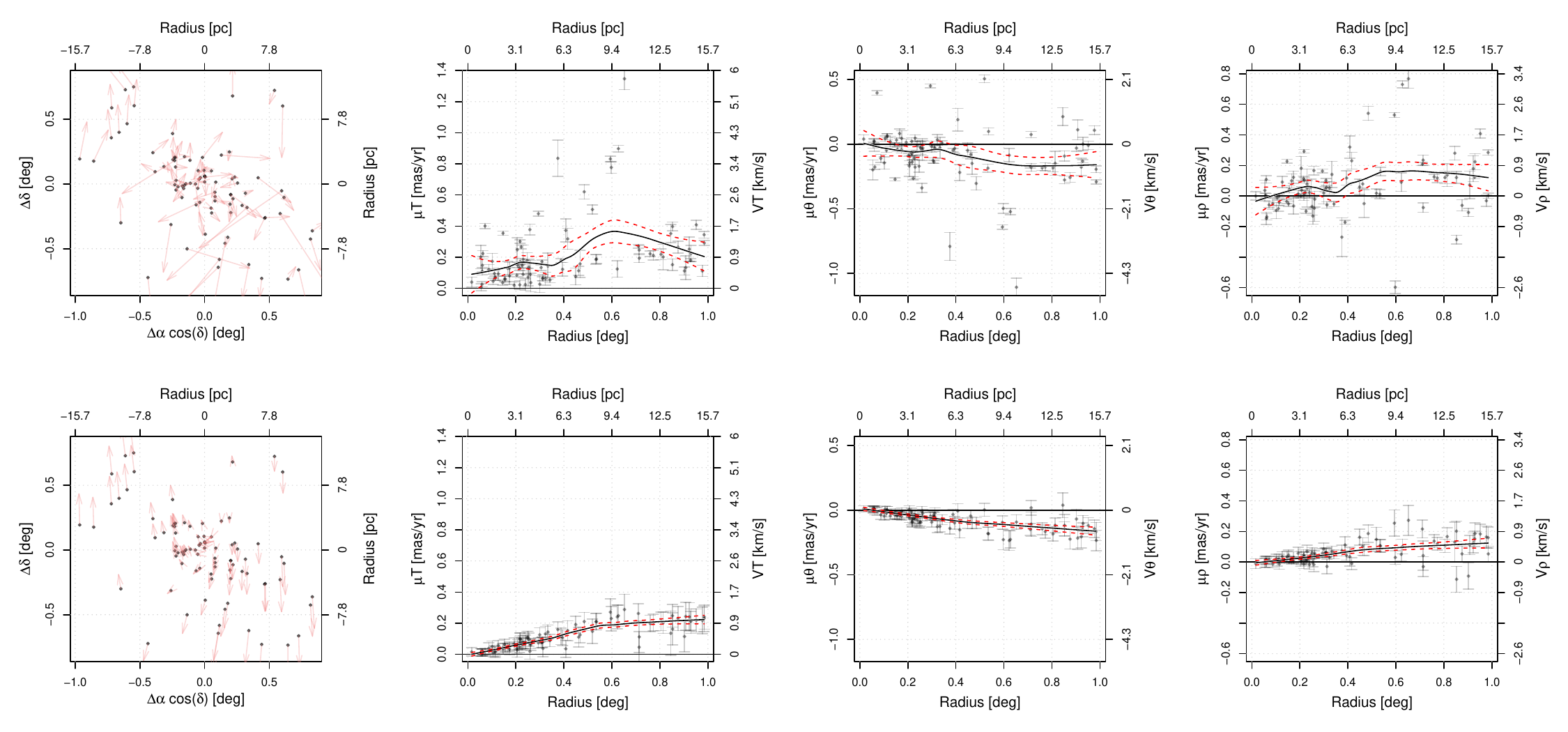}
\caption{Same as Fig.\ \ref{fig:vcurve0}, but for the Ruprecht 161 cluster. }
\label{fig:vcurve46}
\end{figure*}
\begin{figure*}[htb]
\includegraphics[width=\linewidth]{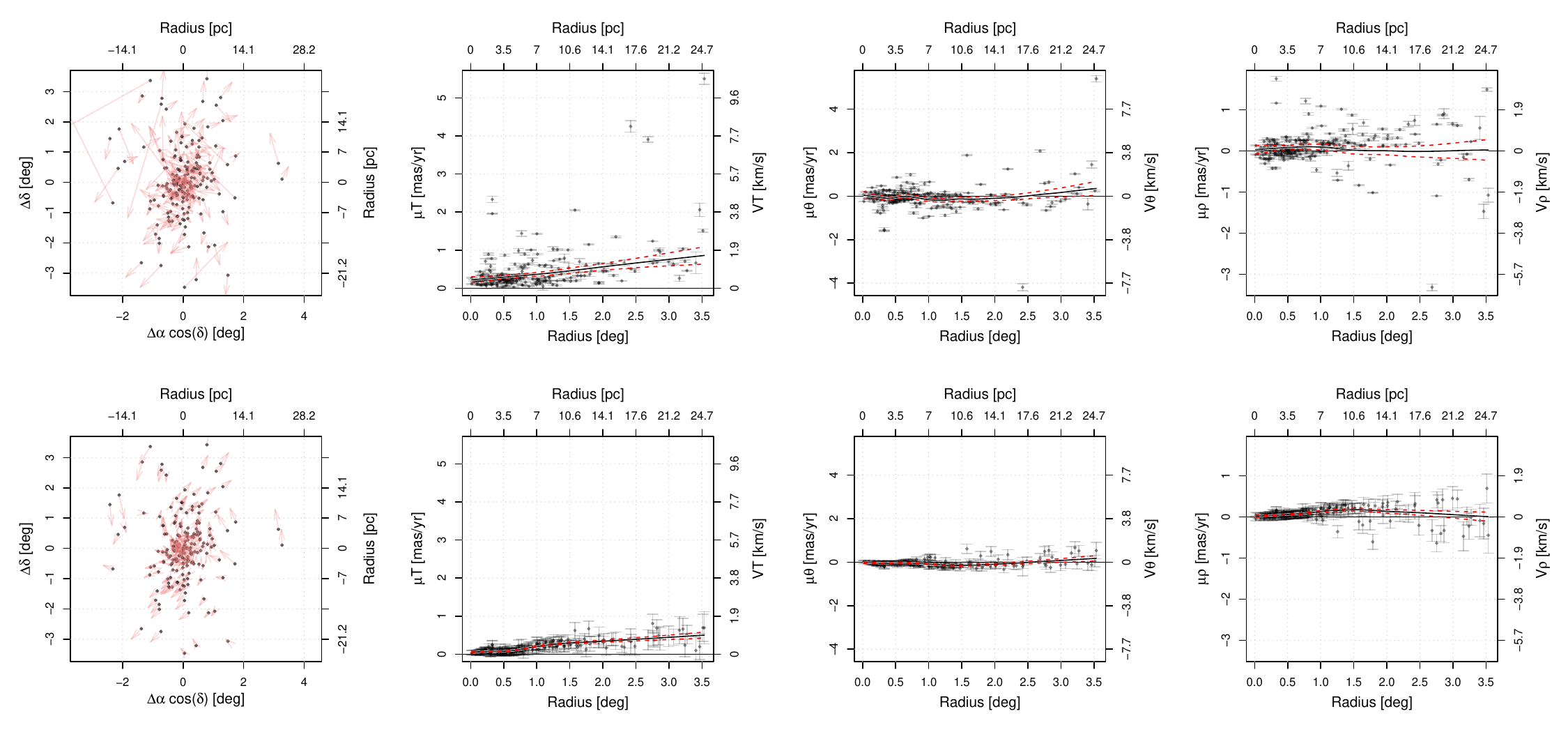}
\caption{Same as Fig.\ \ref{fig:vcurve0}, but for the Stock 1 cluster. }
\label{fig:vcurve49}
\end{figure*}
\begin{figure*}[htb]
\includegraphics[width=\linewidth]{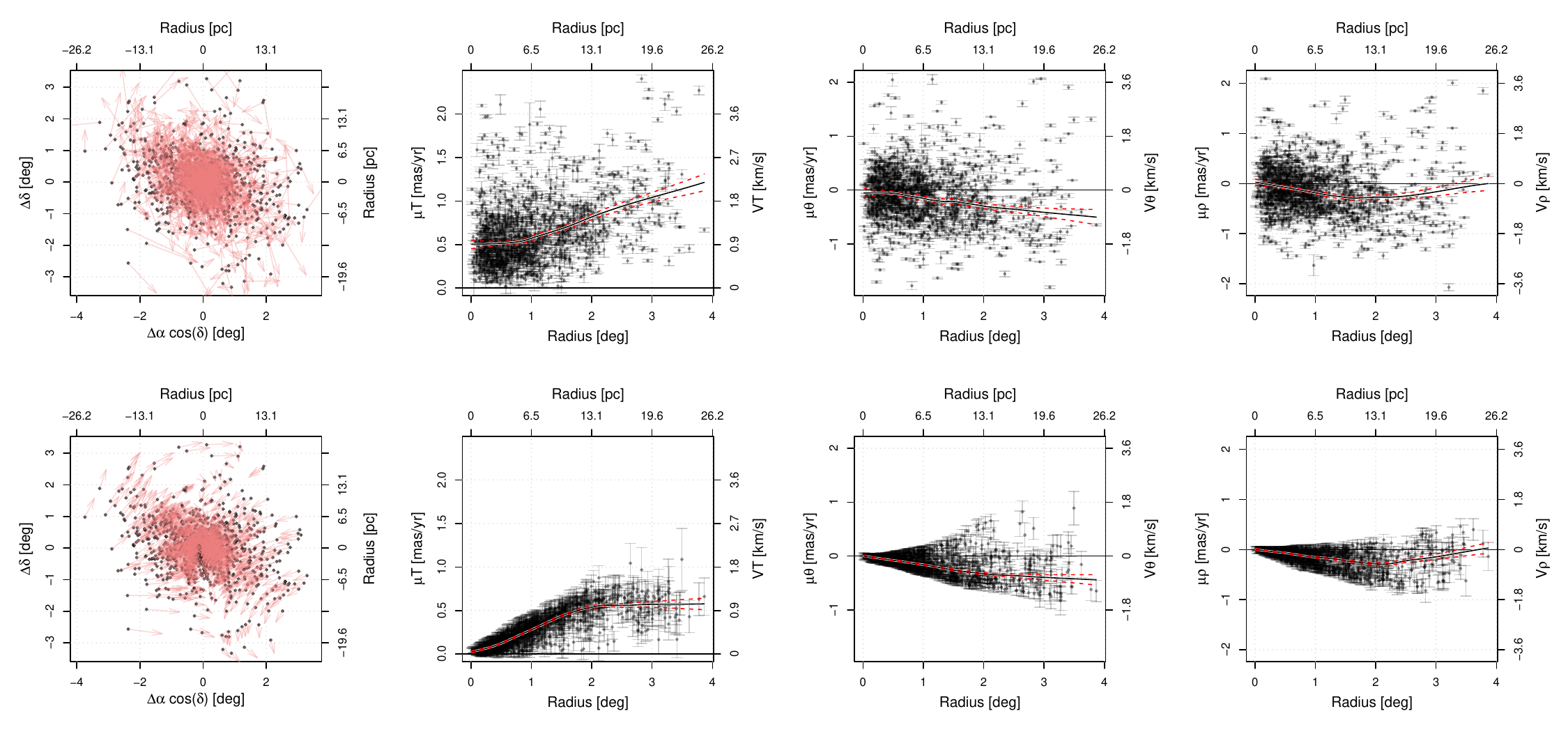}
\caption{Same as Fig.\ \ref{fig:vcurve0}, but for the Stock 2 cluster. }
\label{fig:vcurve50}
\end{figure*}
\begin{figure*}[htb]
\includegraphics[width=\linewidth]{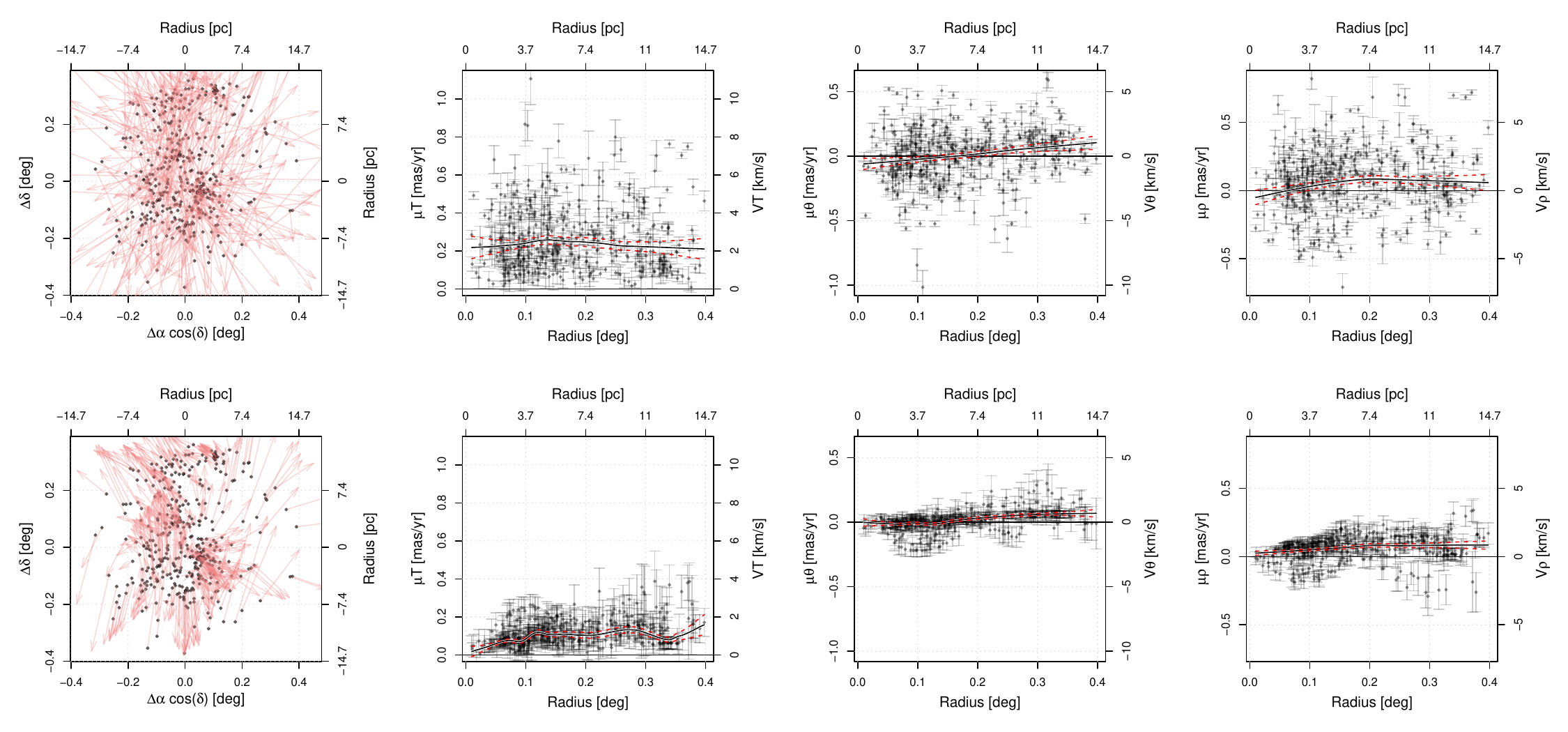}
\caption{Same as Fig.\ \ref{fig:vcurve0}, but for the Stock 8 cluster. }
\label{fig:vcurve51}
\end{figure*}
\begin{figure*}[htb]
\includegraphics[width=\linewidth]{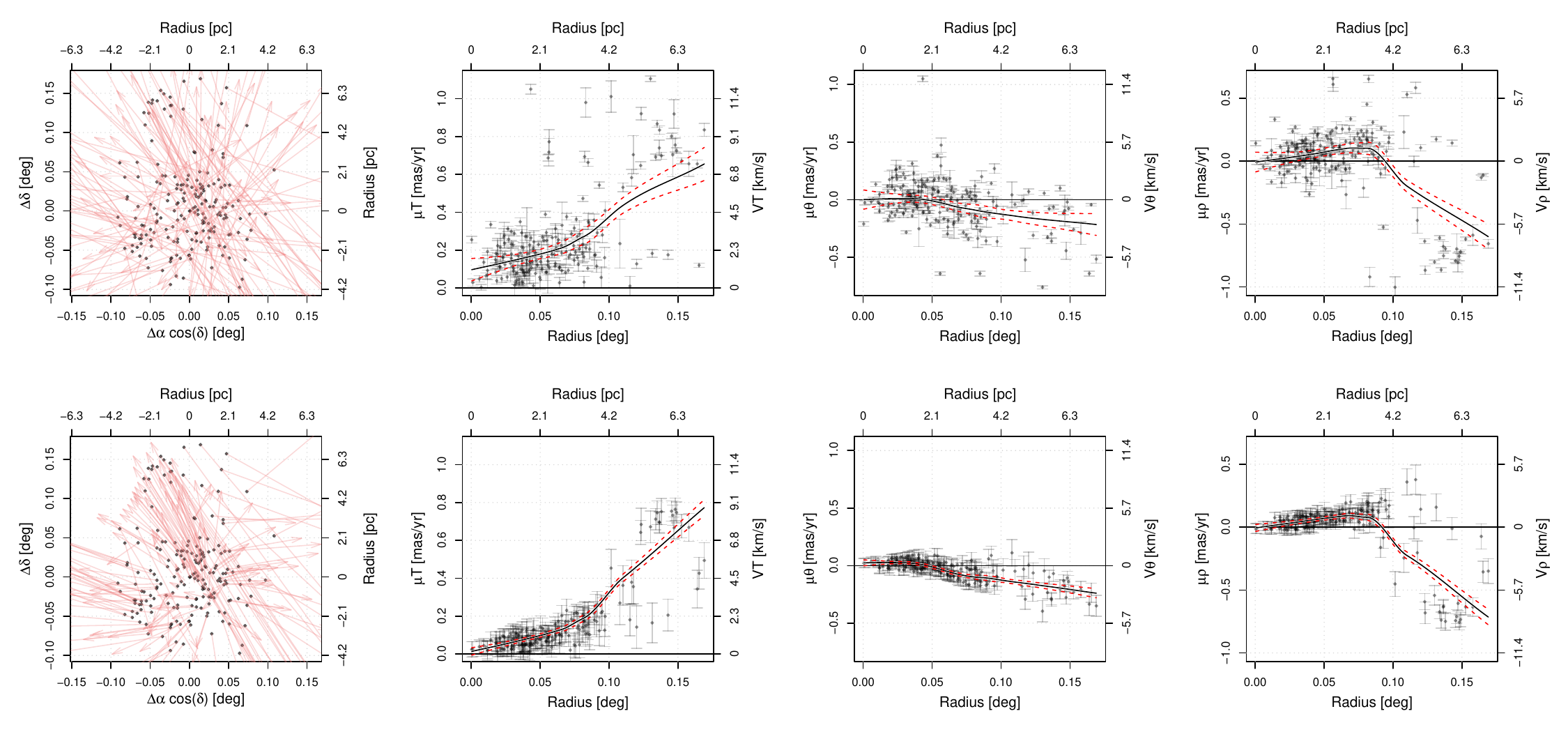}
\caption{Same as Fig.\ \ref{fig:vcurve0}, but for the Trumpler 16 cluster. }
\label{fig:vcurve52}
\end{figure*}
\begin{figure*}[htb]
\includegraphics[width=\linewidth]{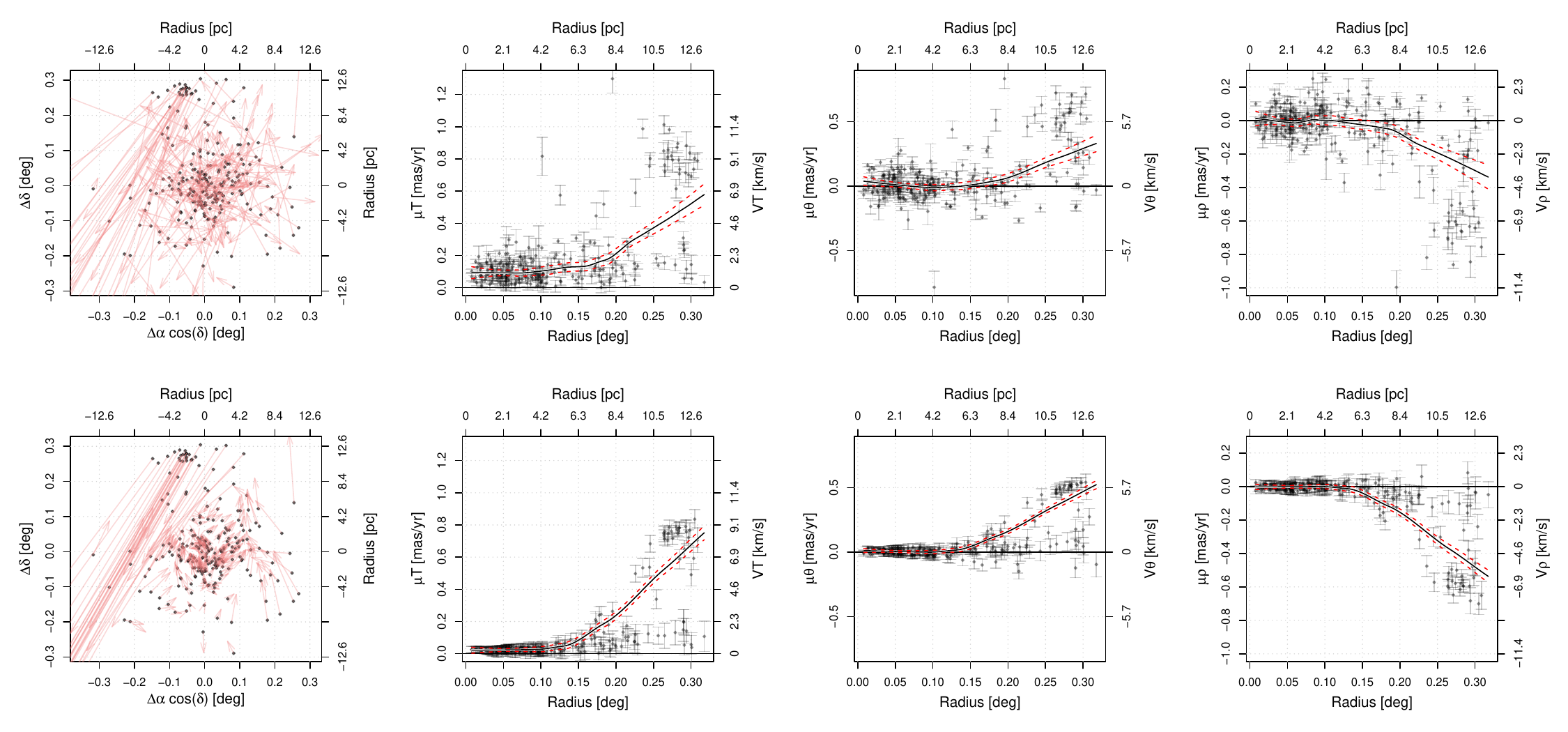}
\caption{Same as Fig.\ \ref{fig:vcurve0}, but for the Trumpler 22 cluster. }
\label{fig:vcurve53}
\end{figure*}
\begin{figure*}[htb]
\includegraphics[width=\linewidth]{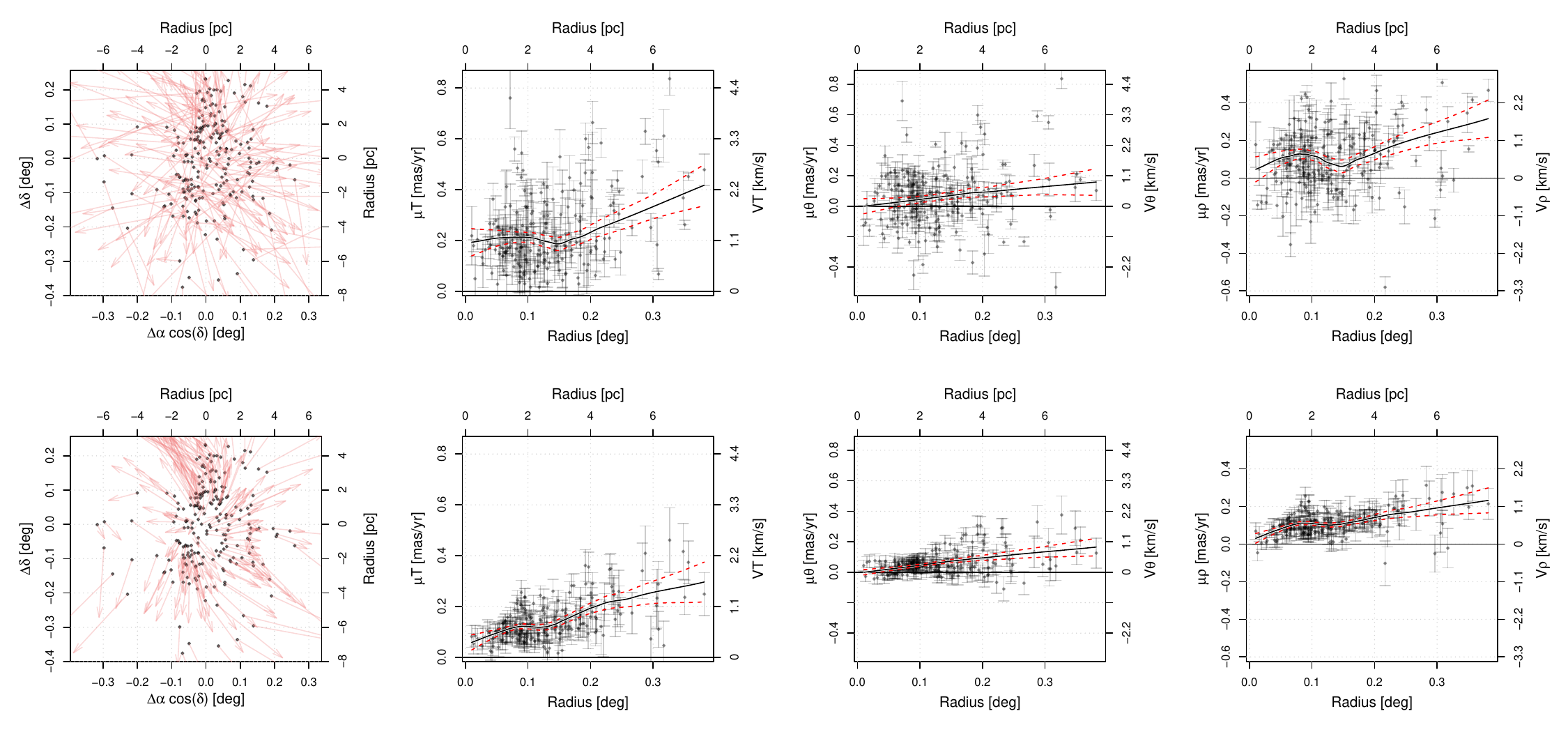}
\caption{Same as Fig.\ \ref{fig:vcurve0}, but for the vdBergh 92 cluster. }
\label{fig:vcurve54}
\end{figure*}
\end{appendix}
\end{document}